\documentstyle[preprint,floats,aps,epsf]{revtex}

\newcommand{\eqnpict}[2]{
   \begin{equation}
   \mbox{\raisebox{-1.625cm}{
      \setlength{\unitlength}{1cm}
      \begin{picture}(0.0,4.0)
         \put(-7.5,1.625){
            \begin{tabular}{cc}
               {$\displaystyle #1$}&{$\displaystyle #2$}
            \end{tabular}
         }
      \end{picture}
   }}
   \end{equation}
}

\newcommand{\gaugeff}[3]{\mbox{$\!\!$\raisebox{-1.65625cm}{
\mbox{
\setlength{\unitlength}{1cm}
\begin{picture}(3.25,3.5)
 \put(0.625,2.125){\mbox{${\displaystyle #1}$}}
 \put(3.0,0.125){\mbox{${\displaystyle #2}$}}
 \put(3.0,3.125){\mbox{${\displaystyle #3}$}}
 \put(-0.25,0.25){
  \mbox{\epsfxsize=3.0cm
   \epsffile{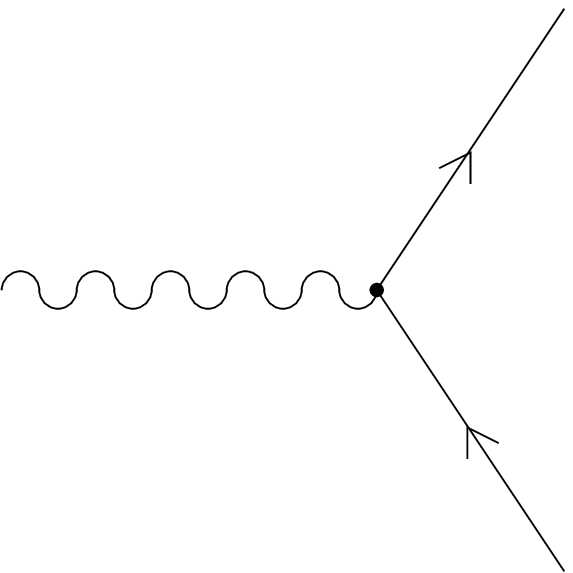}}
 }
\end{picture}
}}}}

\newcommand{\gaugess}[3]{\mbox{$\!\!$\raisebox{-1.65625cm}{
\mbox{
\setlength{\unitlength}{1cm}
\begin{picture}(3.25,3.5)
 \put(0.625,2.125){\mbox{${\displaystyle #1}$}}
 \put(3.0,0.125){\mbox{${\displaystyle #2}$}}
 \put(3.0,3.125){\mbox{${\displaystyle #3}$}}
 \put(-0.25,0.25){
  \mbox{\epsfxsize=3.0cm
   \epsffile{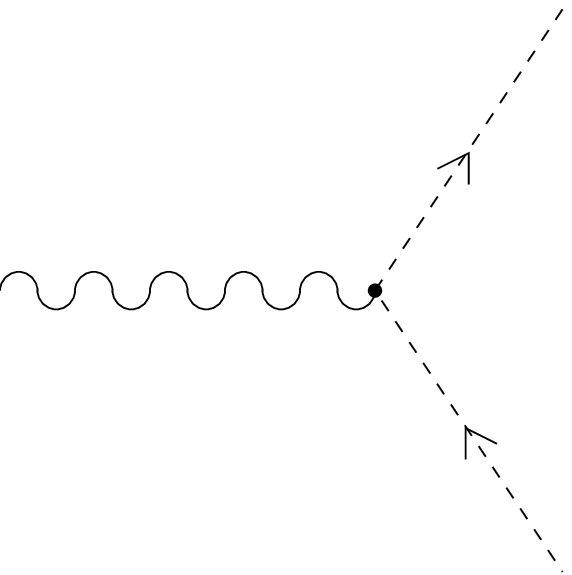}}
 }
\end{picture}
}}}}

\newcommand{\scalarff}[3]{\mbox{$\!\!$\raisebox{-1.65625cm}{
\mbox{
\setlength{\unitlength}{1cm}
\begin{picture}(3.25,3.5)
 \put(0.625,2.125){\mbox{${\displaystyle #1}$}}
 \put(3.0,0.125){\mbox{${\displaystyle #2}$}}
 \put(3.0,3.125){\mbox{${\displaystyle #3}$}}
 \put(-0.25,0.25){
  \mbox{\epsfxsize=3.0cm
   \epsffile{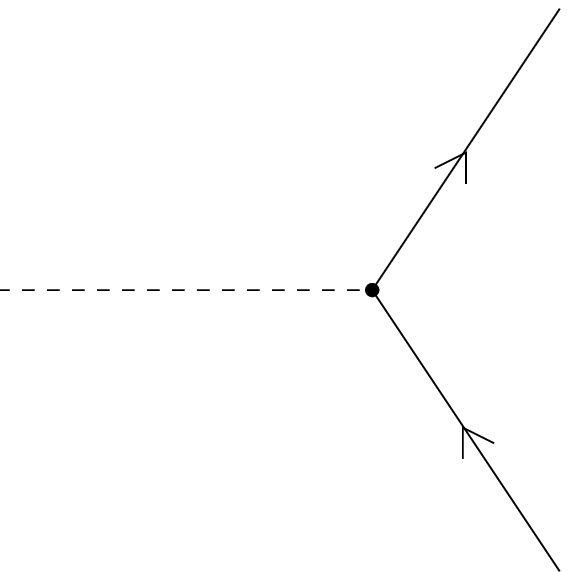}}
 }
\end{picture}
}}}}

\newcommand{\scalarss}[3]{\mbox{$\!\!$\raisebox{-1.65625cm}{
\mbox{
\setlength{\unitlength}{1cm}
\begin{picture}(3.25,3.5)
 \put(0.625,2.125){\mbox{${\displaystyle #1}$}}
 \put(3.0,0.125){\mbox{${\displaystyle #2}$}}
 \put(3.0,3.125){\mbox{${\displaystyle #3}$}}
 \put(-0.25,0.25){
  \mbox{\epsfxsize=3.0cm
   \epsffile{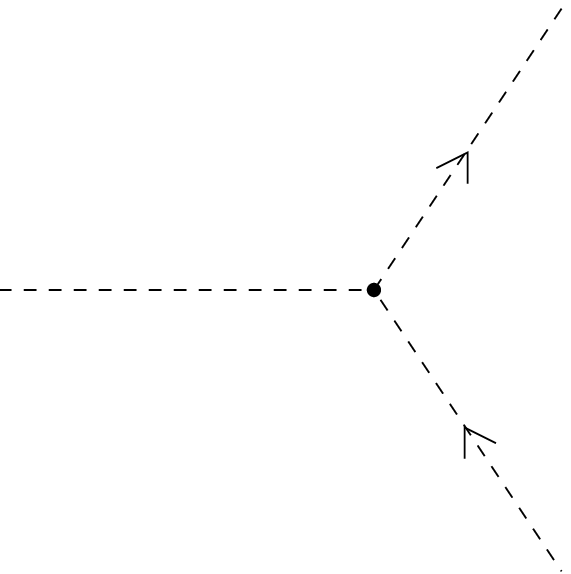}}
 }
\end{picture}
}}}}

\newcommand{\scalarvv}[3]{\mbox{$\!\!$\raisebox{-1.65625cm}{
\mbox{
\setlength{\unitlength}{1cm}
\begin{picture}(3.25,3.5)
 \put(0.625,2.125){\mbox{${\displaystyle #1}$}}
 \put(3.0,0.125){\mbox{${\displaystyle #2}$}}
 \put(3.0,3.125){\mbox{${\displaystyle #3}$}}
 \put(-0.25,0.25){
  \mbox{\epsfxsize=3.0cm
   \epsffile{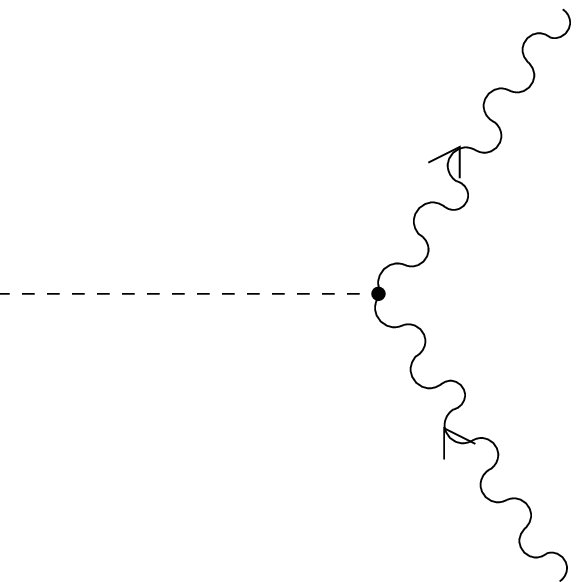}}
 }
\end{picture}
}}}}

\newcommand{\overlaystuff}[4]{\mbox{$\!\!$\raisebox{-1.65625cm}{
\mbox{\setlength{\unitlength}{1cm}
\begin{picture}(4.25,3.5)
 \put(0.0,1.65625){\mbox{{#1}}}
 \put(0.25,1.375){\vector(1,0){1.0}}
 \put(0.625,1.0){\mbox{${\displaystyle #2}$}}
 \put(2.25,1.0){\vector(2,-3){0.5}}
 \put(2.0,0.5){\mbox{${\displaystyle #3}$}}
 \put(2.25,2.5){\vector(2,3){0.5}}
 \put(2.0,2.75){\mbox{${\displaystyle #4}$}}
\end{picture}}
}}}

\begin{document}

\title{
\mbox{}\hfill\mbox{\rm OCIP/C-96-16}\\
\mbox{}\hfill\mbox{\rm UQAM-PHE/96-06}\\
CHARGED HEAVY    LEPTON PRODUCTION   IN SUPERSTRING  INSPIRED  E6 
MODELS} 
%
%
%
\author{M.M.~BOYCE, M.A.~DONCHESKI}
\address{Carleton University, Physics Department, 1125 Colonel By Drive,
Ottawa,\\   Ontario,   K1S  5B6,   Canada}
\author{ H.~K\"{O}NIG  }  \address{UQAM, D\'epartement   de Physique, CP
8888, Succ.  Centre Ville, Montr\'eal,\\ Qu\'ebec, H3C 3P8, Canada}
\maketitle
\begin{abstract}

  The  possibility  of studying  superstring inspired \mbox{${\rm E}_6$}
phenomenology at  high energy hadron  colliders is investigated.  A very
simple low energy rank-5 Supersymmetric (N=1) model is considered, which
consists    of        three    scalar-Higgses,  $H^0_{i=1,2,3}\,$,   two
charged-Higgses, $H^\pm\,$,  one pseudo-scalar-Higgses,  $P^0\,$, and an
extra  vector boson, the  $Z^\prime$.   The production  of charged heavy
leptons  pairs,    $L^+L^-\,$, by   gluon-gluon  fusion  and   Drell-Yan
mechanisms is discussed.  For gluon-gluon  fusion an enhancement in  the
parton level cross-section is expected due to the heavy (s)fermion loops
which couple to the gluons.  This mechanism is expected to dominate over
Drell-Yan for $L^+L^-$ invariant masses above the $Z^\prime$ mass.

\end{abstract}

\clearpage

\section{Introduction}
\label{sec-intro}

In this paper the production of charged heavy leptons pairs, $L^+L^-\,$,
{\it via} superstring inspired \mbox{${\rm E}_6$} models, at high energy
hadron colliders  will be  investigated  \cite{kn:thesisb,kn:BoyceB}.  A
general overview of \mbox{${\rm E}_6$} models will be presented in which
several simplifying assumptions  will be made in  order  to restrict the
$L^+L^-\,$ production computation to a manageable one.  In particular, a
low energy rank-5 model, arising from E$_6\,$, will be constructed, with
this   specific    application in    mind.   The  $L^+L^-\,$  production
cross-sections will then be computed  followed by a discussion and  then
finally conclusions.

Many aspects of the  rank-5 models, that   will be considered  here, are
covered in  the literature.  Unfortunately,  when trying to  extract the
particular model dependent information,  needed for $L^+L^-$ production,
it appeared that the existing literature was not consistent.  Therefore,
it was felt that in order to avoid any ambiguities that the model should
be carefully reconstructed from  the  ground up.  When  constructing the
model careful attention was paid to being as consistent as possible with
the  literature  concerning: factors  of  two, hypercharge  conventions,
signs,  ambiguous notational subtleties,  etc.  Much  of the analysis of
the model was   done by using Mathematica~\cite{kn:Wolfram} to  generate
the  various   couplings,   mass  matrices, etc.,   directly   from  the
superpotential.  This enabled  easy  comparison with  various literature
sources
\cite{kn:Hewett,kn:EllisI,kn:EllisB,kn:EllisA,kn:BargerA,kn:Gunion}.
Differing  conventions and normalizations  aside,  the most  significant
problem  arose with   the  charged-Higgs,   Eq.~(\ref{eq:charged}),  and
pseudo-scalar-Higgs, Eq.~(\ref{eq:neutral}),   mass terms; a  factor two
was missing  in front of the  $\sin\beta\cos\beta$ terms,  $op$.  $cit$.
For  example, in the case  of the pseudo-scalar-Higgs the aforementioned
authors  disagree to by an  overall  factor of  two in their mass-mixing
matrices but not in their eigenvalues.  As a result, the analysis of the
mass  constraints  in   the  Higgs  sector~\cite{kn:Gunion}   had  to be
re-evaluated,     Figs.~\ref{fig:mztwo}-\ref{fig:mhzerob}.  In  addition
Appendix~\ref{sec-appc} contains a  summary  of the couplings   used for
$L^+L^-$ production  which, in general,  could not be  obtained from the
literature.

\section{Superstring Inspired ${\rm E}_6$ Models}

  The    \mbox{$\mbox{${\rm      SU}({3})_{{\rm  c}}$}\otimes\mbox{${\rm
SU}({2})_{{\rm L}}$}\otimes\mbox{${\rm U}({1})_{{\rm   Y}}$}$}  standard
model (SM)  is a very successful  model \cite{kn:STM}.  It  has thus far
withstood a lot of rigorous  experimental testing; however, despite  its
success the SM has many of problems:
\begin{itemize}
\item no unification of the forces
\item gauge hierarchical and fine tuning problems
\item three generations of quarks and leptons for no particular reason
\item too many parameters to be extracted from experiment
\end{itemize}

  Some of the earlier attempts at unification tried  to unify the strong
and   electroweak     forces   by   embedding  the    \mbox{$\mbox{${\rm
SU}({3})_{{\rm             c}}$}\otimes\mbox{${\rm        SU}({2})_{{\rm
L}}$}\otimes\mbox{${\rm  U}({1})_{{\rm  Y}}$}$}  structure into   higher
groups,  such  as SU(5) and SO(10).    These  ``grand unified theories''
\cite{kn:RosnerA},  or  GUT's,  were   only partially   successful.  The
simplest of  the GUT's  was SU(5) which  seemed  promising  at  the time
because it predicted the ratio  of the \mbox{${\rm SU}({2})_{{\rm  w}}$}
and \mbox{${\rm U}({1})_{{\rm  em}}$} couplings and  the proton lifetime
\cite{kn:Ross}.   However,   the  ordinary SU(5)  GUT   is   no longer a
possibility because more  refined experimental  measurements are now  in
disagreement  with its predictions  for   the couplings and the   proton
lifetime \cite{kn:Hewett}.  In addition this  simple model had too  many
parameters and  no explanation for family  replication.  The next likely
candidate group was  SO(10) \cite{kn:Fritzsch},  although the three  (or
more) copies  of the generational structure  still had to be inserted by
hand.

   Difficulties with the SM  and  GUT models concerning gauge  hierarchy
and  fine tuning   problems    led  to theoretical  remedies   such   as
technicolour  and  supersymmetry  (SUSY) \cite{kn:Mohapatra}.   The most
appealing  of   these  theories  was   SUSY \cite{kn:SUSY},    which had
generators that   related particles of   different  spin  in  the   same
supermultiplet.  The locality  of these generators leads to supergravity
models.  SUSY  (and its extended versions) however,  did not have enough
room for all of the  SM particles~\cite{kn:Ross}.  To solve this problem
direct  product structures were    made with SUSY and   Yang-Mills gauge
groups.    These structures are now   commonly  referred to as  ``SUSY''
models~\cite{kn:Haber,kn:Nilles}.  Of course the price paid for this was
a large particle spectrum   (at least twice  that  of  the SM) and   the
problem of family replication still remained.

   In the early  1970's some interest was  sparked in \mbox{${\rm E}_6$}
as a GUT when it  was discovered that all  the then known generations of
fermions  could   be  placed   in   a    single  {\bf 27}    dimensional
representation.   This  (``topless'') model  \cite{kn:Topless} was quite
popular because the newly discovered  $\tau$ lepton and $b$ quark  could
also be fitted neatly into the {\bf  27}; there was  no need for a third
generation.  However  this   model was quickly  disallowed,  as   it was
experimentally \cite{kn:Topless} shown that the  $\tau$ and $b$ belonged
to a third generation, and the idea of \mbox{${\rm E}_6$} as a GUT died.

\begin{figure}[ht]
$${\bf            27}=\left\{\fbox{$\left(\begin{array}{c}           u\\
  d\end{array}\right)_{\rm    L}       \left(\begin{array}{c}    \nu_e\\
  e\end{array}\right)_{\rm  L}  u_{\rm L}^c d_{\rm  L}^c e_{\rm L}^c$}\;
  \nu_{e_{\rm       L}}^c    d_{\rm      L}^\prime    d_{\rm  L}^{\prime
  c}\left(\begin{array}{c}                                \nu_e^\prime\\
  e^\prime\end{array}\right)_{\rm    L}\left(\begin{array}{c} e^\prime\\
  \nu_e^\prime\end{array}\right)_{\rm L}^c\nu_{e_{\rm  L}}^{\prime\prime
  c}\right\}$$
$${\bf           27}=\left\{\fbox{$\left(\begin{array}{c}            c\\
  s\end{array}\right)_{\rm        L}\left(\begin{array}{c}     \nu_\mu\\
  \mu\end{array}\right)_{\rm L} c_{\rm L}^c s_{\rm L}^c\mu_{\rm L}^c$}\;
  \nu_{\mu_{\rm L}}^c     s_{\rm         L}^\prime s_{\rm     L}^{\prime
  c}\left(\begin{array}{c}                              \nu_\mu^\prime\\
  \mu^\prime\end{array}\right)_{\rm             L}\left(\begin{array}{c}
  \mu^\prime\\   \nu_\mu^\prime\end{array}\right)_{\rm L}^c\nu_{\mu_{\rm
  L}}^{\prime\prime c}\right\}$$
$${\bf             27}=\left\{\fbox{$\left(\begin{array}{c}          t\\
  b\end{array}\right)_{\rm       L}\left(\begin{array}{c}     \nu_\tau\\
  \tau\end{array}\right)_{\rm  L}  t_{\rm  L}^c  b_{\rm  L}^c  \tau_{\rm
  L}^c$}\;  \nu_{\tau_{\rm L}}^c   b_{\rm  L}^\prime b_{\rm   L}^{\prime
  c}\left(\begin{array}{c}\nu_\tau^\prime\\
  \tau^\prime\end{array}\right)_{\rm            L}\left(\begin{array}{c}
  \tau^\prime\\                   \nu_\tau^\prime\end{array}\right)_{\rm
  L}^c\nu_{\tau_{\rm L}}^{\prime\prime c}\right\}$$
\caption[\mbox{${\rm  E}_6$} particle content]{\footnotesize \mbox{${\rm
   E}_6$} particle content.$^a$ The SM particles  are shown in the boxes
   on  the left and their  ``exotic'' counterparts outside  the boxes on
   the right.  Although  the exotics are labeled  in  away that suggests
   they have  the same quantum   numbers as the non-exotics, in  general
   they  need not.  The  labeling for these  particles in the literature
   has not been  settled upon and  varies quite significantly from paper
   to  paper \cite{kn:Hewett}.  Here the  labeling  scheme was chosen to
   reflect a specific \mbox{${\rm  E}_6$} model that will be constructed
   in  this paper.   In   particular, all  the  exotics will   carry the
   ``expected'' quantum numbers as their   non-exotic counter parts  do,
   with the exception being L=0 for the primed and double primed ones.}
\vspace{3mm}
   {\footnoterule\footnotesize $\mbox{}^a$Note:     Embedded in the {\bf
   27}'s   is   the  symmetry  group   \mbox{${\rm SU}({2})_{{\rm  I}}$}
   \cite{kn:Hewett} due to an ambiguity in the particle assignments
   {\tiny
   $
    \left\{
      \left(\begin{array}{@{}c@{}}\nu_l\\ l\end{array}\right)_{\rm L}
       d_{\rm L}^c
    \right\}
    \;\Longleftrightarrow\;
    \left\{
      \left(\begin{array}{@{}c@{}}\nu_l^\prime\\ 
            l^\prime\end{array}\right)_{\rm L}
       d_{\rm L}^{\prime c}
    \right\}
   $}
   and
   $
    \{\nu_{l_{\rm L}}^c\}
    \;\mbox{\tiny $\Longleftrightarrow$}\;
    \{\nu_{l_{\rm L}}^{\prime\prime c}\}\,
   $
   [{\it  cf}. Fig.~\ref{fig:Wilson}(d)].  This ambiguity can  easily be
   seen       {\it         via}     the         decomposition       {\bf
   27}=$\sum_\oplus$(SO(10),SU(5)).
   }
\label{fig:esix}
\end{figure}
  In  late   1984  Green and  Schwarz    \cite{kn:Green} showed  that 10
dimensional string theory  is anomaly free  if its gauge group is either
\mbox{${\rm E}_8\otimes{\rm E}_8^\prime$} or  SO(32). The group that had
received the   most     attention   was   \mbox{${\rm    E}_8\otimes{\rm
E}_8^\prime$} as it led to chiral fermions,  similar to those in the SM,
whereas SO(32) did not.  Furthermore, it was shown that compactification
down to 4 dimensions (assuming N=1 SUSY) can  lead to \mbox{${\rm E}_6$}
as an ``effective'' GUT group.  Each family of SM  particles now sits in
its own {\bf 27}, Fig.~\ref{fig:esix}.  The  generational problem may be
solved because  it   is expected  that  any reasonable  compactification
scheme should generate the appropriate number of copies of the {\bf 27}.
For instance  in  a  Calabi-Yau  compactification scheme \cite{kn:Kaku},
$$\mbox{${\rm    E}_8\otimes{\rm    E}_8^\prime$}\;\longrightarrow\;{\rm
SU(3)}\otimes{\rm   E}_6\otimes{\rm   E}_8^\prime\;,$$ the  number    of
generations  is  related to the topology   of the compactified space.  A
further   assertion that  the   matter fields  remain supersymmetrically
degenerate ensures proper management of any  gauge hierarchical and fine
tuning problems.     It is   assumed   that  the hidden    sector, ${\rm
E}_8^\prime$,  which couples to  the matter fields of \mbox{${\rm E}_6$}
by  gravitational interactions will  provide a mechanism for lifting the
degeneracy.

  So the inspiration for using  \mbox{${\rm E}_6$} is  that if it proves
to be a possible GUT  then it opens up  the possibility of finding a TOE
({\bf T}heory {\bf O}f {\bf E}verything).  However, it should be pointed
out  that \mbox{${\rm E}_6$} is not  the only possible stop $en$ $route$
to the SM, but it is the most  studied~\cite{kn:Hewett}.  It is for this
reason  that the  low  energy  phenomenology resulting from  \mbox{${\rm
E}_6$} will be considered.

\subsection{\mbox{${\rm E}_6$} Phenomenology}
\subsubsection{An extra \mbox{${\rm Z}_{\rm E}\,$}}
  In order to produce the SM gauge structure, \mbox{${\rm E}_6$} must be
broken.  Also, to handle any hierarchical and fine tuning problems, SUSY
must be preserved \cite{kn:Kaku}.   This restriction makes the task more
difficult, using  most na\mbox{\"{\i}ve} breaking schemes.  The solution
to the problem was found by using a Wilson-loop mechanism \cite{kn:Kaku}
over the non-simply-connected-compactified-string-manifold to factor out
the  various subgroups   of \mbox{${\rm   E}_6$}.  Fig.~\ref{fig:Wilson}
shows some of the possible, popular,  rank 5 and  rank 6 groups that can
be produced by this scheme.
\begin{figure}[ht]
\begin{center}
 \begin{tabular}{lc@{$\,$}c@{$\,$}l}
    (a) &  ${\rm  E}_6$   &  $\longrightarrow$    &   \mbox{$\mbox{${\rm
        SU}({3})_{{\rm       c}}$}\otimes\mbox{${\rm      SU}({2})_{{\rm
        L}}$}\otimes\mbox{${\rm U}({1})_{{\rm
        Y}}$}$}$\otimes$\mbox{${\rm U}({1})_{{\rm Y_E}}$}
        \\
    (b) & ${\rm E}_6$ & $\longrightarrow$ & 
            \raisebox{-4.5mm}{$\left.\mbox{\begin{tabular}{@{}l}
    	       	      SO(10)$\otimes$U(1)$_\psi$ \\
    	       	      \mbox{
                            \setlength{\unitlength}{1mm}
                            \begin{picture}(6,7)
                               \put(0,7){\line(0,-1){6}}
                               \put(0,1){\vector(1,0){6}}
                            \end{picture}
                      } SU(5)$\otimes$U(1)$_\chi$
    	      	 \end{tabular}}\right\}$
    	       	$\stackrel{\mbox{\scriptsize ER5M}}{\longrightarrow}$
            \begin{tabular}[t]{@{}c}
    	       	 \mbox{$\mbox{${\rm                       SU}({3})_{{\rm
                  c}}$}\otimes\mbox{${\rm                 SU}({2})_{{\rm
                  L}}$}\otimes\mbox{${\rm                  U}({1})_{{\rm
                  Y}}$}$}$\otimes$U(1)$_\theta$\\
              {\tiny ($i.e.$,
              U(1)$_\psi\otimes$U(1)$_\chi\rightarrow$U(1)$_\theta$
              in the large VEV limit.)}
            \end{tabular}
          }                                                      \\
    (c) & \mbox{${\rm E}_6$} & $\longrightarrow$ &   
            $\mbox{${\rm SU}({3})_{{\rm c}}$}\otimes\mbox{${\rm
                   SU}({2})_{{\rm L}}$}\otimes\mbox{${\rm SU}({2})_{{\rm
                   R}}$}\otimes
            \mbox{\hspace{-14.5mm}}
            \underbrace{\mbox{${\rm U}({1})_{{\rm
                        L}}$}\otimes\mbox{${\rm U}({1})_{{\rm R}}$}}_{
            \mbox{\hspace{16mm}}
            \raisebox{3mm}{\rm\scriptsize ER5M}
            \mbox{
                  \setlength{\unitlength}{1mm}
                  \begin{picture}(6,7)
                     \put(0,7){\line(0,-1){6}}
                     \put(0,1){\vector(1,0){6}}
                  \end{picture}
            }\mbox{${\rm U}({1})_{{\rm V=L+R}}$}
          }$                                                     \\
    (d) & \mbox{${\rm E}_6$} & $\longrightarrow$ & 
            $\mbox{${\rm SU}({3})_{{\rm c}}$}\otimes\mbox{${\rm
                   SU}({2})_{{\rm L}}$}\otimes
                   \mbox{${\rm U}({1})_{{\rm Y}}$}\otimes
            \mbox{\hspace{-5.5mm}}
            \underbrace{\mbox{${\rm SU}({2})_{{\rm I}}$}
                        \otimes{\rm U}(1)^\prime}_{
            \mbox{\hspace{8mm}}
            \raisebox{3mm}{\rm\scriptsize ER5M}
            \mbox{
                  \setlength{\unitlength}{1mm}
                  \begin{picture}(6,7)
                     \put(0,7){\line(0,-1){6}}
                     \put(0,1){\vector(1,0){6}}
                  \end{picture}
             } \mbox{${\rm SU}({2})_{{\rm I}}$}
          }$
 \end{tabular}
\end{center}
 \caption[\mbox{${\rm E}_6$} Wilson-loop-breaking schemes]{\footnotesize
   \mbox{${\rm  E}_6$} Wilson-loop-breaking  schemes   \cite{kn:Hewett}.
   (a)  shows   a  rank-5  model  and    (b) through (d)    show  rank-6
   models.      Scheme (a) gives  the     SM  plus  an extra \mbox{${\rm
   U}({1})_{{\rm Y_E}}$}.  Schemes (b) through (d) can produce effective
   rank-5   models,   ER5M,    by    taking a       large   VEV  limit.}
   \label{fig:Wilson}
\end{figure}
As  it can be  seen, the  various  breaking schemes  always give rise to
extra  vector  bosons   beyond the   SM:   in  fact it  is   unavoidable
\cite{kn:HewettA,kn:London,kn:Cohen,kn:Rosner}.  Here, only the simplest
of these  models  [Fig.~\ref{fig:Wilson}(a)]  which generates an   extra
vector boson, the \mbox{${\rm Z}_{\rm E}\,$}, will be considered.

\subsubsection{The Supermatter Fields}

  The   most general   superpotential     that   is   invariant    under
\mbox{$\mbox{${\rm SU}({3})_{{\rm c}}$}\otimes\mbox{${\rm SU}({2})_{{\rm
L}}$}\otimes\mbox{${\rm U}({1})_{{\rm Y}}$}$} and renormalizable for the
fields given in Fig.~\ref{fig:esix}  is of the form  (neglecting various
isospin contractions and generational indices) \cite{kn:Hewett},
\begin{equation}
W=W_0 + W_1 + W_2 + W_3
\label{eq:supot}
\end{equation}
$$
\begin{array}{@{}l@{}l@{}l@{}}
W_0 &=& \lambda_1\mbox{$\Phi_{{{{\rm
        R}^\prime}}}$}\mbox{$\Phi_{{Q}}$}\mbox{$\Phi_{{  u_{\rm L}^c}}$}
        +                                   \lambda_2\mbox{$\Phi_{{{{\rm
        L}^\prime}}}$}\mbox{$\Phi_{{Q}}$}\mbox{$\Phi_{{  d_{\rm L}^c}}$}
        +                                   \lambda_3\mbox{$\Phi_{{{{\rm
        L}^\prime}}}$}\mbox{$\Phi_{{L}}$}\mbox{$\Phi_{{ e_{\rm  L}^c}}$}
        + \lambda_4\mbox{$\Phi_{{{{\rm R}^\prime}}}$}\mbox{$\Phi_{{{{\rm
        L}^\prime}}}$}\mbox{$\Phi_{{\nu_{e_{\rm L}}^{\prime\prime c}}}$}
        +   \lambda_5\mbox{$\Phi_{{  d_{\rm  L}^\prime}}$}\mbox{$\Phi_{{
        d_{\rm      L}^{\prime           c}}}$}\mbox{$\Phi_{{\nu_{e_{\rm
        L}}^{\prime\prime c}}}$} \\
W_1 &=& \lambda_6\mbox{$\Phi_{{    d_{\rm    L}^\prime}}$}\mbox{$\Phi_{{
        u_{\rm   L}^c}}$}\mbox{$\Phi_{{   e_{\rm          L}^c}}$}     +
        \lambda_7\mbox{$\Phi_{{L}}$}\mbox{$\Phi_{{  d_{\rm    L}^{\prime
        c}}}$}\mbox{$\Phi_{{Q}}$} +   \lambda_8\mbox{$\Phi_{{\nu_{e_{\rm
        L}}^c}}$}\mbox{$\Phi_{{     d_{\rm   L}^\prime}}$}\mbox{$\Phi_{{
        d_{\rm L}^c}}$} \\
W_2 &=& \lambda_9                 \mbox{$\Phi_{{                  d_{\rm
        L}^\prime}}$}\mbox{$\Phi_{{Q}}$}\mbox{$\Phi_{{Q}}$}            +
        \lambda_{10}\mbox{$\Phi_{{         d_{\rm             L}^{\prime
        c}}}$}\mbox{$\Phi_{{    u_{\rm  L}^c}}$}\mbox{$\Phi_{{    d_{\rm
        L}^c}}$} \\
W_3 &=& \lambda_{11}        \mbox{$\Phi_{{{{\rm           R}^\prime}}}$}
        \mbox{$\Phi_{{L}}$}  \mbox{$\Phi_{{\nu_{e_{\rm L}}^{\prime\prime
        c}}}$}\,.
\end{array}
$$
$\mbox{$\Phi_{{A}}$} = \Phi(A,\tilde{A})$  is the  superfield, such that
$A={{\rm R}^\prime},Q, u_{\rm L}^c,$\ldots$\,$, and
$$
 \begin{array}{cccc}
   \mbox{$\Phi_{{Q}}$}   = \left(\begin{array}{@{}c@{}}
                     {\mbox{$\Phi_{{ u  }}$}}\\
                     {\mbox{$\Phi_{{ d  }}$}}
                 \end{array}\right)_{\rm L}\,,  &
   \mbox{$\Phi_{{L}}$}   = \left(\begin{array}{@{}c@{}}
                      {\mbox{$\Phi_{{\nu_e }}$}}\\
                      {\mbox{$\Phi_{{ e  }}$}}
                 \end{array}\right)_{\rm L}\,,  &
   \mbox{$\Phi_{{{{\rm L}^\prime}}}$} = \left(\begin{array}{@{}c@{}}
                      {\mbox{$\Phi_{{\nu_e^\prime}}$}}\\
                      {\mbox{$\Phi_{{ e^\prime }}$}}
                 \end{array}\right)_{\rm L}\,,  &
   \mbox{$\Phi_{{{{\rm R}^\prime}}}$} = \left(\begin{array}{@{}c@{}}
                      {\mbox{$\Phi_{{ e^\prime }}$}}\\
                      {\mbox{$\Phi_{{\nu_e^\prime}}$}}
                 \end{array}\right)_{\rm L}^c\,, 
 \end{array}
$$
for the first generation of the {\bf 27}'s, and  similarly for the other
generations.   The    Yukawa   couplings,   $\lambda_i$'s,  also   carry
generational     which   have  been    suppressed;   the couplings   are
inter-generational as well  as intra-generational.  The  superpotential,
W, summarizes the entire  possible spectrum of  low energy physics which
can occur within the context of an \mbox{${\rm E}_6$} framework.

  Notice that $W$ was only  required to be invariant  under the SM gauge
group.   Further constraints from \mbox{${\rm  E}_6$} model building may
cause some of the $\lambda_i$  terms to disappear.  Furthermore, not all
of  these   terms can simultaneously     exist  without giving  rise  to
$\Delta$L$\neq0$ and  $\Delta$B$\neq0$ interactions;  \mbox{${\rm E}_6$}
models say  nothing about the assignments of  baryon (B) and  lepton (L)
number until they   are connected to  SM representations.   As a  result
various scenarios may occur,

\vspace{1ex}
\begin{tabular}
   {@{ $\bullet$ }l@{: }l@{}r@{ $\;$\mbox{$\Longrightarrow$}$\;$ }l}
Leptoquarks 
   & B($q^\prime_{\rm L}$)=&$\frac{1}{3}\,$, L($q^\prime_{\rm L}$)=1 &
   \mbox{$\lambda_{{9}}$}=\mbox{$\lambda_{{10}}$}=0 \\
Diquarks    
   & B($q^\prime_{\rm L}$)=&$-\,\frac{2}{3}\,$, L($q^\prime_{\rm L}$)=0&
   \mbox{$\lambda_{{6}}$}=\mbox{$\lambda_{{7}}$}=\mbox{$
   \lambda_{{8}}$}=0 \\
Quarks      
   & B($q^\prime_{\rm L}$)=&$\frac{1}{3}\,$, L($q^\prime_{\rm L}$)=0 &
   \mbox{$\lambda_{{6}}$}=\mbox{$\lambda_{{7}}$}=\mbox{$
   \lambda_{{8}}$}=\mbox{$\lambda_{{9}}$}=\mbox{$\lambda_{{10}}$}=0
\end{tabular}
\vspace{1ex}

\noindent
where it has  been assumed that   L($\nu_{l_{\rm L}}^c$)= $-1\,$  (these
scenarios assume that there  exist only  three copies  of the  {\bf 27};
more complicated ones  can be constructed  by adding extra  copies).  In
this paper the least exotic of these models, {\it i.e.}, the ``Quarks,''
will  be investigated.  Furthermore,  to avoid  any fine tuning problems
with    the     neutrino     masses,    $$m_{\nu_{e_{\rm L}}^c}\;<<\;m_e
\;\Longleftrightarrow\;
\mbox{$\lambda_{{11}}$}<<\mbox{$\lambda_{{3}}$}\,,$$ it will  be assumed
\mbox{$\lambda_{{11}}$}=0.

  In this model the masses of the particles are generated by letting the
role of the Higgs fields be played by
$$
\begin{array}{ccc}
{\tilde{\rm L}^\prime}= \left(\begin{array}{@{}c@{}}
       {\tilde\nu^{\prime}_{e_{\rm L}}}\\{\tilde  e^\prime_{\rm L}}
    \end{array}\right)\,, &
{\tilde{\rm R}^\prime}= \left(\begin{array}{@{}c@{}}
       {\tilde e^{\prime c}_{\rm L}}\\{\tilde\nu^{\prime c}_{e_{\rm L}}}
    \end{array}\right)
\,,&
\tilde\nu^{\prime\prime c}_{e_{\rm L}}\,,
\end{array}
$$
for each generation.  It is possible to  work in a  basis where only the
third  generation  of   Higgses acquire  a VEV;   the   remainder become
``unHiggses'' \cite{kn:Hewett,kn:EllisI}.     In this basis   the Yukawa
couplings,   $$\lambda_4^{ijk}   \mbox{$\Phi_{{{{\rm    R}^\prime}_i}}$}
\mbox{$\Phi_{{{{\rm                                     L}^\prime}_j}}$}
\mbox{$\Phi_{{\nu^{\prime\prime}_{l_k}}}$} \,,$$ where $i,j,k=1,2,3$ are
generational indices, takes on a much simpler form,
$$\mbox{\footnotesize    {$\lambda_4       \;\in\;    \{\lambda_4^{ijk}|
\lambda_4^{i33}= \lambda_4^{3i3}= \lambda_4^{33i}=0\,,\;\lambda_4^{333}=
\lambda_4^{3jk}=  \lambda_4^{j3k}=   \lambda_4^{jk3}  \neq0 \;{\rm s.t.}
\;i=1,2\;  \& \;j,k=1,2,3\} \,.$}}$$    This basis also eliminates   the
potential  problem  of flavour changing   neutral currents  at  the tree
level.  It is also assumed that the  $\lambda_i$'s are real and that the
couplings to the unHiggses are very small.   The former assumption helps
to further  simplify the model  and reduce any  effects it might have in
the CP violation sector \cite{kn:Hewett}.

\subsection{Heavy Lepton Production} 

   \mbox{${\rm E}_6$} models are very rich in their spectrum of possible
low energy phenomenological predictions.  If any new particles are found
that fit  within this framework then  perhaps it will lead  the way to a
more  unified theory of the fundamental  forces of nature.  However this
is no  small task, for a  full theory would have  to be able to actually
predict the mass spectrum of the particles and the relationships between
various couplings,  and yet  require  very few  parameters.  Superstring
inspired \mbox{${\rm E}_6$}  models are far  from being able to complete
this task.   However, proof that \mbox{${\rm E}_6$}  is an effective GUT
would be a good first step.  But even this would not necessarily qualify
superstrings  to be the  next step for it  is  not totally inconceivable
that  some other theory might   give rise to   \mbox{${\rm E}_6$} as  an
effective GUT --- $caveat$ $emptor$.

\begin{figure}[ht]
  \begin{center}
    \mbox{
      \epsfysize=5.0cm
      \epsffile{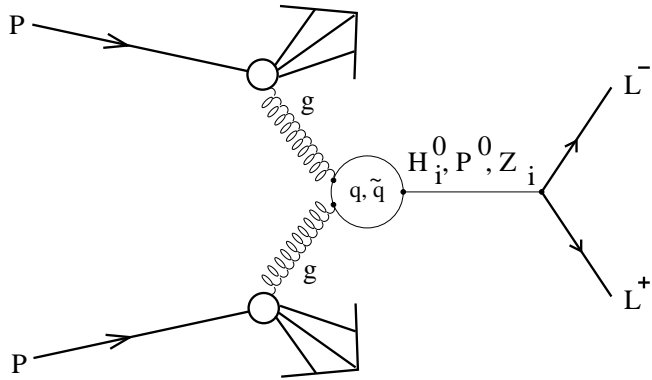}
    }\\\mbox{}\\
  \end{center}
\caption[Gluon-gluon fusion]{\footnotesize Gluon-gluon fusion to two
    heavy leptons, $gg$ $\longrightarrow$  L$^+$L$^-$. The loop contains
    quarks, $q\,$,   which    couple to    vector   bosons, $Z_{i=1,2}$,
    scalar-Higgses, $H^0_{i=1,2,3}$, and a pseudo-scalar-Higgs, $P^0\,$,
    and       squarks,     $\tilde       q\,$,       which     couple to
    scalar-Higgses. \label{fig:ggfuse}}
\end{figure}
   A natural question to ask  would be, ``Where  to look for \mbox{${\rm
E}_6$}  phenomenology?''   High energy   hadron  colliders, such  as the
$\mbox{T{\sc evatron}}$ at Fermilab (1.8 TeV  c.o.m., ${\cal L}\sim 10^2
pb^{-1}/yr\,$, $p\bar{p}$) or the LHC (14 TeV c.o.m., ${\cal L}\sim 10^5
pb^{-1}/yr\,$, $pp$), offer  possibilities of observing phenomena beyond
the SM by  looking  for  the  production of   heavy leptons through    a
mechanism  known as gluon-gluon fusion \cite{kn:Barger,kn:ggfusion}, see
Fig.~\ref{fig:ggfuse}.  This is a  interesting process because there are
enhancements  in the cross-sections  related   to the heavy  (s)fermions
running  around in the loop.   The computation was  done  in the minimal
supersymmetric standard model   (MSSM)  by Cieza Montalvo,  $et$  $al.$,
\cite{kn:Montalvo}  in  which they  predict ${\cal O}(10^5)\,events/yr$.
Therefore for \mbox{${\rm E}_6$} is it expected that the production rate
should in  principle  be higher since  there are  more particles running
around in the loop.

\begin{figure}[ht]
  \begin{center}
    \mbox{
      \epsfysize=5.0cm
      \epsffile{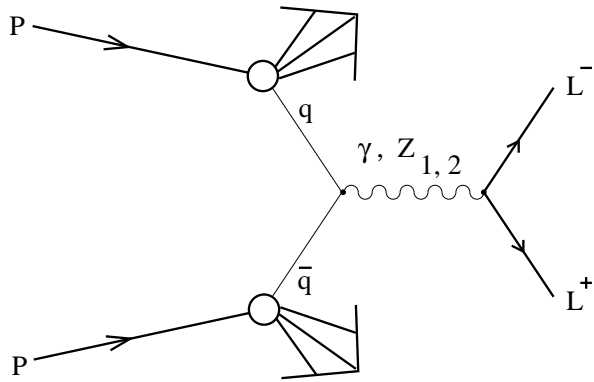}
    }\\
  \end{center}
\caption[Gluon-gluon fusion]{\footnotesize Drell-Yan, direct, production
    to two heavy leptons, $q\bar q$ $\longrightarrow$ L$^+$L$^-$.}
\label{fig:drellyan}
\end{figure}
The  other    contribution  to $L^+L^-$    production  is  the Drell-Yan
mechanism, see Fig.~\ref{fig:drellyan}.  In general, this is expected to
be  small    for $L^+L^-$ invariant   masses  above  the  $Z$ resonances
\cite{kn:Barger,kn:ggfusion}.   Indeed,   for $MSSM$  it   is the  least
dominant effect.  However, since  \mbox{${\rm E}_6$} has an extra vector
boson, $Z^\prime\,$,  that  is   fairly massive, it   is   expected that
Drell-Yan  production  will  compete with gluon-gluon   fusion below the
$Z^\prime$ threshold.

  These processes will be investigated in this paper.

\section{A Low Energy \mbox{${\rm E}_6$} Model}

In section \ref{sec-intro}, a general  overview of superstring  inspired
\mbox{${\rm E}_6$} models was given.  In addition, several comments were
made  about the type of   model that would be   presented.  We shall now
expand on these assumptions to determine their low energy consequences.

 There are  many   ways of  breaking   \mbox{${\rm  E}_6$} down  to   SM
energies. Invariably  these breaking schemes  lead to SM phenomenologies
which contain extra gauge bosons.  Here a rather simple model was chosen
in  which  only  an   extra     $Z$,  the $Z^\prime\,$,   is    produced
\cite{kn:Hewett,kn:EllisA,kn:BargerA}:
$$\mbox{${\rm E}_6$}\;\;\longrightarrow\;\;
\mbox{${\rm SU}({3})_{{\rm c}}$}\otimes
\mbox{${\rm SU}({2})_{{\rm L}}$}\otimes
\mbox{${\rm U}({1})_{{\rm Y}}$}\otimes
\mbox{${\rm U}({1})_{{\rm Y_E}}$}$$ 
\noindent  
({\it cf}. Fig.~\ref{fig:Wilson}). In general  the $Z^\prime\,$ can  mix
with the SM $Z$ to produce the mixed states
\begin{eqnarray}
Z_1&=&\;\;\;\cos\phi\,Z+\sin\phi\,Z^\prime\,,\label{eq:zmixa}\\
Z_2&=&   -\sin\phi\,Z+\cos\phi\,Z^\prime\,,\label{eq:zmixb}
\end{eqnarray}
[{\it cf}. Eq.~(\ref{eq:zzmix})].

  Recall   that    in  order   to      avoid potential   problems   with
flavour-changing-neutral currents, at the tree level, a basis was chosen
in which the third generation of primed-exotic-sleptons were assigned to
play the role of the Higgses \cite{kn:Hewett,kn:EllisI}:
\begin{equation}
\begin{array}{ccc}
{\tilde{\rm L}^\prime}_3= \left(\begin{array}{@{}c@{}}
      {\tilde\nu^{\prime}_{\tau_{\rm L}}}\\{\tilde 
      \tau^\prime_{\rm L}}
    \end{array}\right)\,, &
{\tilde{\rm R}^\prime}_3= \left(\begin{array}{@{}c@{}}
      { \tilde \tau^{\prime c}_{\rm L}}\\
      {\tilde\nu^{\prime c}_{\tau_{\rm L}}}
    \end{array}\right)\,,&
\tilde\nu^{\prime\prime c}_{\tau_{\rm L}}\,,\;\;\;\;\;\;\;
\end{array}
\label{eq:slhgs}
\end{equation}
({\it              i.e.},       $\tilde{\rm        R}_3\equiv{\tilde{\rm
L}}_3^{\mbox{}^{\mbox{\scriptsize $c$}}}$) or by redefining ${\tilde{\rm
L}^\prime}_3\,$,             ${\tilde{\rm        R}^\prime}_3\,$     and
$\tilde\nu^{\prime\prime   c}_{\tau_{\rm  L}}\,$,   in    terms  of  the
complex-isodoublet  fields,    $\Phi_1\,$   and $\Phi_2\,$,     and  the
complex-isoscalar  field, $\Phi_3\,$, respectively, Eq.~(\ref{eq:slhgs})
becomes
\begin{equation}
\begin{array}{ccc}
\Phi_1=\left(\begin{array}{@{}c@{}}
   {\phi^0_1}\\{\phi^-_1}\end{array}\right)\,, &
\Phi_2=\left(\begin{array}{@{}c@{}}
   {\phi^+_2}\\{\phi^0_2}\end{array}\right)\,, &
\Phi_3=\phi^0_3\,.
\end{array}
\label{eq:hgs}
\end{equation}
This assignment was accomplished by setting
\begin{equation}
\left.
\begin{array}{l@{\;\;\;\;\;\;\;\;}r}
\lambda_4^{i33}\,=\lambda_4^{3i3}\,=\lambda_4^{33i}\,=0 & i=1,2\;\;\;\\
\lambda_4^{333}=\lambda_4^{3jk}=\lambda_4^{j3k}=\lambda_4^{jk3}\neq0 
& j,k=1,2,3\label{kn:unyuk}
\end{array}
\right\}
\end{equation}
in  the  superpotential, Eq.~(\ref{eq:supot}),   where the  $ijk$'s  are
generation indices.  Therefore in order to avoid lepton-number violation
the lepton-numbers of all of the primed and double-primed exotic-leptons
must be zero.

  Further restrictions were  placed  on the superpotential by  requiring
that the baryon  and lepton numbers  of the exotic-quarks, $q^\prime\,$,
of the {\bf 27}'s (Fig.~\ref{fig:esix}), are  the same as those of their
non-exotic SM  counterparts  \cite{kn:Hewett}: $$B(q^\prime_{\rm  L})  =
\frac{1}{3},\;L(q^\prime_{\rm        L})=0\;\;\Rightarrow\;           \;
\mbox{$\lambda_{{6}}$} =  \mbox{$\lambda_{{7}}$} =\mbox{$\lambda_{{8}}$}
=\mbox{$\lambda_{{9}}$}=\mbox{$\lambda_{{10}}$}=0\,.$$              Also
$\lambda_{11}$ was set equal to  zero in order  to avoid any fine tuning
problems with the $\nu^c_{l_{\rm L}}$ masses \cite{kn:Hewett}.

  With  all the  aforementioned assumptions about  the Yukawa couplings,
the superpotential now simplifies to
\begin{eqnarray}
W &=& \lambda_1i\Phi_{Q}^T\tau_2\Phi_{{{\rm  R}^\prime}}\Phi_{    u_{\rm
        L}^c}  +\lambda_2i\Phi_{{{\rm  L}^\prime}}^T\tau_2\Phi_{Q}\Phi_{
        d_{\rm               L}^c}                +\lambda_3i\Phi_{{{\rm
        L}^\prime}}^T\tau_2\Phi_{L}\Phi_{   e_{\rm  L}^c}  \nonumber\\&&
        +\lambda_4i\Phi_{{{\rm            R}^\prime}}^T\tau_2\Phi_{{{\rm
        L}^\prime}}\Phi_{\nu_{e_{\rm      L}}^{\prime\prime          c}}
        +\lambda_5\Phi_{  d_{\rm L}^\prime}\Phi_{   d_{\rm    L}^{\prime
        c}}\Phi_{\nu_{e_{\rm L}}^{\prime\prime c}}\label{eq:soup}
\end{eqnarray}
where the $\lambda$'s were chosen to be real, plus similar terms for the
other      generations     and     their        cross-terms.         The
$\mbox{$\Phi_{{A}}$}=\Phi(\psi_A,A)$ are the superfields which contain a
two-component-spinor field,  $\psi_A\,$, and  a   complex-scalar-singlet
field, $A\,$.  Table~\ref{tb:parpr}  summarizes  the particle properties
of this model.
\begin{table}[htb]
\caption[Table of   $\mbox{${\rm  SU}({3})_{{\rm c}}$}\otimes\mbox{${\rm
  SU}({2})_{{\rm          L}}$}\otimes\mbox{${\rm          U}({1})_{{\rm
  Y}}$}\otimes\mbox{${\rm     U}({1})_{{\rm       Y_E}}$}$      Particle
  Properties]{\footnotesize \rule[-1ex]{0ex}{1ex} Table of  $\mbox{${\rm
  SU}({3})_{{\rm         c}}$}\otimes\mbox{${\rm          SU}({2})_{{\rm
  L}}$}\otimes\mbox{${\rm      U}({1})_{{\rm     Y}}$}\otimes\mbox{${\rm
  U}({1})_{{\rm Y_E}}$}$ particle properties.}
{\footnotesize
\def\VS{\mbox{\rule[-3.5ex]{0ex}{8ex}}}
\def\IT{\mbox{\tiny$
   \left(\begin{array}{@{}c@{}}{\;\;1/2}\\{-1/2}\end{array}\right)
$}}
\def\QP{\mbox{\tiny$
   \left(\begin{array}{@{}c@{}}{\;1}\\{\;0}\end{array}\right)
$}}
\def\QM{\mbox{\tiny$
   \left(\begin{array}{@{}c@{}}{\;\;0}\\{-1}\end{array}\right)
$}}
\def\QQ{\mbox{\tiny$
   \left(\begin{array}{@{}c@{}}{\;\;2/3}\\{-1/3}\end{array}\right)
$}}
\def\TH{\mbox{\bf$3$}}
\def\TB{\mbox{\bf$\bar 3$}}
\def\ON{\mbox{\bf$1$}}
\def\CR{\\\hline}
$$
\begin{array}{|l|c|c|c|c|c|c|c|c|c|c|c|}\cline{2-12}
\multicolumn{1}{c|}{\mbox{}}
   &Q &{\rm L} &  u_{\rm L}^c &  d_{\rm L}^c & e_{\rm  L}^c &\nu_{e_{\rm
L}}^c&  d_{\rm L}^\prime& d_{\rm  L}^{\prime c} &{{\rm L}^\prime} &{{\rm
R}^\prime}&\nu_{e_{\rm L}}^{\prime\prime c}\CR
c  &\TH &\ON &\TB    &\TB    &\ON   &\ON&\TH &\TB    &\ON &\ON &\ON \CR
I_3&\IT &\IT & 0     & 0     & 0    &0  & 0  & 0     &\IT &\IT &0\VS\CR
Y  & 1/3&-1  &-4/3   &\;\;2/3& 2    &0  &-2/3&\;\;2/3&-1  & 1  &0   \CR
Y_E& 2/3&-1/3&\;\;2/3&-1/3   & 2/3  &5/3&-4/3&-1/3   &-1/3&-4/3&5/3 \CR
Q  &\QQ &\QM &-2/3   &\;\;1/3&-1\;\;&0  &-1/3&\;\;1/3&\QM &\QP &0\VS\CR
B  & 1/3& 0  &-1/3   &-1/3   & 0    &0  &1/3 &-1/3   & 0  & 0  &0   \CR
L  & 0  & 1  & 0     & 0     &-1    &-1 & 0  & 0     & 0  & 0  &0   \CR
\end{array}
$$}
\label{tb:parpr}
\end{table}

  The   superpotential specifies   all   of the  couplings  between  the
particles  of  the {\bf 27}'s.   According  to appendix B  of  Haber and
Kane~\cite{kn:Haber}, the Yukawa interactions are given by
\begin{equation}
{\cal L}_{\rm Yuk} = -\,\frac{1}{2}\,\left[
\left.\left(\frac{\partial^2W}{\partial A_i\partial A_j}\right)
\right|_{\psi_A's\,=\,0}
\psi_i\psi_j+
\left.\left(\frac{\partial^2W}{\partial A_i\partial A_j}\right)^*
\right|_{\psi_A's\,=\,0}
\bar\psi_i\bar\psi_j
\right]\label{eq:yuk}
\end{equation}
and the scalar interactions are given by
\begin{equation}
V=V_F+V_D+V_{Soft}\,.\label{eq:scalpot}
\end{equation}
In Eq.~(\ref{eq:scalpot}), we have 
\begin{eqnarray}
V_F&=&F_i^*F_i\,,\\
V_D&=&\frac{1}{2}\,[D^aD^a+(D^\prime)^2]\,,
\end{eqnarray}
with
\begin{eqnarray}
F_i&=&\left.\left(\frac{\partial W}{\partial A_i}\right)
\right|_{\psi_A's\,=\,0}\,,\\
D^a&=&gA^*_iT_{ij}^aA_j\,,\\
D^\prime&=&\mbox{\small$\frac{1}{2}$}\, g^\prime Y_i A_i^*A_i\,,
\end{eqnarray}
where $g$ represents an SU(N) coupling constant with generators $T^a\,$,
and $g^\prime$  represents  a U(1)  coupling   constant with hypercharge
$Y$. $V_{Soft}$ is a soft-SUSY-breaking term that was  put in by hand in
order  to  lift the supersymmetric-mass-degeneracy  between $m_{\psi_A}$
and $m_A\,$:
\begin{equation}
V_{Soft}=V_{\tilde q}+V_{\tilde l}+V_{H_{Soft}}\,,
\end{equation}
with
\begin{eqnarray}
V_{\tilde q\;\;\;\;\;\;}\;&\supseteq&\tilde M_Q^2|\tilde Q|^2+
 \tilde M_u^2\tilde u_{\rm R}^* \tilde u_{\rm R} +
 \tilde M_d^2\tilde d_{\rm R}^* \tilde d_{\rm R}+
 \tilde M_{d^\prime_{\rm L}}^2
 \tilde d^{\prime^*}_{\rm L}\tilde d^\prime_{\rm L}+
 \tilde M_{d^\prime_{\rm R}}^2
 \tilde d^{\prime^*}_{\rm R}\tilde d^\prime_{\rm R}+\nonumber\\&&
 2\,\mbox{Re}\,[\lambda_1A_ui\tilde Q^T\tau_2\Phi_2\tilde u_{\rm R}^*+
 \lambda_2A_di\Phi_1^T\tau_2\tilde Q\tilde d_{\rm R}^*+
 \lambda_5A_{d^\prime}\tilde d^{\prime^*}_{\rm R}
 \tilde d^\prime_{\rm L}\Phi_3
 ]\,,\;\;\;\;\;\;\;\;\\
V_{\tilde l\;\;\;\;\;\;\;}\;&\supseteq&\tilde M_L^2|\tilde L|^2+
 \tilde M_e^2\tilde e_{\rm R}^* \tilde e_{\rm R} +
 \tilde M_{\nu_e}^2\tilde \nu_{e_{\rm R}}^* \tilde\nu_{ e_{\rm R}}+
 2\lambda_3A_e\,\mbox{Re}\,[i\Phi_1^T\tau_2\tilde L\tilde e_{\rm R}^*]
 \,,\\
V_{H_{Soft}}&=&\mu_1^2|\Phi_1|^2+\mu_2^2|\Phi_2|^2+\mu_3^2|\Phi_3|^2-
 \mbox{\small$\frac{1}{\sqrt{2}}$}\,
 \lambda A(i\Phi_1^T\tau_2\Phi_2\Phi_3+h.c.)\,.\label{eq:softy}
\end{eqnarray}
The coefficients $\tilde M_A^2\,$, $A_A\,$, $\mu_i^2\,$, and $A$ are the
soft-SUSY-breaking parameters, and $\lambda\equiv\lambda_4^{333}\,$.

  The     Higgs     potential   can    be      extracted  directly  from
Eq.~(\ref{eq:scalpot})               and              is           given
by~\cite{kn:Hewett,kn:BargerA,kn:Gunion}
\begin{eqnarray}
V_H&=&V_{H_{Soft}}+\lambda^2
(|\Phi_1|^2|\Phi_2|^2+|\Phi_1|^2|\Phi_3|^2+|\Phi_2|^2|\Phi_3|^2)
\nonumber\\&&
+\mbox{\small$\frac{1}{8}$}(g^2+g^{\prime^2})(|\Phi_1|^2-|\Phi_2|^2)^2
+\mbox{\small$\frac{1}{72}$}g^{\prime\prime^2}
(|\Phi_1|^2+4|\Phi_2|^2-5|\Phi_3|^2)^2
\nonumber\\&&
+(\mbox{\small$\frac{1}{2}$}g^2-\lambda^2)|\Phi_1^\dagger\Phi_2|^2\,,
\label{eq:fox}
\end{eqnarray}
where $g\,$,  $g^\prime\,$,  and $g^{\prime\prime}$ are  the SU(2)$_{\rm
L}\,$, U(1)$_Y\,$,  and U(1)$_{Y_E}\,$ coupling constants, respectively.
The minimization condition \cite{kn:Hewett}
\begin{equation}
\left.\frac{\partial V_H}{\partial\phi_i^0}\right|_{VEV's}=0\,,
\end{equation}
can   be   used   to  fix  the  $\mu^2_i$     terms  in $V_{H_{Soft}}\,$
[Eq.~(\ref{eq:softy})],  where the  vacuum  expectation values, $VEV$'s,
are given by
\begin{equation}
\langle\Psi\rangle=
\left\{\begin{array}{cl}
{\displaystyle\frac{\nu_i}{\sqrt{2}}}&{\rm if}\;\Psi=\phi^0_i\\
0&{\rm otherwise}
\end{array}\right.\;\;\varepsilon\;\Re\,.
\end{equation}
Therefore, we have
\begin{eqnarray}
\mu^2_i&=&
 \frac{3}{72}\,g^{\prime\prime^2}(v_1^2+4\nu_2^2-5\nu_3^2)Y_{E_i}
 +\,\frac{1}{8}\,(g^2+g^{\prime^2})(\nu_1^2-\nu_2^2)Y_i\nonumber\\&&
 -\sum_{j<k}\tilde\varepsilon_{ijk}\,\left[
 \frac{(v_j^2+v_k^2)}{2}\,\lambda^2-\frac{\nu_j\nu_k}{4\nu_i}
 \,\lambda A\right]
\end{eqnarray}
where    $\tilde\varepsilon_{ijk}=|\varepsilon_{ijk}|\,$.    The kinetic
terms for the scalar fields are given by \cite{kn:Cheng}
\begin{equation}
{\cal L}_{K.E.}\supseteq
|{\cal D_\mu}\Phi_i|^2\stackrel{\mbox{\tiny or}}{=}
|(\partial_\mu-{\textstyle\frac{i}{2}}G_\mu)\Phi_i|^2\,,\label{eq:hke}
\end{equation}
with ({\it cf}. \cite{kn:Hewett})
\begin{eqnarray}
G_\mu&=&(g\tau_3\sin\theta_W+g^\prime Y\cos\theta_W)A_\mu
+(g\tau_3\cos\theta_W-g^\prime Y\sin\theta_W)Z_\mu\nonumber\\&&
+\mbox{\footnotesize$\sqrt{2}$}g[\tau_+W^-_\mu+\tau_-W^+_\mu]
+g^{\prime\prime}Y_E Z^\prime_\mu\,,\label{eq:hkea}
\end{eqnarray}
where          $\tau_\pm=\frac{1}{2}(\tau_1\pm                i\tau_2)$,
$\tau_i|\Phi_3\rangle\equiv0\,$,          and           $g^\prime\approx
g^{\prime\prime}$~\cite{kn:Hewett,kn:EllisB}.  The $\tau_i$'s    are the
Pauli matrices acting in isospin space.

  The $\Phi_i$ fields have  complex components, $\phi^a_i\,$, which were
chosen       to    be    of      the      form~\cite{kn:EllisI,kn:Cheng}
({\it cf}.~\cite{kn:GunionA})
\begin{equation}
\phi^a_i=\frac{1}{\sqrt{2}}\,(\phi^a_{iR}+i\phi^a_{iI})\,,
\label{eq:pfff}
\end{equation}
where $\phi^a_{iR}/\sqrt{2}\,$ and $\phi^a_{iI}/\sqrt{2}\,$ are the real
and  imaginary parts, respectively.   Therefore, the $\{\Phi_i\}$ fields
have a total 10 degrees of  freedom: four are  ``eaten'' to give masses
to  the $W^\pm$,   $Z$,    and $Z^\prime$  bosons,  and    the remainder
yield~\cite{kn:Hewett} two    charged-Higgs    bosons, $H^\pm$,      one
pseudo-scalar-Higgs  boson,  $P^0\,$,   and three  scalar-Higgs  bosons,
$H^0_{i=1,2,3}\,$.  The mass terms for  the Higgs fields can be obtained
from the second order terms of  the expansion of $V_H(\phi_k)$ about its
minimum \cite{kn:Cheng},
\begin{equation}
V_H(\phi_k)\;\supseteq\;\frac{1}{2}\,{\cal M}_{ij}^2\,
(\phi_i-\langle\phi_i\rangle)(\phi_j-\langle\phi_j\rangle)\,,
\end{equation}
where  
\begin{equation}
{\cal M}_{ij}^2=\left.\frac{\partial^2V_H}{\partial\phi_i\partial\phi_j}
\right|_{VEV's}
\end{equation}
is the Higgs-mass-mixing matrix.  Therefore the mass terms for the Higgs
fields are simply
\begin{eqnarray}
{\cal L}_{\cal M}&\supseteq&
-\,(\phi_2^{+^*} \phi_1^-)\,{\cal M}^2_{H^\pm}
   \left(\begin{array}{@{}c@{}}{\phi_2^+}\\{\phi_1^{-^*}}
        \end{array}\right)
-\frac{1}{2}\,
(\phi_{2I}^0\:\phi_{1I}^0\:\phi_{3I}^0)\,{\cal M}^2_{P^0}
   \left(\begin{array}{@{}c@{}}
      {\phi_{2I}^0}\\{\phi_{1I}^0}\\{\phi_{3I}^0}
   \end{array}\right)\nonumber\\&&
-\frac{1}{2}\,
(\phi_{1R}^0-\nu_1\;\phi_{2R}^0-\nu_2\;\phi_{3R}^0-\nu_3)\,
{\cal M}^2_{H^0_i}\label{kn:simply}
   \left(\begin{array}{@{}c@{}}
      {\phi_{1R}^0-\nu_1}\\{\phi_{2R}^0-\nu_2}\\{\phi_{3R}^0-\nu_3}
   \end{array}\right)\,.
\end{eqnarray}
The mass-mixing matrices are
\begin{eqnarray}
{\cal M}^2_{H^\pm}&=&
\frac{1}{2}\left(
\begin{array}{ll}
(\frac{1}{2} g^2-\lambda^2)\nu_1^2+\lambda A\,
\mbox{$\displaystyle\frac{\nu_{13}}{\nu_2}$}&
(\frac{1}{2} g^2-\lambda^2)\nu_{12}+\lambda A\nu_3\\
(\frac{1}{2} g^2-\lambda^2)\nu_{12}+\lambda A\nu_3&
(\frac{1}{2} g^2-\lambda^2)\nu_2^2+
\mbox{$\displaystyle\lambda A\,\frac{\nu_{23}}{\nu_1}$}
\end{array}
\right)\,,
\\\nonumber\\
{\cal M}_{P^0}^2&=&\frac{\lambda A\nu_3}{2}
\left(
\begin{array}{ccc}
{\displaystyle\frac{\nu_1}{\nu_2}}&
1&
{\displaystyle\frac{\nu_1}{\nu_3}}\\
1&
{\displaystyle\frac{\nu_2}{\nu_1}}&
{\displaystyle\frac{\nu_2}{\nu_3}}\\
{\displaystyle\frac{\nu_1}{\nu_3}}&
{\displaystyle\frac{\nu_2}{\nu_3}}&
{\displaystyle\frac{\nu_{12}}{\nu_3^2}}
\end{array}
\right)\,,
\\\nonumber\\
{\cal M}_{H^0_i}^2&=&\frac{1}{2}
\left(
\begin{array}{lll}
B_1\nu_1^2+\lambda A\,\mbox{$\displaystyle\frac{\nu_{23}}{\nu_1}$}&
B_2\nu_{12}-\lambda A\nu_3&
B_3\nu_{13}-\lambda A\nu_2\\
B_2\nu_{21}-\lambda A\nu_3&
B_4\nu_2^2+\lambda A\,\mbox{$\displaystyle\frac{\nu_{13}}{\nu_2}$}&
B_5\nu_{23}-\lambda A\nu_1\\
B_3\nu_{31}-\lambda A\nu_2&
B_5\nu_{32}-\lambda A\nu_1&
B_6\nu_3^2+\lambda A\,\mbox{$\displaystyle\frac{\nu_{12}}{\nu_3}$}
\end{array}
\right)\label{eq:mhimat}
\end{eqnarray}
where $\nu_{ij}=\nu_i\nu_j\,$. In Eq.~(\ref{eq:mhimat})
\begin{equation}
\mbox{\small$\left.
\begin{array}{@{}lll@{}}
B_1=\mbox{$\textstyle\frac{{1}}{{2}}$}(g^2+ g^{\prime^2})+
    \mbox{$\textstyle\frac{{1}}{{18}}$} 
    g^{\prime\prime^2}&
B_2=2\lambda^2+\mbox{$\textstyle\frac{{2}}{{9}}$} g^{\prime\prime^2}-
     \mbox{$\textstyle\frac{{1}}{{2}}$}(g^2+ g^{\prime^2})&
B_3=2\lambda^2-\mbox{$\textstyle\frac{{5}}{{18}}$} g^{\prime\prime^2}\\
&B_4=\mbox{$\textstyle\frac{{1}}{{2}}$}(g^2+ g^{\prime^2})+
     \mbox{$\textstyle\frac{{8}}{{9}}$} 
     g^{\prime\prime^2}&
B_5=2\lambda^2-\mbox{$\textstyle\frac{{10}}{{9}}$} g^{\prime\prime^2}\\
&&B_6=\mbox{$\textstyle\frac{{25}}{{18}}$} g^{\prime\prime^2}
\end{array}\right\}$}\,.
\end{equation}
The  physical states are obtained  by  diagonalizing the terms in ${\cal
L}_{\cal M}\,$. The eigenvectors for the charged and pseudo-scalar Higgs
terms are respectively,
\begin{eqnarray}
H^\pm&=&\;\;\;\;
\cos\beta\,\phi_2^\pm+\sin\beta\,\phi_1^\pm\,,\\
G^\pm&=&-
\sin\beta\,\phi_2^\pm+\cos\beta\,\phi_1^\pm\,,
\end{eqnarray}
and
\begin{eqnarray}
P^0&=&\sqrt{\frac{\lambda A\nu_3}{2m_{P^0}^2}}\,
\left[
\sqrt{\frac{\nu_1}{\nu_2}}\,\phi_{2I}^0+
\sqrt{\frac{\nu_2}{\nu_1}}\,\phi_{1I}^0+
\sqrt{\frac{\nu_{12}}{\nu_3^2}}\,\phi_{3I}^0
\right]\,,\\
G^0_1&=&\frac{\nu_2^2\nu_3}{\sqrt{
(\nu_2^2+\nu_3^2)(\nu_{12}^2+\nu^2\nu_3^2)
}}\,
\left[
\frac{\nu_3}{\nu_2}\,\phi_{2I}^0-
\frac{\nu_1}{\nu_3}\left(1+\frac{\nu_3^2}{\nu_2^2}\right)\phi_{1I}^0+
\phi_{3I}^0
\right]\,,\\
G^0_2&=&\frac{\nu_3}{\sqrt{\nu_2^2+\nu_3^2}}\,
\left[-\,\frac{\nu_2}{\nu_3}\,\phi_{2I}^0+\phi_{3I}^0\right]\,,
\end{eqnarray}
where $\phi^\pm_i=(\phi^\mp_i)^*\,$, $\nu^2 =  \nu_1^2 + \nu_2^2\,$, and
$\tan\beta\equiv\nu_2/\nu_1\,$. Here, the  $H^\pm$ are the charged-Higgs
states with masses
\begin{equation}
m^2_{H^\pm}=\frac{\lambda A\nu_3}{\sin(2\beta)}
+\left(1-2\,\frac{\lambda^2}{g^2}\right)m_W^2\,,
\label{eq:charged}
\end{equation}
the $P^0$ is the pseudo-scalar-Higgs state with mass
\begin{equation}
m_{P^0}^2=\frac{\lambda A\nu_3}{\sin(2\beta)}\,
\left(1+\frac{\nu^2}{4\nu_3^2}\sin^2(2\beta)\right)\,,
\label{eq:neutral}
\end{equation}
and the $G^\pm$ and $G^0_{1,2}$ are the Goldstone-boson states with zero
mass.  The  scalar-Higgs term can  be diagonalized analytically, however
the result is, in  general, not very enlightening.   For our purposes it
suffices  to  resort to  numerical   techniques.  In the  unitary  gauge
(U-gauge)  the Goldstone modes vanish,  {\it i.e.}, the $G{\rm 's}=0\,$,
and the fields become physical. This allows the change of basis:
\begin{eqnarray}
\phi^\pm_1&=&\sin\beta H^\pm\,,\label{eq:basa}\\
\phi^\pm_2&=&\cos\beta H^\pm\,,\\
\phi^0_{1I}&=&\kappa\nu_{23}\,P^0\,,\\
\phi^0_{2I}&=&\kappa\nu_{13}\,P^0\,,\\
\phi^0_{3I}&=&\kappa\nu_{12}\,P^0\,,\\
\phi^0_{iR}&=&\nu_i+\sum_{j=1}^3U_{ij}H^0_j\,,\label{eq:basb}
\end{eqnarray}
where  $\kappa=1/\sqrt{v_1^2v_2^2+v^2v_3^2}\,$.   The  $U_{ij}$  are the
elements  of the inverse of  the matrix that  was used in the similarity
transformation to diagonalized the  scalar-Higgs-mass term.  With  these
transformations at hand it is now a straightforward matter to get all of
the masses and couplings for the various particles in this model.

  The mass terms  for the gauge fields can  be found by transforming the
kinetic terms   for  the  $\Phi_i$ fields,  Eq.~(\ref{eq:hke}),   to the
U-gauge basis, Eqs.~(\ref{eq:basa})-(\ref{eq:basb}), yielding:
\begin{equation}
{\cal L}_{K.E.}^{\Phi_i}\supseteq
m_W^2\,W^+_\mu W^{-^\mu}+
\frac{1}{2}\,(Z\;Z^\prime)_\mu\,{\cal M}^2_{Z-Z^\prime}\,
\left(\begin{array}{@{}c@{}}{Z}\\{Z^\prime}\end{array}\right)^\mu\,.
\label{eq:zzmix}
\end{equation}
As a consequence, the $W$ mass is
\begin{equation}
m_W^2=\frac{1}{4} g^2\nu^2\,,
\end{equation}
and                   the                       $Z-Z^\prime$-mass-mixing
matrix is~\cite{kn:Hewett,kn:EllisB,kn:EllisA,kn:BargerA}
\begin{equation}
{\cal M}^2_{Z-Z^\prime}=
\left(
\begin{array}{cc}
m_Z^2&\delta m^2\\
\delta m^2&m_{Z^\prime}^2
\end{array}
\right)\,,\label{eq:mzzmat}
\end{equation}
with matrix elements:
\begin{eqnarray}
m_{Z^{\;}}^2&=&\frac{1}{4}(g^2+g^{\prime^2})\,\nu^2\,,\\
m_{Z^\prime}^2&=&\frac{1}{36} 
   g^{\prime\prime^2}(\nu_1^2+16\nu_2^2+25\nu_3^2)\,,\\
\delta m^2&=&\frac{1}{12}\,\sqrt{g^2+ g^{\prime^2}}\, 
   g^{\prime\prime}(4\nu_2^2-\nu_1^2)\,.
\end{eqnarray}
Diagonalization of ${\cal M}^2_{Z-Z^\prime}$ yields the mass eigenstates
given by Eqs.~(\ref{eq:zmixa}) and~(\ref{eq:zmixb}) with eigenvalues
\begin{eqnarray}
m_{Z_1}^2&=&m_Z^2\cos^2\phi+\delta m^2\sin(2\phi)+
m_{Z^\prime}^2\sin^2\phi\,,\\
m_{Z_2}^2&=&m_Z^2\sin^2\phi-\delta m^2\sin(2\phi)+
m_{Z^\prime}^2\cos^2\phi\,,
\end{eqnarray}
and mixing angle
\begin{equation}
\tan(2\phi)=\frac{2\delta m^2}{m_Z^2-m_{Z^\prime}^2}\,.
\end{equation}
Notice that  in the  large $\nu_3$ limit,  $\phi\;\rightarrow\;\pi/2\,$,
and             therefore           $Z_1\;\rightarrow\;Z\,$,         and
$Z_2\;\rightarrow\;Z^\prime\,$.  In  fact for the  range of $VEV$'s that
will be considered here,  $m_{Z_1}<m_{Z_2}\,$.  Therefore, $Z_1$ will be
designated  the role of the observed  $Z$, at facilities  such as LEP or
SLC.

  The  mass terms for the fermions,  and hence the Yukawa couplings, can
be found by evaluating ${\cal L}_{Yuk}^{\mbox{}}\,$, Eq.~(\ref{eq:yuk}),
in the   U-gauge basis  and  then using  Appendix A   of Haber  and Kane
\cite{kn:Haber},   to      convert   to    four     component     spinor
notation.\footnote{{\it cf}. Eq.~(\ref{eq:yields}).  For a more explicit
example   see section 4.2  of   Gunion and  Haber \cite{kn:GunionA}} The
result is
\begin{equation}
{\cal L}_{\rm Yuk}\supseteq
-\frac{1}{\sqrt{2}}\,\left\{
\lambda_1\nu_2\,\bar u u+
\lambda_2\nu_1\,\bar d d+
\lambda_3\nu_1\,\bar e e+
\lambda_4\nu_3\,\bar e^\prime e^\prime+
\lambda_5\nu_3\,\bar d^\prime d^\prime
\right\}\,.\label{eq:yucky}
\end{equation}
Therefore, the Yukawa couplings for the first generation are given by:
\begin{eqnarray}
\lambda_1&=&\frac{g\,m_u}{\sqrt{2}\, m^{\mbox{}}_W\sin\beta}\,,
            \label{eq:ycpa}\\
\lambda_2&=&\frac{g\,m_d}{\sqrt{2}\, m^{\mbox{}}_W\cos\beta}\,,\\
\lambda_3&=&\frac{g\,m_e}{\sqrt{2}\, m^{\mbox{}}_W\cos\beta}\,,\\
\lambda_4&=&\frac{\sqrt{2}}{\nu_3}\,m_{e^\prime}\,,\\
\lambda_5&=&\frac{\sqrt{2}}{\nu_3}\,m_{d^\prime}\,,\label{eq:ycpd}
\end{eqnarray}
and similarly for the other generations.

  The sfermion masses  are obtained by evaluating the scalar-interaction
potential, V [Eq.~(\ref{eq:scalpot})],  and then transforming  it to the
U-gauge basis:
\begin{equation}
{\cal L}_{\cal M}\supseteq-V\supseteq
-(\tilde f_L^* \tilde f_R^*){\cal M}^2_{\tilde f}
   \left(\begin{array}{@{}c@{}}
      {\tilde f_L}\\{\tilde f_R}\end{array}\right)\,,
\end{equation}
with
\begin{equation}
{\cal M}^2_{\tilde f}=
\left(
\begin{array}{cc}
M^2_{LL}&M^2_{LR}\\
M^2_{LR}&M^2_{RR}
\end{array}
\right)\,,
\end{equation}
being the sfermion-mass-mixing  matrix. The mass-mixing  matrix elements
are given by:
\begin{eqnarray}
M^{(\tilde u)^2}_{LL}&=&
\tilde M_Q^2+m_u^2+\frac{1}{6}\,
(3-4 x^{\mbox{}}_W)\,m_Z^2\,\cos(2\beta)-
\frac{1}{36}\, 
g^{\prime\prime^2}(\nu_1^2+4\nu_2^2-5\nu_3^2)\,,\;\;\;\;\\
M^{(\tilde u)^2}_{RR}&=&
\tilde M_u^2+m_u^2+\frac{2}{3}\, x^{\mbox{}}_W\,m_Z^2\,\cos(2\beta)-
\frac{1}{36}\, g^{\prime\prime^2}(\nu_1^2+4\nu_2^2-5\nu_3^2)\,,\\
M^{(\tilde u)^2}_{LR}&=&m_u\,(A_u-m_{e^\prime}\cot\beta)\,,
\end{eqnarray}
for the $\tilde u_{L,R}$ squarks;
\begin{eqnarray}
M^{(\tilde d)^2}_{LL}&=&
\tilde M_Q^2+m_d^2-\frac{1}{6}\,
(3-2 x^{\mbox{}}_W)\,m_Z^2\,\cos(2\beta)-
\frac{1}{36}\, 
g^{\prime\prime^2}(\nu_1^2+4\nu_2^2-5\nu_3^2)\,,\;\;\;\;\\
M^{(\tilde d)^2}_{RR}&=&
\tilde M_d^2+m_d^2-\frac{1}{3}\, x^{\mbox{}}_W\,m_Z^2\,\cos(2\beta)+
\frac{1}{72}\, g^{\prime\prime^2}(\nu_1^2+4\nu_2^2-5\nu_3^2)\,,\\
M^{(\tilde d)^2}_{LR}&=&m_d\,(A_d-m_{e^\prime}\tan\beta)\,,
\end{eqnarray}
for the $\tilde d_{L,R}$ squarks;
\begin{eqnarray}
M^{(\tilde d^\prime)^2}_{LL}&=&
\tilde M_{d^\prime_L}^2+m_{d^\prime}^2+
\frac{1}{6}\, x^{\mbox{}}_W\,m_Z^2\,\cos(2\beta)+
\frac{1}{18}\, g^{\prime\prime^2}(\nu_1^2+4\nu_2^2-5\nu_3^2)\,,\;\;\;\;
\,\;\;\;\;\;\;\;\\
M^{(\tilde d^\prime)^2}_{RR}&=&
\tilde M_{d^\prime_R}^2+m_{d^\prime}^2-
\frac{1}{3}\, x^{\mbox{}}_W\,m_Z^2\,\cos(2\beta)+
\frac{1}{72}\, g^{\prime\prime^2}(\nu_1^2+4\nu_2^2-5\nu_3^2)\,,\\
M^{(\tilde d^\prime)^2}_{LR}&=&
m_{d^\prime}\,(A_{d^\prime}-
m_{e^\prime}\,\frac{\nu_{12}}{\nu_3^2})\,,
\end{eqnarray}
for the $\tilde d^\prime_{L,R}$ squarks;
\begin{eqnarray}
M^{(\tilde e)^2}_{LL}&=&
\tilde M_L^2+m_e^2-\frac{1}{2}\,
(1-2 x^{\mbox{}}_W)\,m_Z^2\,\cos(2\beta)+
\frac{1}{72}\, 
g^{\prime\prime^2}(\nu_1^2+4\nu_2^2-5\nu_3^2)\,,\;\;\;\;\\
M^{(\tilde e)^2}_{RR}&=&
\tilde M_e^2+m_e^2- x^{\mbox{}}_W\,m_Z^2\,\cos(2\beta)-
\frac{1}{36}\, g^{\prime\prime^2}(\nu_1^2+4\nu_2^2-5\nu_3^2)\,,\\
M^{(\tilde e)^2}_{LR}&=&m_e\,(A_e-m_{e^\prime}\tan\beta)\,,
\end{eqnarray}
for the $\tilde e_{L,R}$ sleptons;
\begin{eqnarray}
M^{(\tilde \nu_e)^2}_{LL}&=&
\tilde M_L^2+m_{\nu_e}^2+\frac{1}{2}\,m_Z^2\,\cos(2\beta)+
\frac{1}{72}\, g^{\prime\prime^2}(\nu_1^2+4\nu_2^2-5\nu_3^2)\,,\;\;\;\;
\,\;\;\;\;\;\;\;\;\;\\
M^{(\tilde \nu_e)^2}_{RR}&=&
\tilde M_{\nu_e}^2-
\frac{5}{72}\, g^{\prime\prime^2}(\nu_1^2+4\nu_2^2-5\nu_3^2)\,,\\
M^{(\tilde \nu_e)^2}_{LR}&=&0\,,
\end{eqnarray}
for   the  $\tilde \nu_{L,R}$ sleptons,   and  similarly for  the other
generations.  The mass eigenstates are given by
\begin{equation}
\left(\begin{array}{c}
\tilde f_1\\\tilde f_2
\end{array}\right)=
\left(\begin{array}{rr}
\cos\theta_{\tilde f}&\sin\theta_{\tilde f}\\
-\sin\theta_{\tilde f}&\cos\theta_{\tilde f}
\end{array}\right)
\left(\begin{array}{c}
\tilde f_L\\\tilde f_R
\end{array}\right)\label{eq:smix}
\end{equation}
with mass eigenvalues
\begin{eqnarray}
m_{f_1}^2&=&M^2_{LL}\cos^2\theta_{\tilde f}+
M^2_{LR}\sin(2\theta_{\tilde f})+M^2_{RR}\sin^2\theta_{\tilde f}\,,\\
m_{f_2}^2&=&M^2_{LL}\sin^2\theta_{\tilde f}-
M^2_{LR}\sin(2\theta_{\tilde f})+M^2_{RR}\cos^2\theta_{\tilde f}\,,
\end{eqnarray}
and mixing angle
\begin{equation}
\tan(2\theta_{\tilde f})=\frac{M^2_{LR}}{M^2_{LL}-M^2_{RR}}\,.
\end{equation}
Notice that for fairly large $\nu_3$
\begin{equation}
\tan(2\theta_{\tilde f})\;\sim\;
{\cal O}\left(\frac{m_fA_f}{\nu_3^2}\right)\,,
\end{equation}
where  the soft  terms  have been assumed  to  be large and  degenerate.
Therefore,  in  general,  the mixing   is only expected   to  affect the
sfermions that have fairly heavy fermion partners.
   
  The supersymmetric partner, or   spartner, degrees of freedom  for the
neutral-Higgs    fields  (neutral-Higgsinos),   $\nu_{\tau_L}^\prime\,$,
$\nu_{\tau_L}^{\prime^c}\,$   and  $\nu_{\tau_L}^{\prime   \prime^c}\,$,
along with the spartner degrees of freedom for  the neutral gauge fields
(neutral-gauginos),  $\tilde\gamma\,$,  $\tilde   Z\,$,   and    $\tilde
Z^\prime\,$, mix to form  a (6$\times$6) neutralino,  $\tilde \chi^0\,$,
mass-mixing  matrix.\footnote{A detailed  study  of  the $\tilde \chi^0$
mass spectrum can    be   found in  \cite{kn:thesisa}.} Similarly    the
charged-Higgsinos,    $\tau_L^\prime$ and $\tau_L^{\prime^c}\,$, and the
charged-gauginos,  $\tilde  W^\pm\,$,   form a  (2$\times$2)   chargino,
$\tilde \chi^\pm\,$, mass-mixing matrix.\footnote{The full form of these
mass matrices  can be found  in Ellis,  $et$ $al.$,~\cite{kn:EllisI} and
the details of how  to obtain them  can be found  in appendix B of Haber
and Kane~\cite{kn:Haber}.}  By  virtue of supersymmetry  the  neutralino
and  chargino  mass-mixing matrices  contain  the  same Yukawa and gauge
couplings as their spartners, modulo soft terms.

  The  real   and  imaginary parts   of  the  sfermion  fields,  $\tilde
\nu_{e_L}^\prime\,$,     $\tilde       \nu_{e_L}^{\prime^c}\,$,  $\tilde
\nu_{e_L}^{\prime\prime^c}\,$,  $\tilde  \nu_{\mu_L}^\prime\,$,  $\tilde
\nu_{\mu_L}^{\prime^c}\,$, and $\tilde  \nu_{\mu_L}^{\prime\prime^c}\,$,
yield   two   separate   (6$\times$6)-mass-mixing    matrices  for   the
neutral-unHiggses [{\it  cf}.  Eq.~(\ref{kn:simply}] for   $\phi^0_{iR}$
and  $\phi^0_{iI}$),  which    contain the  Yukawa  couplings   given in
Eq.~(\ref{kn:unyuk}).    In  general,  these  mass-mixing  matrices  are
expected to lead to very massive unHiggs states \cite{kn:EllisI}.

  The   spartner degrees  of      freedom for  the    neutral-unHiggses,
$\nu_{e_L}^\prime\,$,    $\nu_{e_L}^{\prime^c}\,$,    $\nu_{e_L}^{\prime
\prime^c}\,$,  $\nu_{\mu_L}^\prime\,$,  $\nu_{\mu_L}^{\prime^c}\,$,  and
$\nu_{\mu_L}^{\prime \prime^c}\,$,     form  a  (6$\times$6)-mass-mixing
matrix  for  the neutral-unHiggsinos.  Therefore, the neutral-unHiggsino
mass-mixing  matrix   contains   the same    Yukawa  couplings as  their
neutral-unHiggs partners.
 
  The            sfermion      fields            $\tilde{e}^\prime_L\,$,
$\tilde{e}^{\prime^c}_L\,$,         $\tilde\mu^\prime_L\,$,          and
$\tilde\mu^{\prime^c}_L\,$, yield two  separate (2$\times$2)-mass-mixing
matrices for the charged-unHiggses  [{\it cf}. Eq.~(\ref{kn:simply}] for
$\phi^{\pm}_i$).   These   matrices have    a  large number  of  unknown
parameters  and    quite   naturally  acquire   a    very    large  mass
[{\it cf}.~\cite{kn:EllisI}].

  Finally,   the spartner degrees   of freedom for the charged-unHiggses
give diagonalized mass   eigenstates [see  Eq.~(\ref{eq:yucky})]   which
correspond to the charged heavy leptons.

\section{$L^+L^-$ Production in \mbox{${\rm E}_6$}}
\label{sec-cross}

\subsection{Gluon-Gluon Fusion}

\begin{figure}[htbp]
$$
\begin{array}{lclc}
  a) & 
      \raisebox{-1.693cm}{
       \mbox{
       \setlength{\unitlength}{1cm}
       \begin{picture}(5.743,3.6)
        \put(-0.257,0.086){
         \mbox{\epsfxsize=6.0cm
          \epsffile{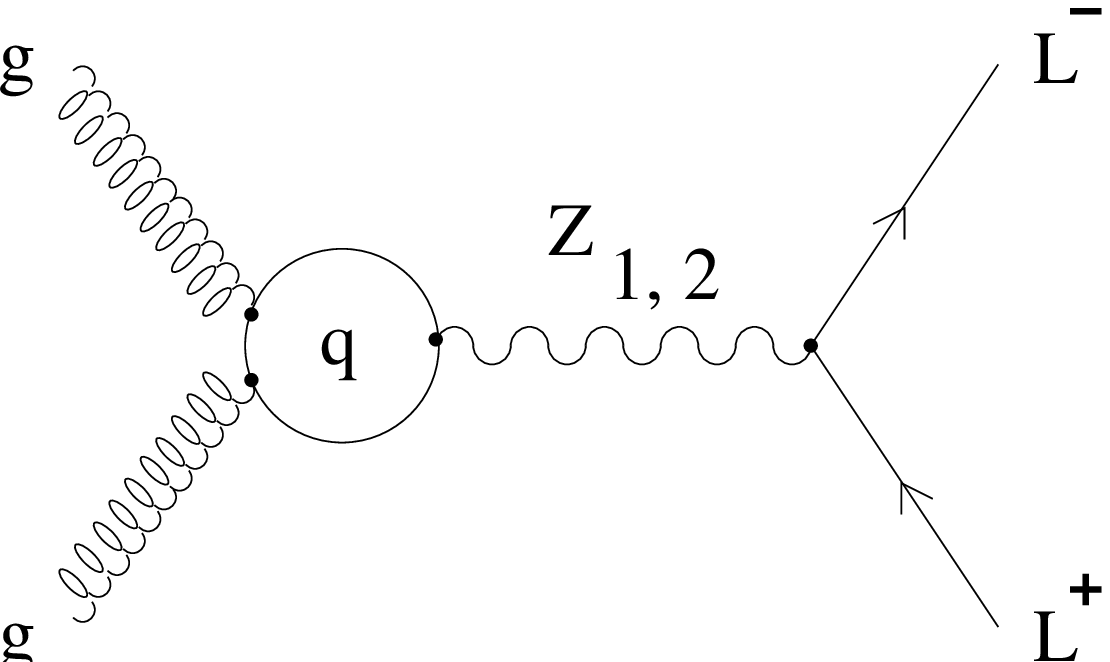}}
        }
       \end{picture}
      }}
     & 
  b) &
      \raisebox{-1.693cm}{
       \mbox{
       \setlength{\unitlength}{1cm}
       \begin{picture}(5.743,3.6)
        \put(-0.257,0.086){
         \mbox{\epsfxsize=6.0cm
          \epsffile{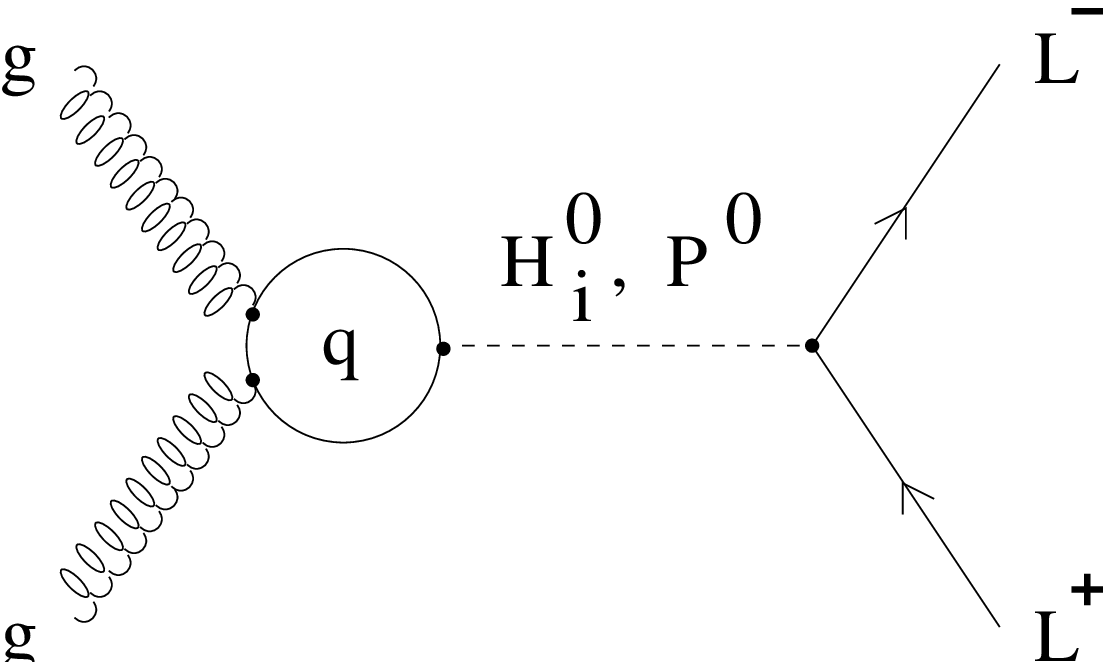}}
        }
       \end{picture}
      }} 
     \\ &&& \\
  c) & 
      \raisebox{-1.693cm}{
       \mbox{
       \setlength{\unitlength}{1cm}
       \begin{picture}(5.743,3.6)
        \put(-0.257,0.086){
         \mbox{\epsfxsize=6.0cm
          \epsffile{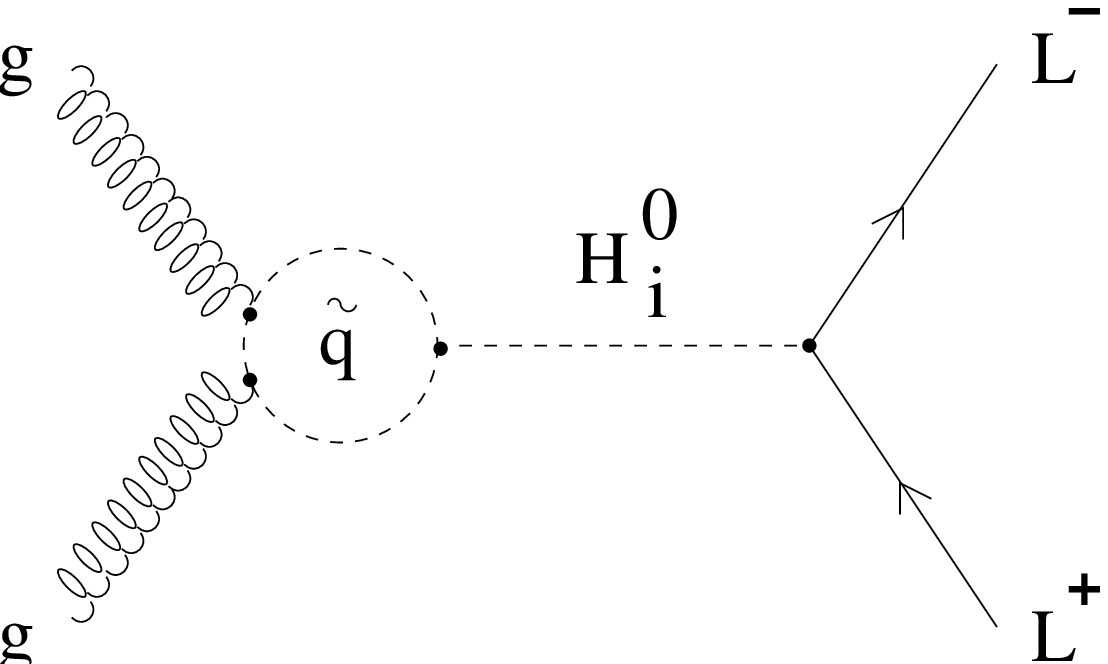}}
        }
       \end{picture}
      }}
     & 
  d) &
      \raisebox{-1.693cm}{
       \mbox{
       \setlength{\unitlength}{1cm}
       \begin{picture}(5.743,3.6)
        \put(-0.257,0.086){
         \mbox{\epsfxsize=6.0cm
          \epsffile{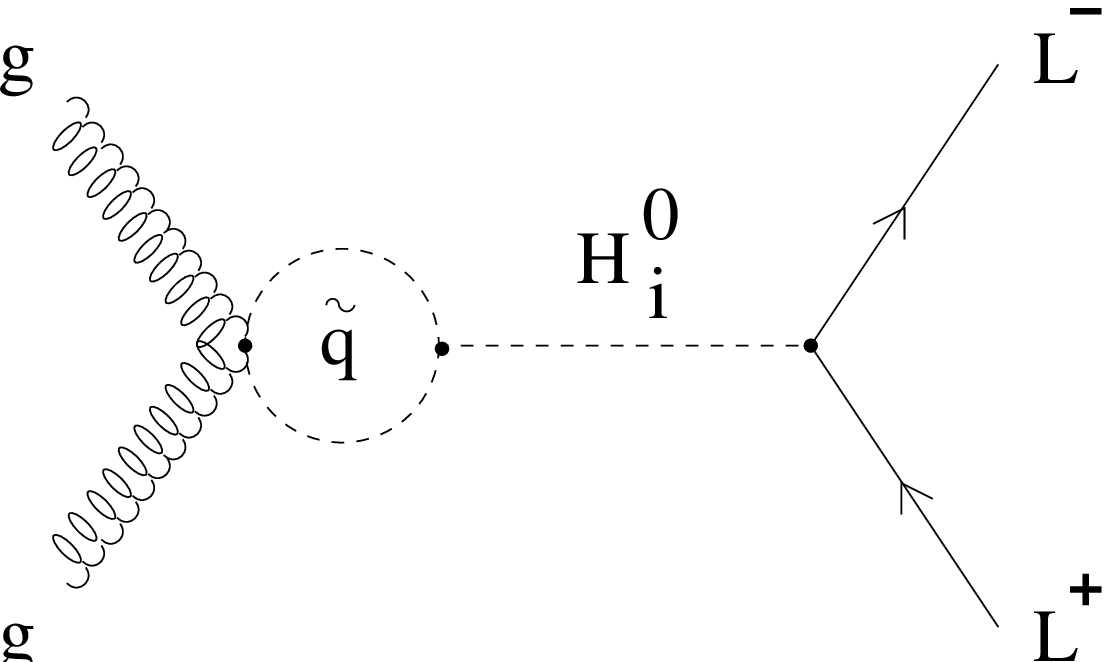}}
        }
       \end{picture}
      }}
\end{array}
$$
\caption[$gg\rightarrow L^+L^-$ Feynman  Diagrams]{\footnotesize Feynman
         diagrams for gluon-gluon fusion to charged heavy leptons.}
\label{fig:fusion}
\end{figure}

   Fig.~\ref{fig:fusion} shows  the Feynman diagrams  used for computing
the parton level gluon-gluon fusion to heavy leptons matrix elements.
It was found that the \mbox{${\rm E}_6$} matrix element computations are
very similar to the  corresponding $MSSM$ calculation by Cieza Montalvo,
$et$ $al.$,   \cite{kn:Montalvo}  ({\it cf}.~\cite{kn:KonigA})  and can,
with some care, be extracted from their paper.   The matrix elements are
as follows:
\begin{enumerate}
\item
For the $Z_{1,2}$ exchange diagram shown in Fig.~\ref{fig:fusion}(a)
\begin{equation}
\hat\sigma^{qZ_{1,2}}_{L^\pm}=
\frac{\alpha^2\alpha_s^2}{128\pi x_W^2}\,
\frac{m_{L}^2}{m_W^4}\,\beta_L
|\sum_{i=1}^2[\tilde C^{L^\pm Z_i}_L-\tilde C^{L^\pm Z_i}_R]\,
\xi^{Z_i}(\hat s)
\sum_q[\tilde C^{qZ_i}_L-\tilde C^{qZ_i}_R](1+2\lambda_qI_q)|^2
\label{eq:hatqz}
\end{equation}
where the left-right Fermion, $f\,$, couplings are given by
\begin{equation}
\left(\begin{array}{c}
   \tilde C^{fZ_1}_{L,R}\\\tilde C^{fZ_2}_{L,R}
\end{array}\right)=
\left(\begin{array}{rr}
   \cos\phi & \sin\phi \\
  -\sin\phi & \cos\phi
\end{array}\right)
\left(\begin{array}{c}
   C^{fZ}_{L,R}\\ C^{fZ^\prime}_{L,R}
\end{array}\right)\,,
\end{equation}
such that
\begin{eqnarray}
C^{fZ}_{L,R}        &=& T_{3_{L,R}}-e_f x^{\mbox{}}_W\,,
\label{eq:hatqzc}\\
C^{fZ^\prime}_{L,R} &=& 
\frac{1}{2}\left(\frac{ g^{\prime\prime}}{g}\right)
y^\prime_{f_{L,R}}\,\sqrt{1- x^{\mbox{}}_W}\,,\label{eq:hatqzd}
\end{eqnarray}
where   $T_{3_R}  = -T_{3_L}^c$     is the  isospin,  $y^\prime_{f_R}  =
-y^\prime_{f^c_L}$  is the $Y_E$-hypercharge, and  $e_f$ is the electric
charge.
\item
For  the    $H^0_{1,2,3}$ and  $P^0$   exchange   diagrams   shown  in
Fig.~\ref{fig:fusion}(b)
\begin{eqnarray}
\hat\sigma^{qH^0_{i}{\rm 's}}_{L^\pm}&=&
\frac{\alpha^2\alpha_s^2}{512\pi x_W^2}\,
\frac{m_{L}^2}{m_W^4}\,\beta_L^3
|\sum_{i=1}^3K^{L^\pm H^0_i}\zeta^{H^0_i}(\hat s)
\sum_qK^{qH^0_i}m_q^2[2+(4\lambda_q-1)I_q]|^2\,,\nonumber\\&&\\
\hat\sigma^{qP^0_{\;}\;\;}_{L^\pm}&=&
\frac{\alpha^2\alpha_s^2}{512\pi x_W^2}\,
\frac{m_{L}^2}{m_W^4}\,\beta_L(K^{L^\pm P^0})^2
|\zeta^{P^0}(\hat s)|^2
|\sum_qK^{qP^0}m_q^2I_q|^2\,,
\end{eqnarray}
where  the  couplings   $K^{fH^0_i}$    and $K^{fP^0}$   are  given   by
Eqs.~(\ref{eq:rata})-(\ref{eq:ratb}).
\item
For       the    $H^0_{1,2,3}$     exchange    diagrams       shown   in
Figs.~\ref{fig:fusion}(c) and \ref{fig:fusion}(d)
\begin{equation}
\hat\sigma^{\tilde qH^0_{i}{\rm 's}}_{L^\pm}=
\frac{\alpha^2\alpha_s^2m_L\beta_L}{512\pi x_W^2(1- x^{\mbox{}}_W)^2}\,|
\sum_{i=1}^3K^{L^\pm H^0_i}\zeta^{H^0_i}(\hat s)
\sum_{\tilde q}\sum_{k=1}^2\tilde K^{\tilde q H^0_i}_k
(1+2\lambda_{\tilde q_k}I_{\tilde q_k})|^2\,,
\end{equation}
where the  sfermion  mass eigenstates, $\tilde  f_{1,2}\,$, couplings to
the $H^0_i$ are given by
\begin{eqnarray}
\tilde K^{\tilde f H^0_i}_1&=&
K^{\tilde f H^0_i}_{LL}\cos^2\theta_{\tilde f}+
K^{\tilde f H^0_i}_{LR}\sin2\theta_{\tilde f}+
K^{\tilde f H^0_i}_{RR}\sin^2\theta_{\tilde f}
\,,\label{eq:kcupa}\\
\tilde K^{\tilde f H^0_i}_2&=&
K^{\tilde f H^0_i}_{LL}\sin^2\theta_{\tilde f}-
K^{\tilde f H^0_i}_{LR}\sin2\theta_{\tilde f}+
K^{\tilde f H^0_i}_{RR}\cos^2\theta_{\tilde f}
\,,\label{eq:kcupb}
\end{eqnarray}
where  $K^{\tilde   f  H^0_i}_{AB}$  ($A,B=L,R$)  are  the corresponding
couplings for  the sfermion  helicity  states, $\tilde f_{L,R}\,$,  with
mixing  angle  $\theta_{\tilde  f}\,$.   The  $K^{\tilde f  H^0_i}_{AB}$
couplings are given by Eqs.~(\ref{eq:cupd})-~(\ref{eq:cupdy}).
\item
For     the    $q(Z_{1,2}$-$P^0)$  interference   terms,    {\it    via}
Figs.~\ref{fig:fusion}(a) and~\ref{fig:fusion}(b),
\begin{eqnarray}
\hat\sigma^{\tilde q(Z_i{\rm 's}-P^0)}_{L^\pm}&=&
\frac{-\alpha^2\alpha_s^2}{128\pi x_W^2}\,\frac{m_L^2}{m_W^4}\,
\beta_L\,K^{L^\pm P^0}\,\mbox{Re}\,\{\,\zeta^{P^0}(\hat s)
\sum_{i=1}^2\xi^{Z_i}(\hat s)^*\,
[\tilde C^{L^\pm Z_i}_L-\tilde C^{L^\pm Z_i}_R]\nonumber\\&&\times
\sum_qK^{qP^0}m_q^2\,I_q\sum_{q^\prime}\,
[\tilde C^{q^\prime Z_i}_L-\tilde C^{q^\prime Z_i}_R]\,
(1+2\lambda_{q^\prime}I_{q^\prime}^*)\,\}\,.
\end{eqnarray}
\item
For  the  $(\tilde  q -   q)H^0_{1,2,3}$ interference  terms, {\it  via}
Figs.~\ref{fig:fusion}(b)-\ref{fig:fusion}(d),
\begin{eqnarray}
\hat\sigma^{(\tilde q-q)\tilde H^0_i{\rm 's}}_{L^\pm}&=&
\frac{-\alpha^2\alpha_s^2}{256\pi x_W^2(1- x^{\mbox{}}_W)^2}
\left(\frac{m_L^{\mbox{}}}{m_Z^{\mbox{}}}\right)^2
\beta_L^3\,\mbox{Re}\,\{
\sum_{i=1}^3K^{L_\pm H^0_i}\zeta^{H^0_i}(\hat s)
\;\;\;\;\;\;\;\;\;\;\;\;\nonumber\\&&\times
\sum_qK^{qH^0_i}m_q^2[2+(4\lambda_q-1)I_q]
\sum_{j=1}^3K^{L_\pm H^0_j}\zeta^{H^0_j}(\hat s)^*
\nonumber\\&&\times
\sum_{\tilde q}\sum_{k=1}^2\tilde K^{\tilde qH^0_j}_k
(1+2\lambda_{\tilde q}I_{\tilde q}^*)\}\,.
\end{eqnarray}
\end{enumerate}
In the aforementioned list of cross-section equations,
\begin{eqnarray}
\lambda_p&=&\frac{m_p^2}{\hat s}\,,\\
I_p&\equiv& I_p(\lambda_p)=\int_0^1\frac{dx}{x}\ln\left[
1-\frac{(1-x)x}{\lambda_p}\right]\nonumber\\&&\nonumber\\&=&
\left\{\begin{array}{l@{\;\;{\rm if}\;\;}l}
-2\left[\sin^{-1}\left(\frac{1}{2\sqrt{\lambda_p}}\right)\right]^2&
\lambda_p\,>\,\mbox{$\textstyle\frac{{1}}{{4}}$}\\
\mbox{$\textstyle\frac{{1}}{{2}}$}\,
\ln^2\left(\frac{r_+}{r_-}\right)-
\mbox{$\textstyle\frac{{\pi^2}}{{2}}$}+
i\pi\,\ln\left(\frac{r_+}{r_-}\right)&
\lambda_p\,<\,\mbox{$\textstyle\frac{{1}}{{4}}$}
\end{array}\right.\,,
\end{eqnarray}
with $r_\pm=1\pm\sqrt{1-4\lambda_p}\,$   such  that  $p\,  \varepsilon\,
\{f,\tilde f\}\,$,
\begin{eqnarray}
\;\;\;\;\xi^{Z_i}(\hat s)&=&
\frac{\hat s-m_{Z_i}^2}{\hat s-m_{Z_i}^2
+i\,m_{Z_i}^{\mbox{}}\Gamma_{Z_i}^{\mbox{}}}\,,\\
\zeta^{H^0_i,\,P^0}(\hat s)&=&
\frac{1}{\hat s-m_{H^0_i,P^0}^2
+i\,m_{H^0_i,\,P^0}^{\mbox{}}\Gamma_{H^0_i,\,P^0}^{\mbox{}}}\,,
\end{eqnarray}
where    the  $\Gamma_{V,\phi}$'s    computations   are   summarized  in
\S~\ref{sec-wit}, and
\begin{equation}
\beta_L=\sqrt{1-\frac{4m_L^2}{\hat s}}\,.
\end{equation}
The details  of   the various  components   that have   gone into   this
computation can be  found in appendix~\ref{sec-appc}.  Before the parton
level  cross-section can be used  to compute the heavy lepton production
rates some assumptions about the parameters and masses in the model must
be made.

\begin{figure}[hbtp]
\begin{center}
    \mbox{\epsfxsize=144mm\epsffile{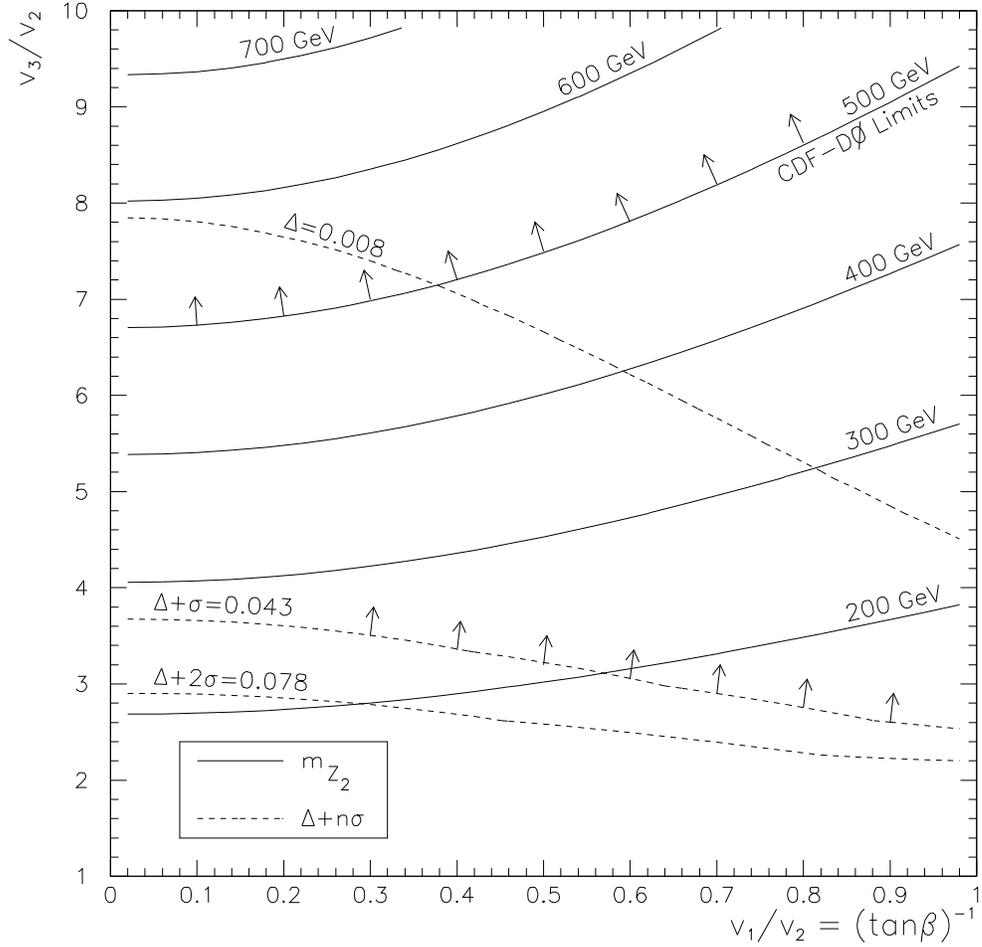}}
\end{center}
\caption[$(v_3/v_2,v_1/v_2)$ Parameter Space  Constraints]{\footnotesize
     Plot  of  $m_{Z_2}^{\mbox{}}$ and   $\Delta$  contour  lines  as  a
     function of $v_3/v_2$  and $v_1/v_2\,$.  The $\Delta$ contour lines
     are shown  at  the $0\sigma\,$, $1\sigma\,$, and  $2\sigma$ levels.
     The arrows point  to toward the allowed regions  on the plot  ({\it
     cf}.~\cite{kn:EllisA}).  The  $m_{Z_2}^{\mbox{}}=500\,GeV$     line
     shows the  $CDF$  and  $D0\!\!\!/$ constraints,  assuming  standard
     couplings \cite{kn:Shochet}.}
\label{fig:mztwo}
\end{figure}

     The first thing that has to be constrained are the $VEV\,$'s. It is
reasonable                   to                 assume              that
$v_1/v_2\,\raisebox{-0.625ex}{$\stackrel{<}{\sim}$}\,1$,           since
$m_b<<m_t\,$,    for  any    reasonable  range  of     Yukawa  couplings
\cite{kn:EllisA,kn:Gunion}.  Now the  ratios $v_1/v_2$ and $v_3/v_2$ can
be constrained by looking at how the variation in the $Z_1$ ({\it i.e.},
the      ``$Z$'')     mass        affects      $\bar      x^{\mbox{}}_W$
($\equiv\sin^2\bar\theta_W^{\mbox{}}$) such that \cite{kn:EllisA}
\begin{equation}
\sin^2\bar\theta_W^{\mbox{}}\equiv 1-\frac{m_W^2}{m_{Z_1}^2}\;\;<\;\;
\sin^2\theta_W^{\mbox{}}\equiv 
\left.\frac{ g^{\prime^2}}{g^2+ g^{\prime^2}}
\right|_{\mu= m^{\mbox{}}_W}\,.
\end{equation}
$   x^{\mbox{}}_W(  m^{\mbox{}}_W)$   can  be    found  by  evolving   $
x^{\mbox{}}_W(m^{\mbox{}}_Z)\approx0.2319\pm0.0005$ \cite{kn:PDG}   down
to $ m^{\mbox{}}_W\,$   \cite{kn:Barger}, which  gives $  x^{\mbox{}}_W(
m^{\mbox{}}_W)        \approx         0.233\pm0.035\,$,             with
$\alpha^{-1}(m^{\mbox{}}_Z)   \approx      127.9\pm0.1$   \cite{kn:PDG},
$m^{\mbox{}}_Z \approx   (91.187\pm0.007)  \,GeV$ \cite{kn:PDG},   and $
m^{\mbox{}}_W   \approx (80.23\pm0.18)   \,GeV$       \cite{kn:Shochet}.
Therefore   given     $\bar x^{\mbox{}}_W      \approx  0.2247\pm0.0019$
\cite{kn:PDG} yields
\begin{equation}
\Delta\equiv x^{\mbox{}}_W-\bar x^{\mbox{}}_W\approx0.008\pm0.035\,.
\end{equation}
Fig.~\ref{fig:mztwo} shows the  $\Delta$ contour line  as a function  of
$v_3/v_2$  and $v_1/v_2$, along with  its $1\sigma$  and $2\sigma$ level
contour lines.  Also   shown are the $m^{\mbox{}}_{Z_2}$ contour  lines.
Taking    a    $1\sigma$      level   constraint    implies     $v_3/v_2
\,\raisebox{-0.625ex}{$\stackrel{>}{\sim}$}\,  {\cal    O}({3.5})$ ({\it
cf}.~\cite{kn:Hewett,kn:EllisA})         and          $m^{\mbox{}}_{Z_2}
\,\raisebox{-0.625ex}{$\stackrel{>}{\sim}$}\,   {\cal   O}({200})\,GeV$.
Unfortunately these  constraints  are not  that tight  due to  the large
uncertainty in   $\alpha(m^{\mbox{}}_Z)$.  A stronger constraint  can be
found    by     using the  $CDF$   and       $D0\!\!\!/$ limits on   the
$m^{\mbox{}}_{Z_2}$  mass \cite{kn:Shochet}, assuming SM-like couplings,
Fig.~\ref{fig:mztwo}.  This    constraint  is  fairly   reasonable since
$Y_E{\rm 's}\sim{\cal  O}(Y)$'s ({\it cf}.  table~\ref{tb:parpr}).  With
these constraints $v_3/v_2 \,\raisebox{-0.625ex}{$\stackrel{>}{\sim}$}\,
{\cal                      O}({7.5})$                                and
$m^{\mbox{}}_{Z_2}\,\raisebox{-0.625ex}{$\stackrel{>}{\sim}$}\,    {\cal
O}({500})\,GeV$.

\begin{figure}[hbtp]
\begin{center}
   \mbox{\epsfxsize=144mm\epsffile{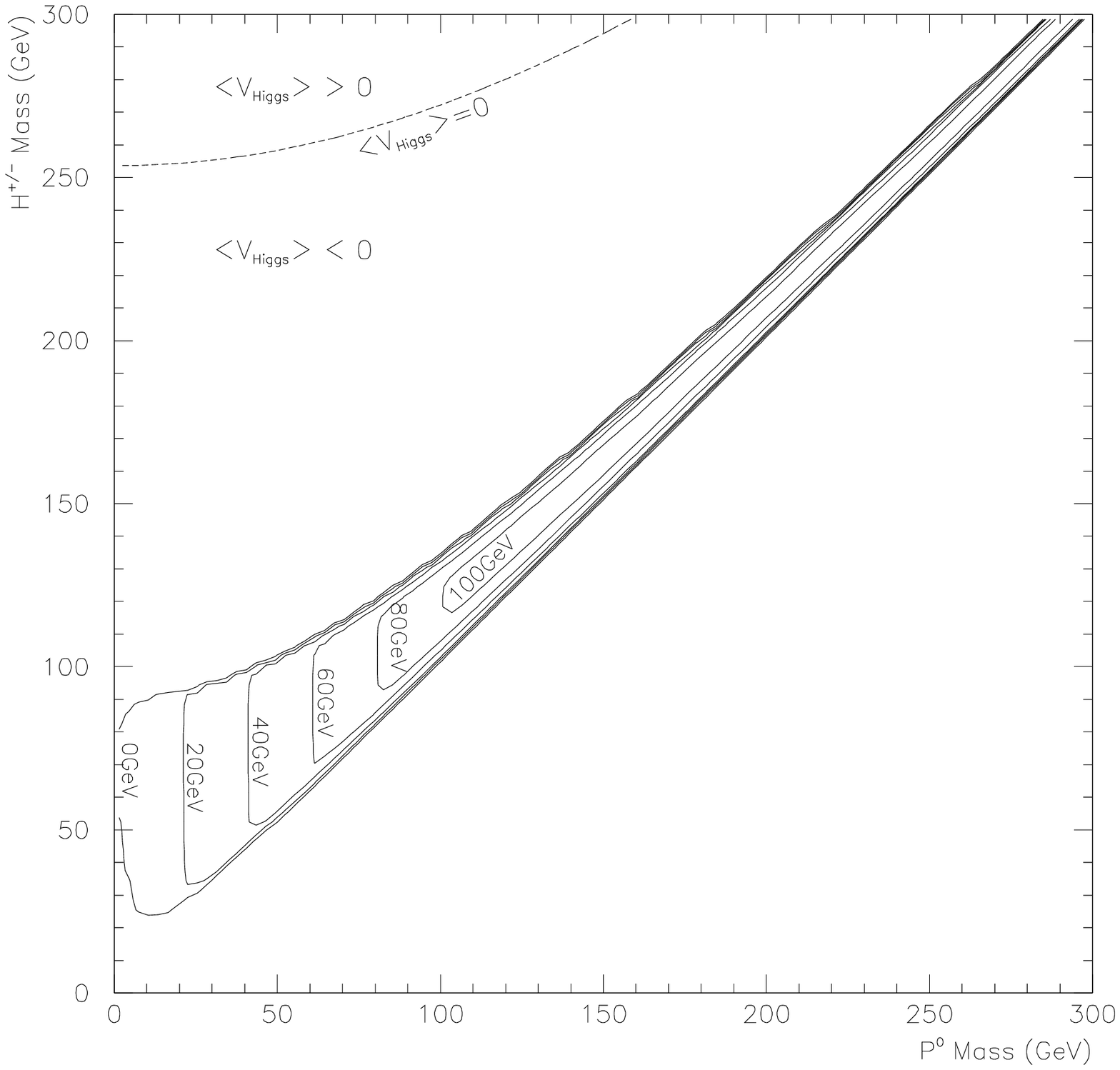}}
\end{center}
\caption[$m_{H^0_1}^{\mbox{}}(m^{\mbox{}}_{P^0},m^{\mbox{}}_{H^\pm})$
         contour       plot     for        $v_1/v_2=0.02$            and
         $v_3/v_2=6.7\,$]{\footnotesize  A plot  of   the  $H^0_1$  mass
         contour  lines as     a  function of  $m^{\mbox{}}_{P^0}$   and
         $m^{\mbox{}}_{H^\pm}$,  for  $v_1/v_2=0.02$ and $v_3/v_2=6.7\,$
         ($m^{\mbox{}}_{Z_2}\approx496\,GeV$).  The dashed curve in  the
         upper left-hand corner  is  a plot  of  the zero of   the Higgs
         potential above which it becomes positive.}
\label{fig:mhzeroa}
\end{figure}

\begin{figure}[hbtp]
\begin{center}
   \mbox{\epsfxsize=144mm\epsffile{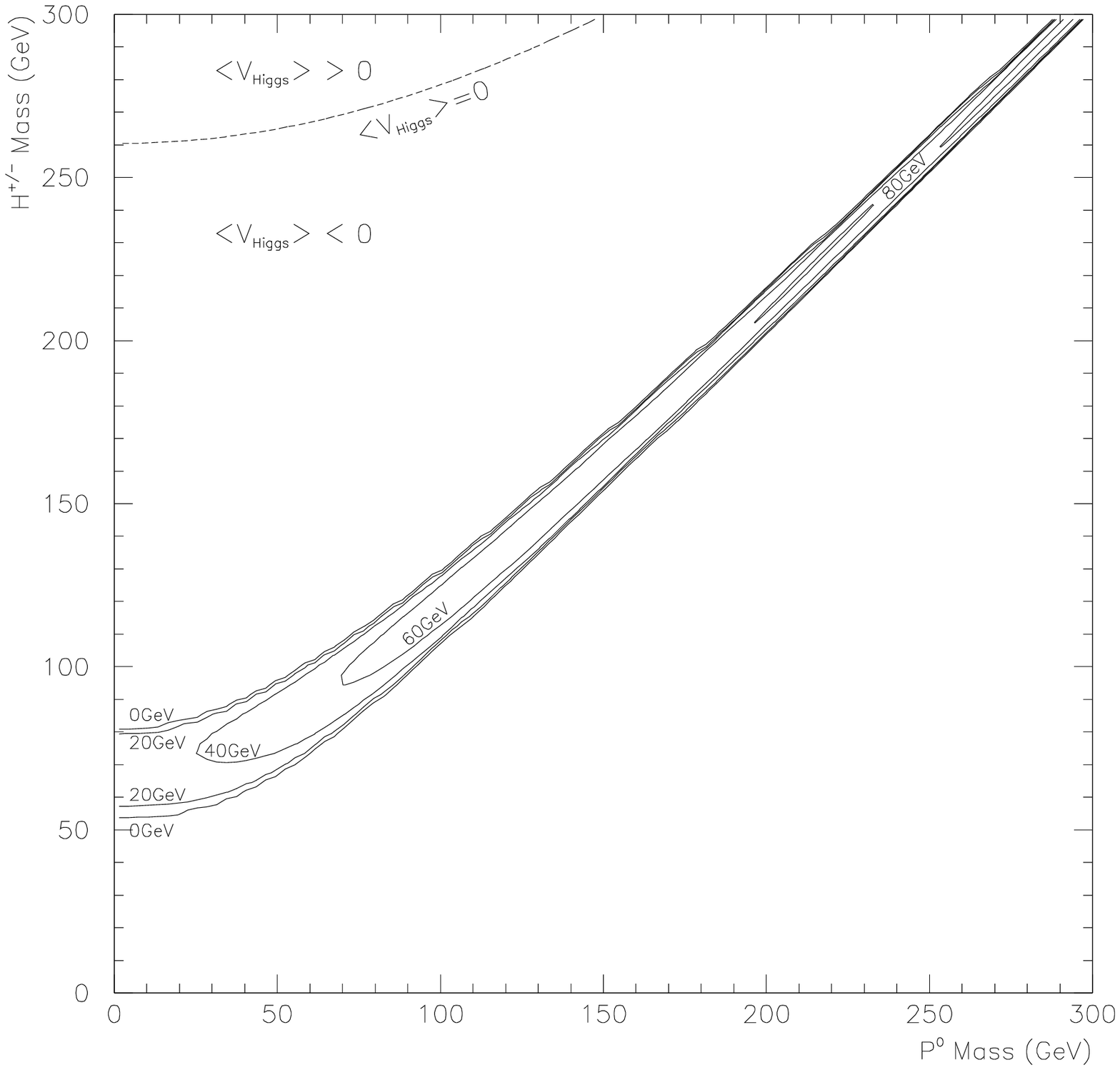}}
\end{center}
\caption[$m_{H^0_1}^{\mbox{}}(m^{\mbox{}}_{P^0},m^{\mbox{}}_{H^\pm})$
         contour       plot       for         $v_1/v_2=0.02$         and
         $v_3/v_2=7.7\,$]{\footnotesize  A  plot   of the $H^0_1$   mass
         contour   lines  as   a   function of  $m^{\mbox{}}_{P^0}$  and
         $m^{\mbox{}}_{H^\pm}$, for $v_1/v_2=0.5$ and    $v_3/v_2=7.7\,$
         ($m^{\mbox{}}_{Z_2}\approx509\,GeV$).  The dashed curve in  the
         upper  left-hand  corner is a  plot  of the  zero  of the Higgs
         potential above which it becomes positive.}
\label{fig:mhzeroc}
\end{figure}

\begin{figure}[hbtp]
\begin{center}
   \mbox{\epsfxsize=144mm\epsffile{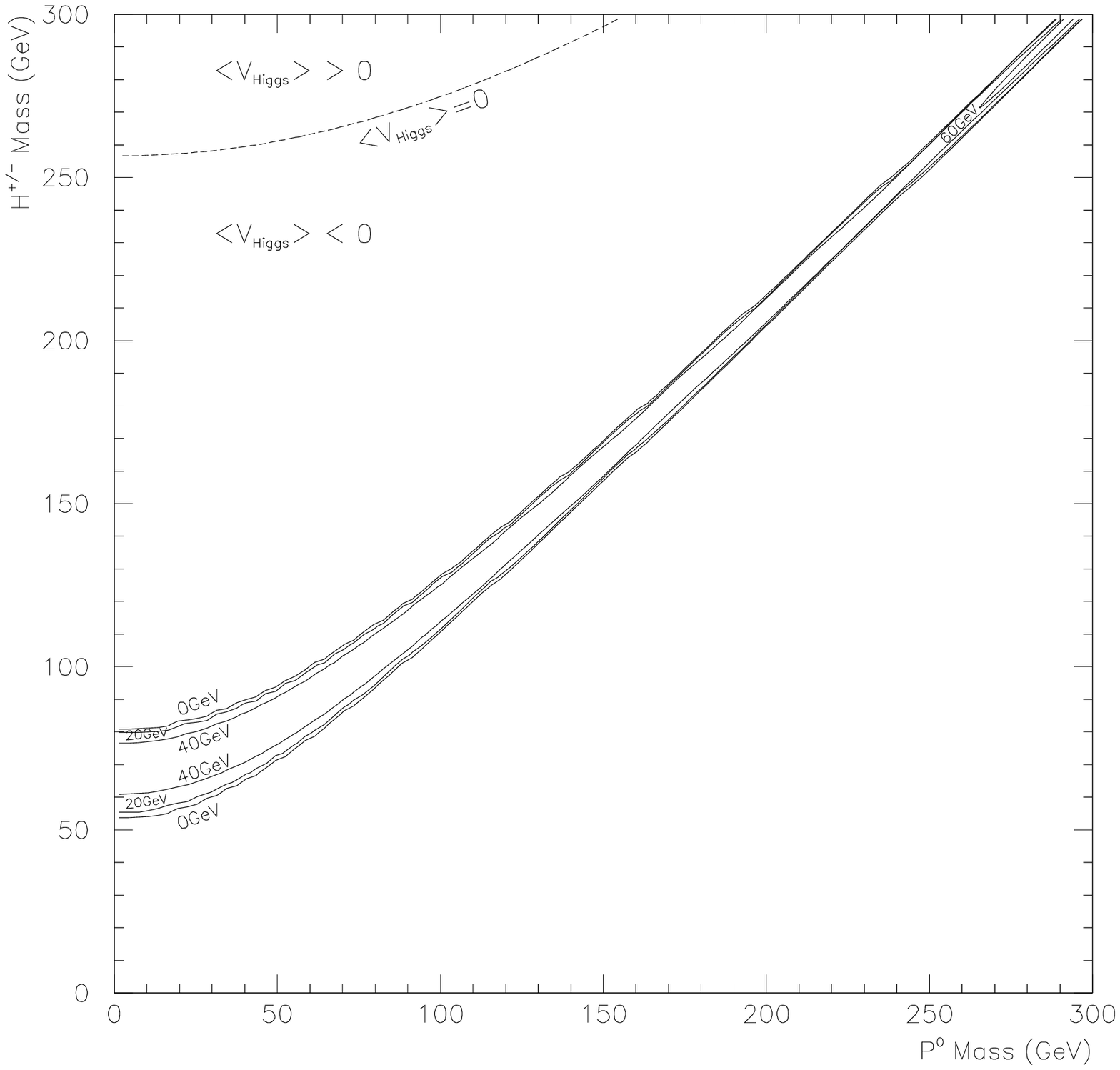}}
\end{center}
\caption[$m_{H^0_1}^{\mbox{}}(m^{\mbox{}}_{P^0},m^{\mbox{}}_{H^\pm})$
         contour              plot            for   $v_1/v_2=0.9$    and
         $v_3/v_2=9.1\,$]{\footnotesize   A  plot of   the  $H^0_1$ mass
         contour   lines  as a     function of  $m^{\mbox{}}_{P^0}$  and
         $m^{\mbox{}}_{H^\pm}$, for $v_1/v_2=0.9$    and $v_3/v_2=9.1\,$
         ($m^{\mbox{}}_{Z_2}\approx499\,GeV$).  The dashed curve in  the
         upper left-hand corner  is  a plot  of  the zero of  the  Higgs
         potential above which it becomes positive.}
\label{fig:mhzerob}
\end{figure}

  Figs.~\ref{fig:mhzeroa} through   \ref{fig:mhzerob} show $H^0_1$  mass
contour   plots    as    function     of     $m^{\mbox{}}_{P^0}$     and
$m^{\mbox{}}_{H^\pm}$            for     $(v_1/v_2,v_3/v_2)=(0.02,6.7)$,
$(v_1/v_2,v_3/v_2)=(0.5,7.7)$,     and    $(v_1/v_2,v_3/v_2)=(0.9,9.1)$,
respectively, such  that $m_{Z_2}$  lies roughly  around  the  $CDF$ and
$D0\!\!\!/$ limits.   These figures are  a fairly good representation of
the behavior of the $m_{H^0_1}^{\mbox{}}$ contour lines as a function of
$v_1/v_2$.  For fixed $v_1/v_2$,  the  contour lines change very  little
({\it     i.e.},     $\,\raisebox{-0.625ex}{$\stackrel{<}{\sim}$}\,{\cal
O}({5})$\%)                          for                          ${\cal
O}({10})\,\raisebox{-0.625ex}{$\stackrel{>}{\sim}$}\,
v_3/v_2\,\raisebox{-0.625ex}{$\stackrel{>}{\sim}$}\, {\cal O}({4.5})\,$.
This region  corresponds to the  $(v_1/v_2,v_3/v_2)$ parameter space for
$m_{Z_2}\,\raisebox{-0.625ex}{$\stackrel{>}{\sim}$}\,              {\cal
O}({300})\,GeV$  depicted in Fig.~\ref{fig:mztwo}.  Further  examination
of  the    other scalar-Higgses  shows  $m_{H^0_3}^{\mbox{}}$  is fairly
insensitive   to      variations     in    $m^{\mbox{}}_{P^0}$       and
$m^{\mbox{}}_{H^\pm}$  for  fixed   $m^{\mbox{}}_{Z_2}$ and is  slightly
sensitive    to variations in   $v_3/v_2$,   whereas  the behaviour   of
$m_{H^0_2}^{\mbox{}}$ appears to be  quite  sensitive to any  variation.
Fortunately for  the range of VEV's  considered here  ({\it i.e.}, large
$v_3$), the only contributions  to the parton level  cross-sections turn
out to be the diagrams which contain the  $Z_i$ and $H^0_3$ propagators;
the other  terms  are, in    general, suppressed by  several  orders  of
magnitude.\footnote{in   the     large   $v_3$  limit    the   couplings
$P^0L^+L^-\rightarrow0$,       {\it      via}       Eqs.~(\ref{eq:cupc})
and~(\ref{eq:ratb}),            and              $H^0_iL^+L^-\rightarrow
-(m^{\mbox{}}_{L}/v_3)\,\delta_{3i}\,$,  {\it via}  Eqs.~(\ref{eq:cupb})
and     (\ref{eq:ratae}),    and     Eq.~(4.12)     of    Hewett     and
Rizzo~\cite{kn:Hewett}  for  the $U_{3i}$'s   in this limit,   to ${\cal
O}({1/v_3})$.} Therefore  the  heavy lepton  production cross-section is
insensitive  to         variations  in      $m^{\mbox{}}_{P^0}$      and
$m^{\mbox{}}_{H^\pm}\,$.  Here the $P^0$ mass will be set to $200\,GeV$.
The   corresponding  $H^\pm$  mass was   chosen    to be $215\,GeV$  for
Fig.~\ref{fig:mhzeroa}    and     $212\,GeV$ for Figs.~\ref{fig:mhzeroc}
and~\ref{fig:mhzerob},  which lies  within the  allowed  regions on  the
$m_{H^0_1}^{\mbox{}}$ contour    plots.   Based  on   the  very  limited
experimental    constraints      that do     exist    for supersymmetric
models~\cite{kn:PDG} these appear to be very conservative choices.  They
also lead to fairly reasonable values for the $H^0_i$ masses.

%
%
%

%
%
%

%
%
%

   The next parameters that need to be fixed are the soft terms. Exactly
how these  terms should  behave at  low  energy  is not clear.    At the
moment,   their   behaviour   is  very    model   dependent and   unless
supersymmetric particles  are  found  this  situation  will most  likely
remain so.  Here    the soft terms  will  be  treated  parametrically as
function of a  single parameter $  m^{\mbox{}}_S\,$.   In particular the
soft terms  will   be assumed  to  be degenerate,  $\tilde  M_{f}\approx
A_f\approx m^{\mbox{}}_S\,$, with the exception of the $\lambda$ and $A$
terms    which  were fixed by   selecting    the $m^{\mbox{}}_{P^0}$ and
$m^{\mbox{}}_{H^\pm}$ masses.   The selection of  the soft terms in this
way,           including           $\lambda$      and     $A$        for
$m_{P^0,\,H^\pm}^{\mbox{}}\,\raisebox{-0.625ex}{$\stackrel{<}{\sim}$}\,
{\cal O}({1})\,TeV$,  typify   the generic  outcome,  for the   sfermion
masses,        of        most       SUSY-breaking    models,        {\it
cf}.~\cite{kn:EllisB,kn:Gunion}.    In       these      models    ${\cal
O}({0.2})TeV\,\raisebox{-0.625ex}{$\stackrel{<}{\sim}$}\,
m^{\mbox{}}_S\,\raisebox{-0.625ex}{$\stackrel{<}{\sim}$}\,         {\cal
O}({10})\,TeV$.  How  low $  m^{\mbox{}}_S$ can  be  pushed down depends
upon the  choice of VEV's ($v_3$  in particular, since it  is relatively
large).    For     the  VEV's    used     here   it    was    found    $
m^{\mbox{}}_S\,\raisebox{-0.625ex}{$\stackrel{>}{\sim}$}\,         {\cal
O}({400-450})\,GeV$.  In general the sfermions with light spartners have
masses        $\,\raisebox{-0.625ex}{$\stackrel{>}{\sim}$}\,{\cal     O}
({m^{\mbox{}}_S})\,$,  which   are  roughly   degenerate  (within ${\cal
O}({50})\,GeV$) with their mass-eigenstate partners.  The stops, $\tilde
t^\prime_{1,2}\,$, and  the  exotic  squarks, $\tilde q^\prime_{1,2}\,$,
have  splittings     $\,\raisebox{-0.625ex}{$\stackrel{>}{\sim}$}\,{\cal
O}({\mbox{\footnotesize$\frac{{1}}{{2}}$}\,  m_{t,\,q^\prime}})\,$,  for
fermion                           masses                          ${\cal
O}({200})\,\raisebox{-0.625ex}{$\stackrel{<}{\sim}$}\,
m_{t,\,q^\prime}\,\raisebox{-0.625ex}{$\stackrel{<}{\sim}$}\,      {\cal
O}({600})\,GeV$,  for low   values  of    $  m^{\mbox{}}_S\,$.  As     $
m^{\mbox{}}_S$ approaches ${\cal  O}({1})\,TeV$ all of the sfermion mass
become degenerate and $\approx{\cal O}({ m^{\mbox{}}_S})\,$.

  Finally there is the matter  of fixing the heavy  fermion masses.  The
heavy quark masses, $m_{q^\prime}\,$, will be assumed degenerate as will
the    charged   heavy   leptons  masses,  $m_{l^\prime}\,$,   such that
$m_{q^\prime,\,l^\prime}\,\raisebox{-0.625ex}{$\stackrel{>}{\sim}$}\,
{\cal  O}({0.1})TeV   $~\cite{kn:PDG}.   The $e^{\prime^\pm}$   will  be
designated to play the role of the charged heavy leptons, $L^\pm\,$.

\begin{figure}[hbtp]
\begin{center}
   \mbox{\epsfxsize=144mm\epsffile{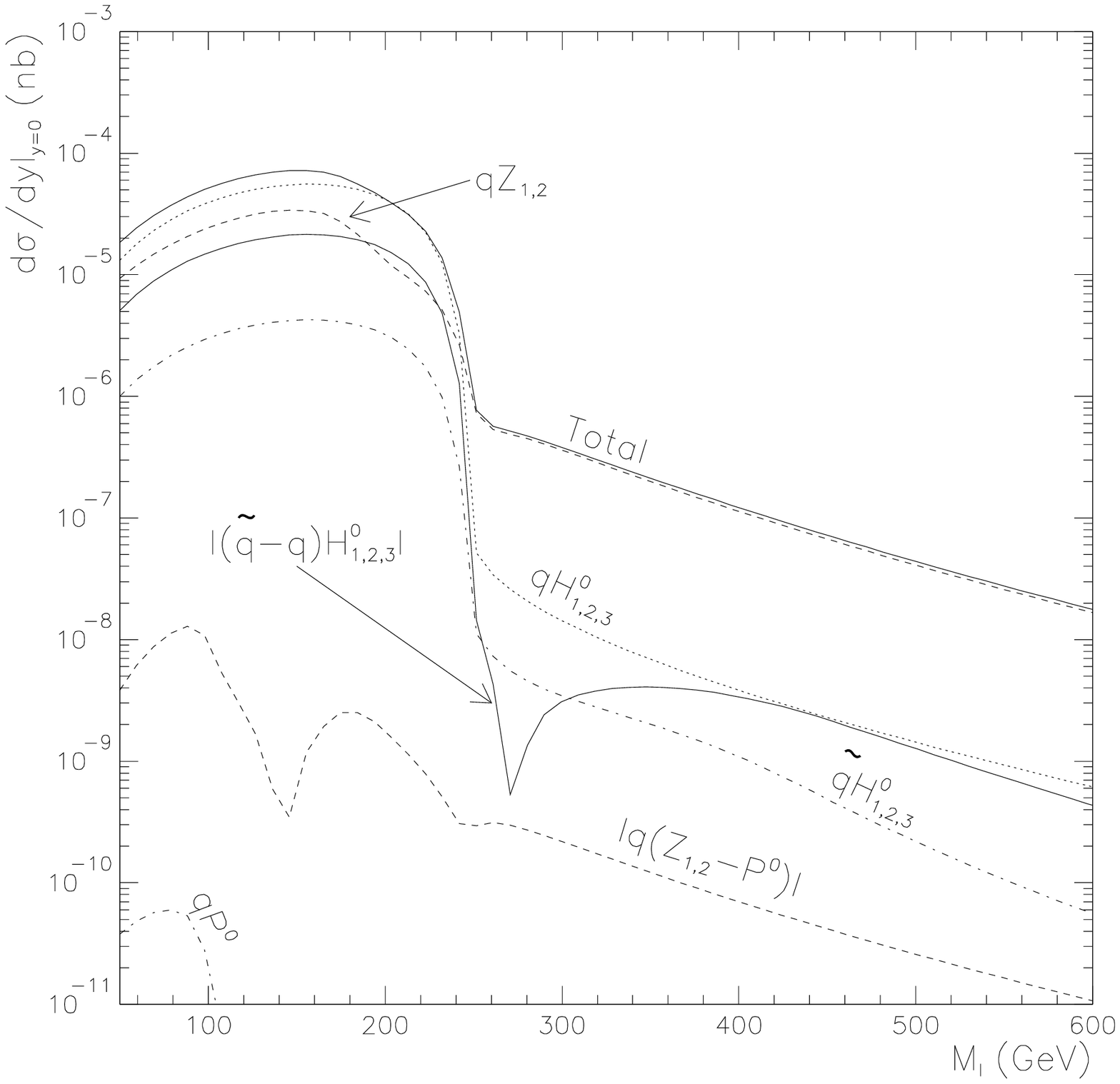}}
\end{center}
\caption[$\frac{d\sigma}{dy}|_{y=0}$     at  $LHC$ for $v_1/v_2=0.02\,$,
         $v_3/v_2=6.7\,$,     $     m^{\mbox{}}_S=400\,GeV$,    and    $
         m^{\mbox{}}_{q^\prime}=200\,GeV$]{\footnotesize        Rapidity
         distribution at $y=0$   for charged heavy  lepton production at
         $LHC$ ($14\,TeV$)  as  a function  of heavy lepton  mass, where
         $v_1/v_2=0.02\,$,          $v_3/v_2=6.7\,$,       and         $
         m^{\mbox{}}_S=400\,GeV$.    The  mass  spectrum  for the non-SM
         particles    involved     in  these        processes       are,
         $m^{\mbox{}}_{Z_2}\approx496\,GeV$
         ($\Gamma^{\mbox{}}_{Z_2}\approx20.9\,GeV$),
         $m^{\mbox{}}_{P^0}\approx200\,GeV$
         ($\Gamma^{\mbox{}}_{P^0}\approx16.4\,GeV$),
         $m^{\mbox{}}_{H^\pm}\approx215\,GeV$,
         $m^{\mbox{}}_{H^0_1}\approx94.3\,GeV$
         ($\Gamma^{\mbox{}}_{H^0_1}\approx7.50\times10^{-3}\,GeV$),
         $m^{\mbox{}}_{H^0_2}\approx200\,GeV$
         ($\Gamma^{\mbox{}}_{H^0_2}\approx16.5\,GeV$),
         $m^{\mbox{}}_{H^0_3}\approx495\,GeV$
         ($\Gamma^{\mbox{}}_{H^0_3}\approx0.230\,GeV$),                $
         m^{\mbox{}}_{q^\prime}=200\,GeV$.}
\label{fig:decaya}
\end{figure}

\begin{figure}[hbtp]
\begin{center}
   \mbox{\epsfxsize=144mm\epsffile{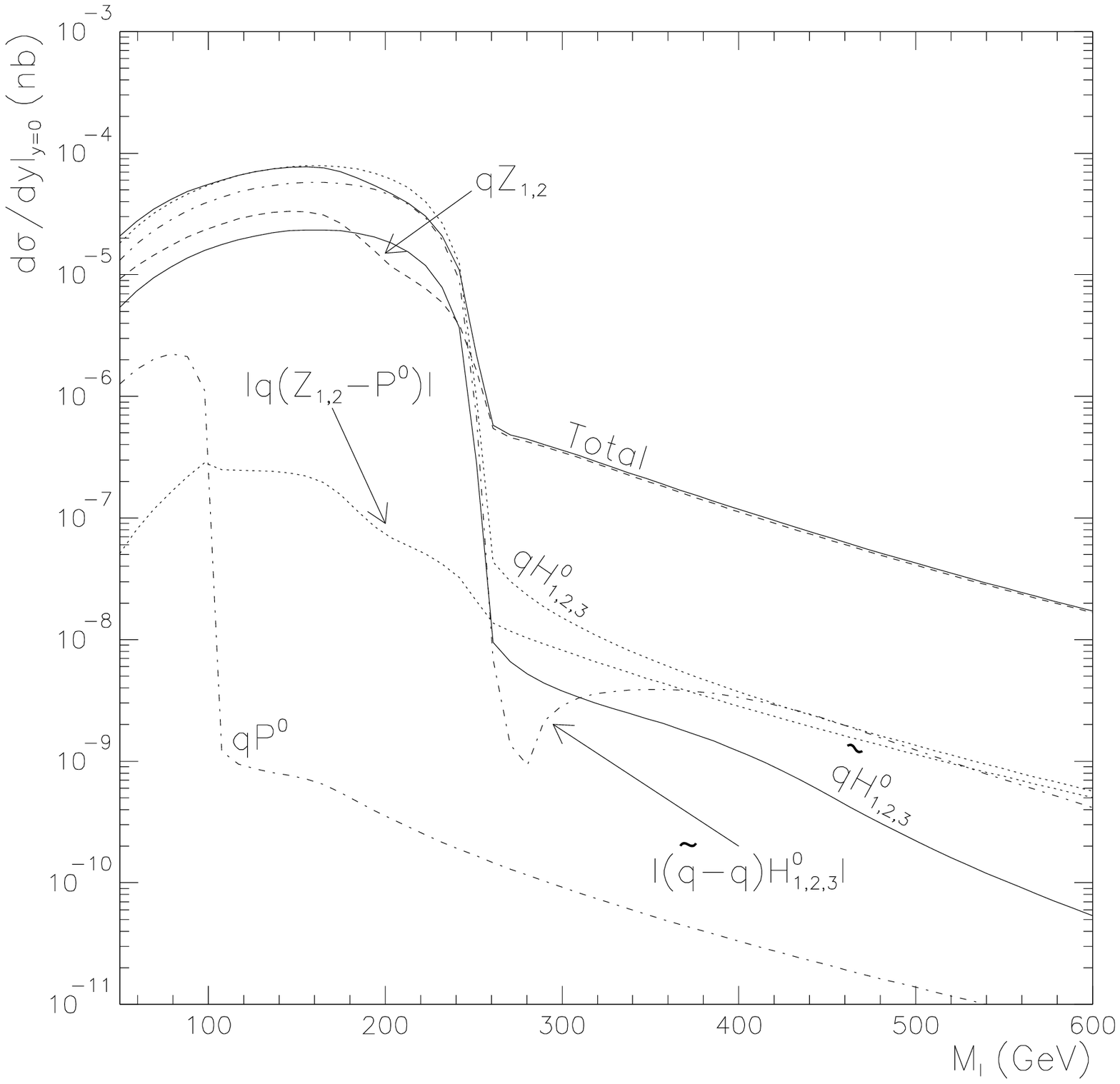}}
\end{center}
\caption[$\frac{d\sigma}{dy}|_{y=0}$    at  $LHC$ for   $v_1/v_2=0.5\,$,
         $v_3/v_2=7.7\,$,      $    m^{\mbox{}}_S=400\,GeV$,     and   $
         m^{\mbox{}}_{q^\prime}=200\,GeV$]{\footnotesize        Rapidity
         distribution at $y=0$   for charged heavy  lepton production at
         $LHC$ ($14\,TeV$) as  a function  of  heavy lepton mass,  where
         $v_1/v_2=0.5\,$,        $v_3/v_2=7.7\,$,        and           $
         m^{\mbox{}}_S=400\,GeV$.  The mass   spectrum for   the  non-SM
         particles   involved    in     these  processes            are,
         $m^{\mbox{}}_{Z_2}\approx509\,GeV$
         ($\Gamma^{\mbox{}}_{Z_2}\approx21.5\,GeV$),
         $m^{\mbox{}}_{P^0}\approx200\,GeV$
         ($\Gamma^{\mbox{}}_{P^0}\approx2.52\times10^{-2}\,GeV$),
         $m^{\mbox{}}_{H^\pm}\approx212\,GeV$,
         $m^{\mbox{}}_{H^0_1}\approx75.4\,GeV$
         ($\Gamma^{\mbox{}}_{H^0_1}\approx3.65\times10^{-3}\,GeV$),
         $m^{\mbox{}}_{H^0_2}\approx212\,GeV$
         ($\Gamma^{\mbox{}}_{H^0_2}\approx7.49\times10^{-2}\,GeV$),
         $m^{\mbox{}}_{H^0_3}\approx507\,GeV$
         ($\Gamma^{\mbox{}}_{H^0_3}\approx0.198\,GeV$),                $
         m^{\mbox{}}_{q^\prime}=200\,GeV$.}
\label{fig:decayb}
\end{figure}

\begin{figure}[hbtp]
\begin{center}
   \mbox{\epsfxsize=144mm\epsffile{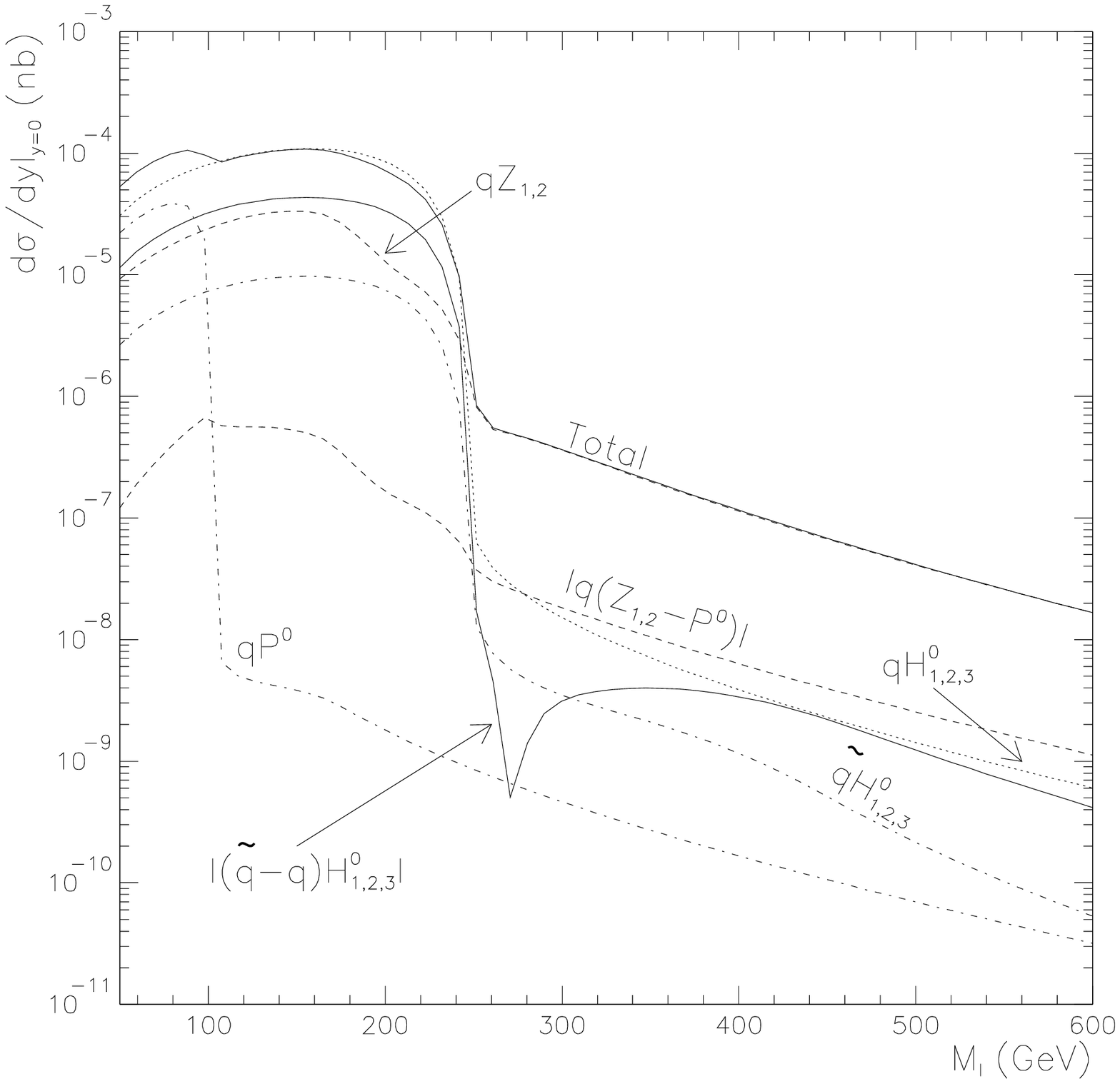}}
\end{center}
\caption[$\frac{d\sigma}{dy}|_{y=0}$   at    $LHC$ for  $v_1/v_2=0.9\,$,
         $v_3/v_2=9.1\,$,  $    m^{\mbox{}}_S=400\,GeV$,       and     $
         m^{\mbox{}}_{q^\prime}=200\,GeV$]{\footnotesize        Rapidity
         distribution at  $y=0$  for charged heavy lepton  production at
         $LHC$ ($14\,TeV$) as   a function of  heavy  lepton mass, where
         $v_1/v_2=0.9\,$,           $v_3/v_2=9.1\,$,      and          $
         m^{\mbox{}}_S=400\,GeV$.   The  mass  spectrum  for the  non-SM
         particles    involved        in     these    processes     are,
         $m^{\mbox{}}_{Z_2}\approx499\,GeV$
         ($\Gamma^{\mbox{}}_{Z_2}\approx20.8\,GeV$),
         $m^{\mbox{}}_{P^0}\approx200\,GeV$
         ($\Gamma^{\mbox{}}_{P^0}\approx8.13\times10^{-3}\,GeV$),
         $m^{\mbox{}}_{H^\pm}\approx212\,GeV$,
         $m^{\mbox{}}_{H^0_1}\approx52.3\,GeV$
         ($\Gamma^{\mbox{}}_{H^0_1}\approx1.87\times10^{-3}\,GeV$),
         $m^{\mbox{}}_{H^0_2}\approx216\,GeV$
         ($\Gamma^{\mbox{}}_{H^0_2}\approx1.37\times10^{-2}\,GeV$),
         $m^{\mbox{}}_{H^0_3}\approx498\,GeV$
         ($\Gamma^{\mbox{}}_{H^0_3}\approx0.130\,GeV$),                $
         m^{\mbox{}}_{q^\prime}=200\,GeV$.}
\label{fig:decayc}
\end{figure}

\begin{figure}[hbtp]
\begin{center}
   \mbox{\epsfxsize=144mm\epsffile{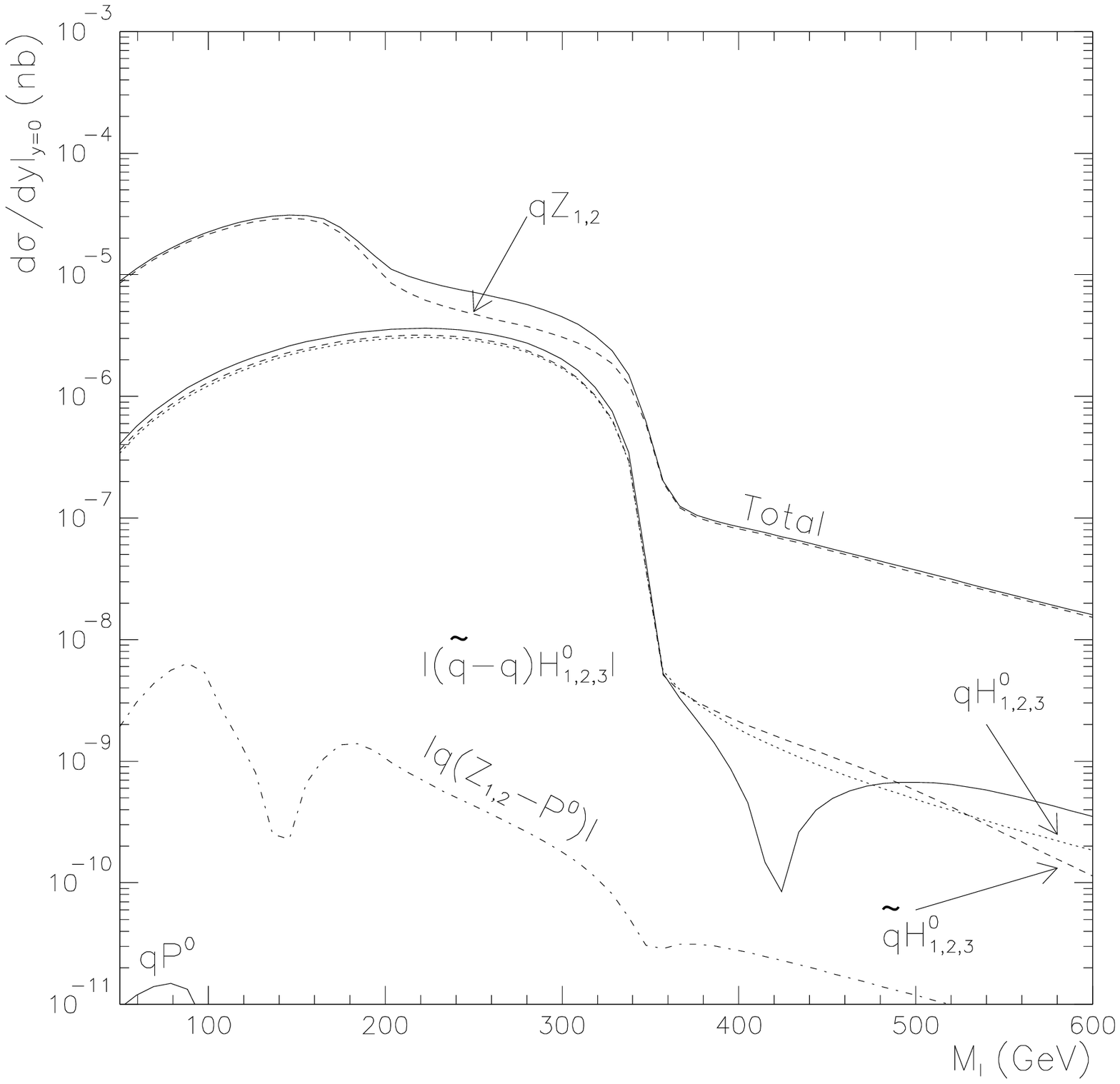}}
\end{center}
\caption[$\frac{d\sigma}{dy}|_{y=0}$    at  $LHC$  for $v_1/v_2=0.02\,$,
         $v_3/v_2=9.5\,$,    $   m^{\mbox{}}_S=450\,GeV$,      and     $
         m^{\mbox{}}_{q^\prime}=200\,GeV$]{\footnotesize        Rapidity
         distribution at  $y=0$ for  charged heavy lepton  production at
         $LHC$ ($14\,TeV$) as a   function of heavy lepton  mass,  where
         $v_1/v_2=0.02\,$,          $v_3/v_2=9.5\,$,      and          $
         m^{\mbox{}}_S=450\,GeV$.    The  mass spectrum for  the  non-SM
         particles    involved      in     these     processes      are,
         $m^{\mbox{}}_{Z_2}\approx700\,GeV$
         ($\Gamma^{\mbox{}}_{Z_2}\approx31.9\,GeV$),
         $m^{\mbox{}}_{P^0}\approx200\,GeV$
         ($\Gamma^{\mbox{}}_{P^0}\approx16.5\,GeV$),
         $m^{\mbox{}}_{H^\pm}\approx215\,GeV$,
         $m^{\mbox{}}_{H^0_1}\approx94.6\,GeV$
         ($\Gamma^{\mbox{}}_{H^0_1}\approx7.49\times10^{-3}\,GeV$),
         $m^{\mbox{}}_{H^0_2}\approx200\,GeV$
         ($\Gamma^{\mbox{}}_{H^0_2}\approx16.5\,GeV$),
         $m^{\mbox{}}_{H^0_3}\approx700\,GeV$
         ($\Gamma^{\mbox{}}_{H^0_3}\approx1.04\,GeV$),                 $
         m^{\mbox{}}_{q^\prime}=200\,GeV$.}
\label{fig:decayd}
\end{figure}

\begin{figure}[hbtp]
\begin{center}
   \mbox{\epsfxsize=144mm\epsffile{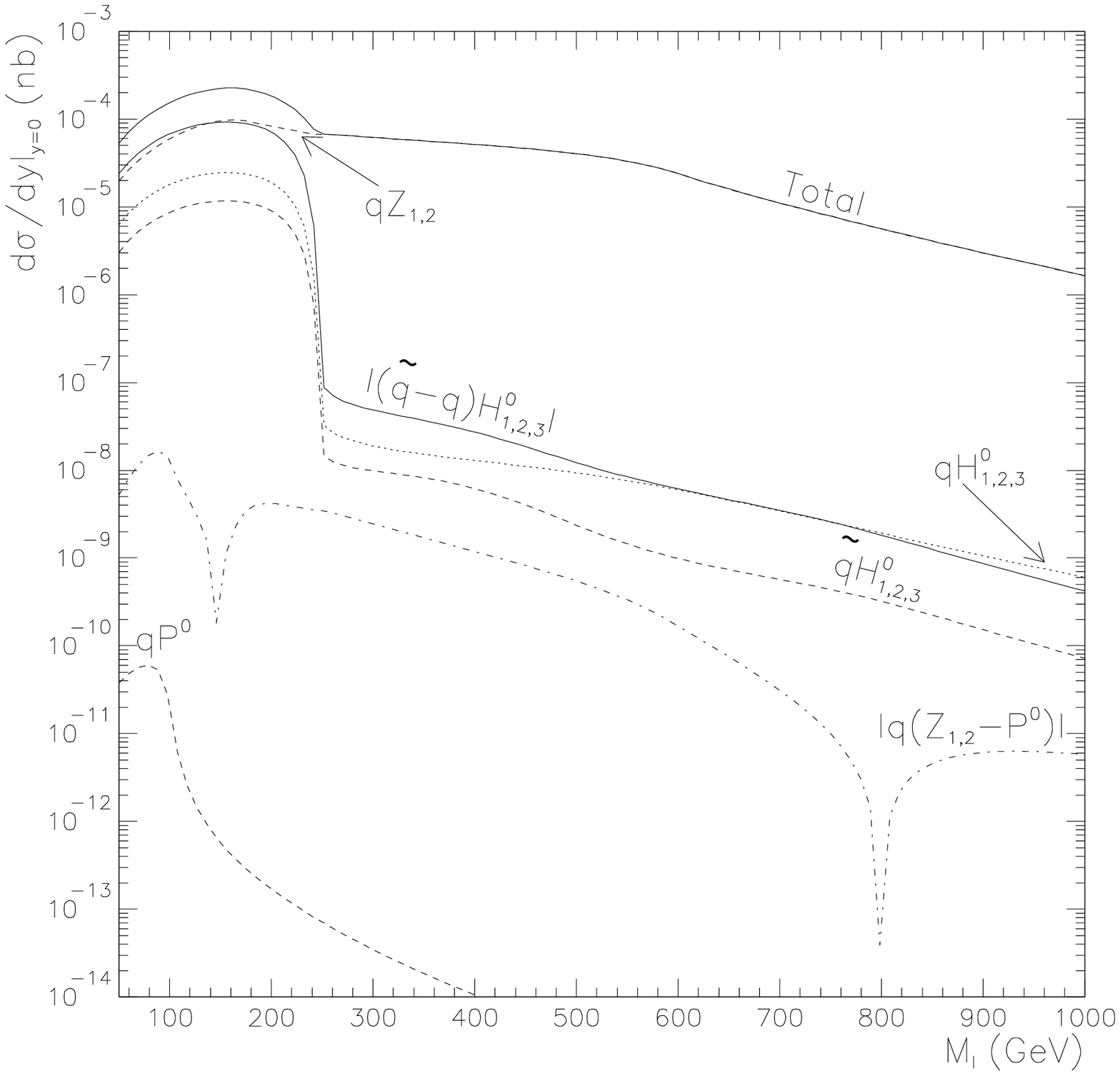}}
\end{center}
\caption[$\frac{d\sigma}{dy}|_{y=0}$   at $LHC$  for   $v_1/v_2=0.02\,$,
         $v_3/v_2=6.7\,$,  $  m^{\mbox{}}_S=400\,GeV$,       and       $
         m^{\mbox{}}_{q^\prime}=600\,GeV$]{\footnotesize        Rapidity
         distribution at $y=0$   for charged heavy  lepton production at
         $LHC$  ($14\,TeV$)  as a function  of  heavy lepton mass, where
         $v_1/v_2=0.02\,$,      $v_3/v_2=6.7\,$,              and      $
         m^{\mbox{}}_S=400\,GeV$.  The mass  spectrum is for the  non-SM
         particles      involved      in     these     processes    are,
         $m^{\mbox{}}_{Z_2}\approx496\,GeV$
         ($\Gamma^{\mbox{}}_{Z_2}\approx19.4\,GeV$),
         $m^{\mbox{}}_{P^0}\approx200\,GeV$
         ($\Gamma^{\mbox{}}_{P^0}\approx16.4\,GeV$),
         $m^{\mbox{}}_{H^\pm}\approx215\,GeV$,
         $m^{\mbox{}}_{H^0_1}\approx94.3\,GeV$
         ($\Gamma^{\mbox{}}_{H^0_1}\approx7.50\times10^{-3}\,GeV$),
         $m^{\mbox{}}_{H^0_2}\approx200\,GeV$
         ($\Gamma^{\mbox{}}_{H^0_2}\approx16.5\,GeV$),
         $m^{\mbox{}}_{H^0_3}\approx495\,GeV$
         ($\Gamma^{\mbox{}}_{H^0_3}\approx0.138\,GeV$),                $
         m^{\mbox{}}_{q^\prime}=600\,GeV$.}
\label{fig:decaye}
\end{figure}

\begin{figure}[hbtp]
\begin{center}
   \mbox{\epsfxsize=144mm\epsffile{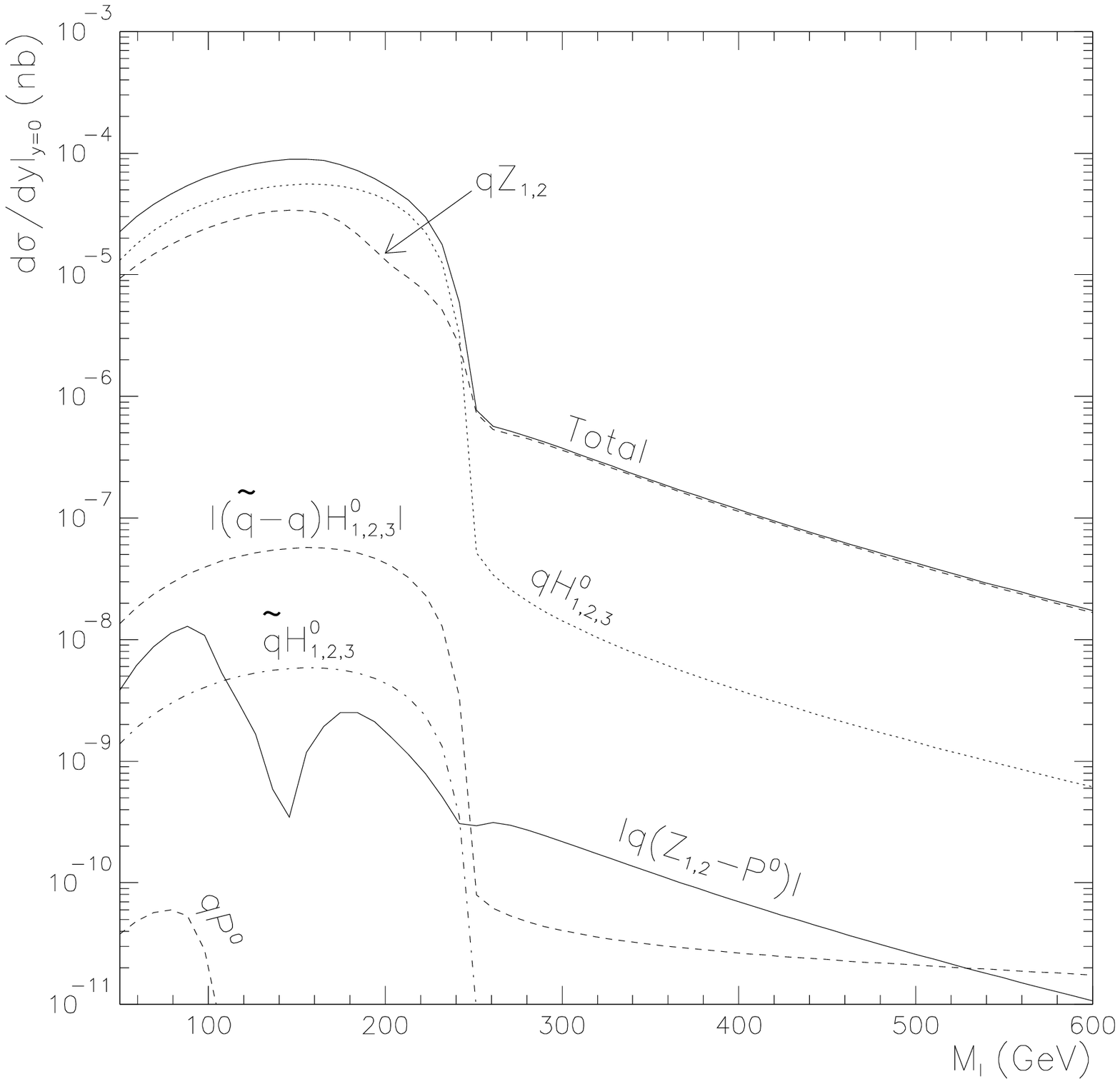}}
\end{center}
\caption[$\frac{d\sigma}{dy}|_{y=0}$   at  $LHC$ for   $v_1/v_2=0.02\,$,
         $v_3/v_2=6.7\,$,  $  m^{\mbox{}}_S=1\,TeV$,         and       $
         m^{\mbox{}}_{q^\prime}=200\,GeV$]{\footnotesize        Rapidity
         distribution at $y=0$ for charged   heavy lepton production  at
         $LHC$ ($14\,TeV$)  as a function  of  heavy lepton mass,  where
         $v_1/v_2=0.02\,$, $v_3/v_2=6.7\,$, and $ m^{\mbox{}}_S=1\,TeV$.
         The mass spectrum  for  the non-SM particles involved  in these
         processes          are,      $m^{\mbox{}}_{Z_2}\approx496\,GeV$
         ($\Gamma^{\mbox{}}_{Z_2}\approx20.9\,GeV$),
         $m^{\mbox{}}_{P^0}\approx200\,GeV$
         ($\Gamma^{\mbox{}}_{P^0}\approx16.4\,GeV$),
         $m^{\mbox{}}_{H^\pm}\approx215\,GeV$,
         $m^{\mbox{}}_{H^0_1}\approx94.3\,GeV$
         ($\Gamma^{\mbox{}}_{H^0_1}\approx7.50\times10^{-3}\,GeV$),
         $m^{\mbox{}}_{H^0_2}\approx200\,GeV$
         ($\Gamma^{\mbox{}}_{H^0_2}\approx16.5\,GeV$),
         $m^{\mbox{}}_{H^0_3}\approx495\,GeV$
         ($\Gamma^{\mbox{}}_{H^0_3}\approx0.230\,GeV$),                $
         m^{\mbox{}}_{q^\prime}=200\,GeV$.}
\label{fig:decayf}
\end{figure}

\begin{figure}[hbtp]
\begin{center}
   \mbox{\epsfxsize=144mm\epsffile{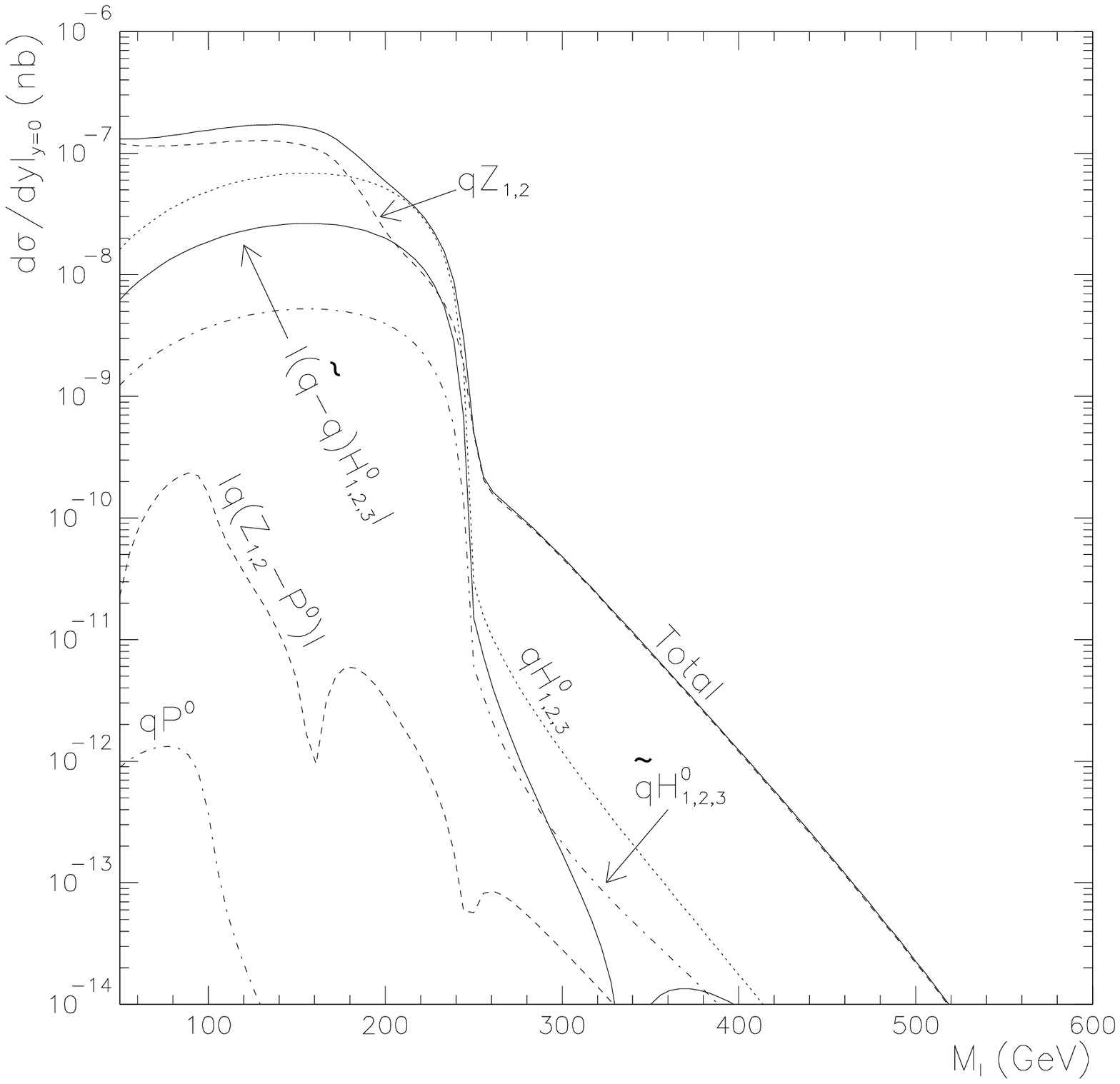}}
\end{center}
\caption[$\frac{d\sigma}{dy}|_{y=0}$ at  the $\mbox{T{\sc evatron}}$ for
         $v_1/v_2=0.02\,$, $v_3/v_2=6.7\,$,   $ m^{\mbox{}}_S=400\,GeV$,
         and  $ m^{\mbox{}}_{q^\prime}=200\,GeV$]{\footnotesize Rapidity
         distribution at $y=0$  for charged  heavy lepton production  at
         the $\mbox{T{\sc evatron}}$ ($1.8\,TeV$) as a function of heavy
         lepton  mass,  where $v_1/v_2=0.02\,$,  $v_3/v_2=6.7\,$, and  $
         m^{\mbox{}}_S=400\,GeV$.  The   mass spectrum  for  the  non-SM
         particles     involved       in  these         processes   are,
         $m^{\mbox{}}_{Z_2}\approx496\,GeV$
         ($\Gamma^{\mbox{}}_{Z_2}\approx20.9\,GeV$),
         $m^{\mbox{}}_{P^0}\approx200\,GeV$
         ($\Gamma^{\mbox{}}_{P^0}\approx16.4\,GeV$),
         $m^{\mbox{}}_{H^\pm}\approx215\,GeV$,
         $m^{\mbox{}}_{H^0_1}\approx94.2\,GeV$
         ($\Gamma^{\mbox{}}_{H^0_1}\approx7.50\times10^{-3}\,GeV$),
         $m^{\mbox{}}_{H^0_2}\approx200\,GeV$
         ($\Gamma^{\mbox{}}_{H^0_2}\approx16.5\,GeV$),
         $m^{\mbox{}}_{H^0_3}\approx495\,GeV$
         ($\Gamma^{\mbox{}}_{H^0_3}\approx0.230\,GeV$),                $
         m^{\mbox{}}_{q^\prime}=200\,GeV$.}
\label{fig:decayg}
\end{figure}

    Figs.~\ref{fig:decaya} through  \ref{fig:decayg}  show the  rapidity
distribution       at      y=0   for    $p{\!\!\!    \raisebox{-0.35ex}{
$\stackrel{\,\mbox{{\tiny  (}-{\tiny  )}}}{p}$    }\!\!     }\rightarrow
gg\rightarrow L^+L^-\,$  as function of $m^{\mbox{}}_{L}\,$, for various
scenarios.  The rapidity distribution, $d\sigma/dy\,$, is related to the
parton level cross-section, $\hat\sigma(gg\rightarrow L^+L^-)$, through
\begin{equation}
\frac{d\sigma}{dy}=\int_{\tau_{min}}^{e^{|y|}}d\tau
G(\sqrt{\tau}e^y,Q^2)G(\sqrt{\tau}e^{-y},Q^2)\hat\sigma(\tau s)\,,
\end{equation}
where   $\tau=\hat  s/s\,$,  $\tau_{min}=4m_L^2/s\,$, $\sqrt{s}$  is the
center-of-mass energy, and $G(x,Q^2)$  is the gluon structure  function.
The values    of  $\sqrt{s}$ have been  set   to  $14\,TeV$ ($LHC$)  for
Figs.~\ref{fig:decaya}-~\ref{fig:decayf},  and $1.8\,TeV$  ($\mbox{T{\sc
evatron}}$) for    Fig.~\ref{fig:decayg}.  In   these  figures the  $SM$
couplings and masses where  extracted from the PDG~\cite{kn:PDG}, except
for $m_t\approx180\pm12\,GeV$~\cite{kn:Top}.  For $G(x,Q^2)$ the leading
order  Duke  and Owens    1.1 ($DO1.1$)   \cite{kn:Duke,kn:Owens}  gluon
distribution was used.   The results   were  compared with the next   to
leading order  $MRSA$~\cite{kn:MRSA} gluon  distribution function, which
yielded a negligible difference.   Although these results include squark
mixing it was  found that there was no  significant change if  mixing is
not included.  Since $d\sigma/dy$ is  flat about $y=0$, the relationship
between  $\frac{d\sigma}{dy}|_{y=0}$   and the  total  cross-section  is
immediate.  Therefore     the  total  event  rate   for    the $p{\!\!\!
\raisebox{-0.35ex}{   $\stackrel{\,\mbox{{\tiny (}-{\tiny )}}}{p}$ }\!\!
}\rightarrow gg\rightarrow L^+L^-$ production mechanism can be estimated
from $y=0\,$,
\begin{equation}
\sigma=\int_{\ln\sqrt{\tau_{min}}}^{-\ln\sqrt{\tau_{min}}}
\frac{d\sigma}{dy}dy\approx -\ln(\tau_{min})\,\sigma\,.
\end{equation}

    Figs.~\ref{fig:decaya}-\ref{fig:decayc}                         show
$\frac{d\sigma}{dy}|_{y=0}$  for  different    VEV's  ratios  along  the
$m^{\mbox{}}_{Z_2}\approx{\cal  O}({500})\,GeV$  contour      line    of
Fig.~\ref{fig:mztwo}.  Notice that  as  $v_1/v_2$ becomes  comparable to
$v_3/v_2$  the  large $v_3$ limit   breaks down and  the generally small
$qP^0$ term starts to contribute (the $q(Z_{1,2}-P^0)$ contribution also
grows  quite   significantly but  remains  a  negligible  contribution).
Therefore for relatively    large  values of  $v_1/v_2$   variations  in
$m^{\mbox{}}_{P^0}$ ($\approx m^{\mbox{}}_{H^\pm}$ up to at least ${\cal
O}({1})\,TeV$) become important.  However  it is more natural to  assume
that the  intragenerational  Yukawa couplings are  of  the same order of
magnitude  and therefore for $v_1/v_2$   to be small.   For  the rest of
these figures then,     it   will be assumed    that   $v_1/v_2=0.02\,$.
Fig.~\ref{fig:decaya} is the figure with the default values.

   Fig.~\ref{fig:decayd} shows what happens when a  larger $Z_2$ mass of
${\cal O}({700\,GeV})$ ({\it i.e.}, $v_3/v_2=9.5$)   is used.  For  this
figure $  m^{\mbox{}}_S$ had to be pushed  up slightly to $450\,GeV$, in
order to produce   physical squark  masses.  The noticeable   difference
between  this and  all   of the other figures     is that the   peak has
broadened.  This is  expected  since the  $Z_2$ can  remain on-shell for
larger values of $m^{\mbox{}}_{L}\,$.  Notice that the $H^0_3$ resonance
cut off   seems      to   follow  the   $Z_2$'s.       More   precisely,
$m^{\mbox{}}_{H^0_3}\approx   m^{\mbox{}}_{Z_2}$  for large $v_3$.  This
becomes  immediately evident when taking  the large  $v_3$ limits of the
$H^0_i$   and  $Z_i$   mass  mixing    matrices,  Eqs.~(\ref{eq:mhimat})
and~(\ref{eq:mzzmat}) respectively:
\begin{equation}
\lim_{v_3\rightarrow\infty}m_{H^0_3}^2 =
\lim_{v_3\rightarrow\infty}m_{Z_2}^2   =
\frac{25}{36}\, g^{\prime\prime^2}\,v_3^2         \approx
\frac{25}{9}\,\frac{(v_3/v_2)^2\, 
x^{\mbox{}}_W}{1+(v_1/v_2)^2}\,m_Z^2\,,
\end{equation}
which  is  in fairly good agreement  with  all of  the figures. Also the
overall  production  is slightly  suppressed due to the smallness of the
gluon distribution function at large momentum fraction.

  Fig.~\ref{fig:decaye} shows what happens when the heavy quark mass was
pushed up to $600\,GeV$.  The effect is quite dramatic.  To see why this
is      so,  notice   the    slight   kink    in      the   curve around
$m^{\mbox{}}_{L}\approx600\,GeV$.  There is also a much more significant
kink in all  of the other  graphs around $200\,GeV$, {\it  i.e.}, around
$m^{\mbox{}}_{L}\approx m^{\mbox{}}_{q^\prime}$.  Further examination of
the parton  level  cross-section shows that  kink occurs  when the heavy
quarks in the loops can no longer be on shell.

   In Fig.~\ref{fig:decayf}  the scalar mass was  pushed up to $1\,TeV$.
Increasing $  m^{\mbox{}}_S$ has caused  the terms involving the squarks
to be supressed by several orders of  magnitude.  The difference between
the heavy and  light squark cases is that,  for heavy squarks, the gluon
luminosity is  relatively small  in  the  kinematical region  where  the
squarks in the   loop are on   shell.   The $qH^0_i$ term   now enhances
$L^+L^-$ production,  below the $m^{\mbox{}}_{H^0_3}$  threshold, as the
destructive interference with $\tilde qH^0_i$ term, {\it i.e.}, $(\tilde
q-q)H^0_i\,$, has been suppressed.

  Finally  Fig.~\ref{fig:decayg}  shows what   happens  at  $\sqrt{s}  =
1.8\,TeV$, the $\mbox{T{\sc  evatron}}$.   The overall topology is   the
same as  depicted  in Fig.~\ref{fig:decaya} but the  $L^+L^-$ production
rate is dramatically reduced: very  little gluon luminosity is available
to produce these heavy particles.

\subsection{Drell-Yan}

\begin{figure}[hbtp]
$$
\mbox{\epsfxsize=6.0cm\epsffile{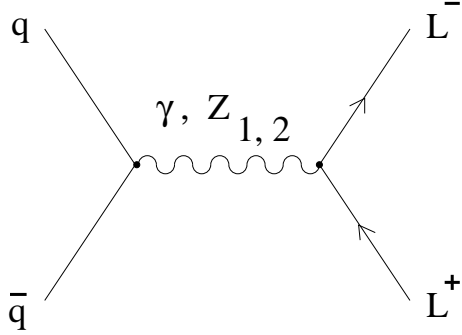}}
$$
\caption[foo]{\footnotesize
         Feynman  diagrams  for  Drell-Yan production of charged  heavy
         leptons.}
\label{fig:drelly}
\end{figure}

  Fig.~\ref{fig:drelly} shows  the  Feynman diagrams used  for computing
the   parton  level Drell-Yan contribution  to   heavy lepton production
\cite{kn:Hewett,kn:HewettB}.
Drell-Yan  production of heavy  leptons  occurs  through an  $s$-channel
$\gamma$,  $Z_1$ or  $Z_2$.   The  differential  cross section for  this
process can be expressed as follows:
\begin{equation}
\frac{\mbox{$d \hat{\sigma}_{L^\pm}$}}{\mbox{$d \Omega$}\;\;\;} = 
\frac{1}{64 \pi^2 \hat{s}} \beta \left\{ \overline{|{\cal M}_\gamma|^2} 
+ \overline{|{\cal M}_{Z_1}|^2} + \overline{|{\cal M}_{Z_2}|^2} + 
2 \mbox{Re}\, (\overline{{\cal M}_\gamma {\cal M}_{Z_1}^\dagger} + 
\overline{{\cal M}_\gamma {\cal M}_{Z_2}^\dagger} + 
\overline{{\cal M}_{Z_1} {\cal M}_{Z_2}^\dagger}) 
\right\}
\end{equation}
The ``direct'' squared matrix elements (the first three terms in the 
above expression) are given by
\begin{eqnarray}
\overline{|{\cal M}_i|^2} & = & 
\frac{\mbox{$2 G_i^4$}}{\mbox{$(\hat{s} - M_i^2)^2 + \Gamma_i^2 M_i^2$}}
\left\{4 v_q^i a_q^i v_L^i a_L^i (2 m_L^2 (\hat{t} - \hat{u}) - 
     (\hat{t}^2 - \hat{u}^2)) \; \right. \\
& + & \left. [(v_q^i)^2 + (a_q^i)^2] 
\left[ [(v_L^i)^2 + (a_L^i)^2] 
           (2 m_L^2( \hat{s} - m_L^2) + \hat{u}^2 + \hat{t}^2) 
     - [(v_L^i)^2 - (a_L^i)^2] (2 m_L^2 \hat{s}) \right] \right\}
 \nonumber
\end{eqnarray}
while the interference terms are given by
\begin{eqnarray}
2 \mbox{Re}\, (\overline{{\cal M}_i {\cal M}_j}) & = & 
\frac{\mbox{$4 G_i^2 G_j^2((\hat{s} - M_i^2)(\hat{s} - M_j^2) + 
       \Gamma_i M_i \Gamma_j M_j)$}}
   {\mbox{$((\hat{s} - M_i^2)^2 + \Gamma_i^2 M_i^2)
         ((\hat{s} - M_j^2)^2 + \Gamma_j^2 M_j^2)$}}
\left\{(v_q^i v_q^j + a_q^i a_q^j) \right.
\nonumber \\& \times&
  \left[(v_L^i v_L^j + a_L^i a_L^j)
     (2 m_L^2( \hat{s} - m_L^2) + \hat{u}^2 + \hat{t}^2)
 - (v_L^i L_q^j - a_L^i a_L^j) (2 m_L^2 \hat{s}) \right] 
\nonumber \\
& + & \left. (v_q^i a_q^j + a_q^i v_q^j)(v_L^i a_L^j + a_L^i v_L^j) 
     (2 m_L^2 (\hat{t} - \hat{u}) - (\hat{t}^2 - \hat{u}^2)) \right\}.
\end{eqnarray}
In the expressions above,  $i$ and $j$ can  be $\gamma$, $Z_1$ or $Z_2$;
$M_i$ and $\Gamma_i$ are the mass and width of the gauge boson; $m_L$ is
the mass of the heavy  lepton; $G_\gamma = e$ and  $G_{Z_1} = G_{Z_2}  =
g/\sqrt{1 - x_W}\,$.  If $i = \gamma$, $v^i_f = Q_f$ and $a^i_f = 0$ for
both quarks ($f = q$) or the heavy lepton ($f = L$).   If $i = Z_{1,2}$,
$v_f^i  =  (\tilde C^{fi}_R + \tilde   C^{fi}_L)/2$ and $a_f^i = (\tilde
C^{fi}_R  -  \tilde C^{fi}_L)/2$,  where again    $f =  (q,L)$  and  the
couplings   $\tilde  C^{fi}_{R,L}$ are   given by Eqs.~(\ref{eq:hatqzc})
and~(\ref{eq:hatqzd}).  Also, $\hat{s}$, $\hat{t}$ and $\hat{u}$ are the
usual parton level process   Mandelstam variables: $(m_L^2 -  \hat{t}) =
\hat{s} (1 - \beta \cos \theta)/2$ and $(m_L^2 - \hat{u}) = \hat{s} (1 +
\beta \cos \theta)/2$, where $\theta$  is the angle between the outgoing
$L^-$ and the incoming quark $q$.

\begin{figure}[hbtp]
$$\mbox{\epsfxsize=144mm\epsffile{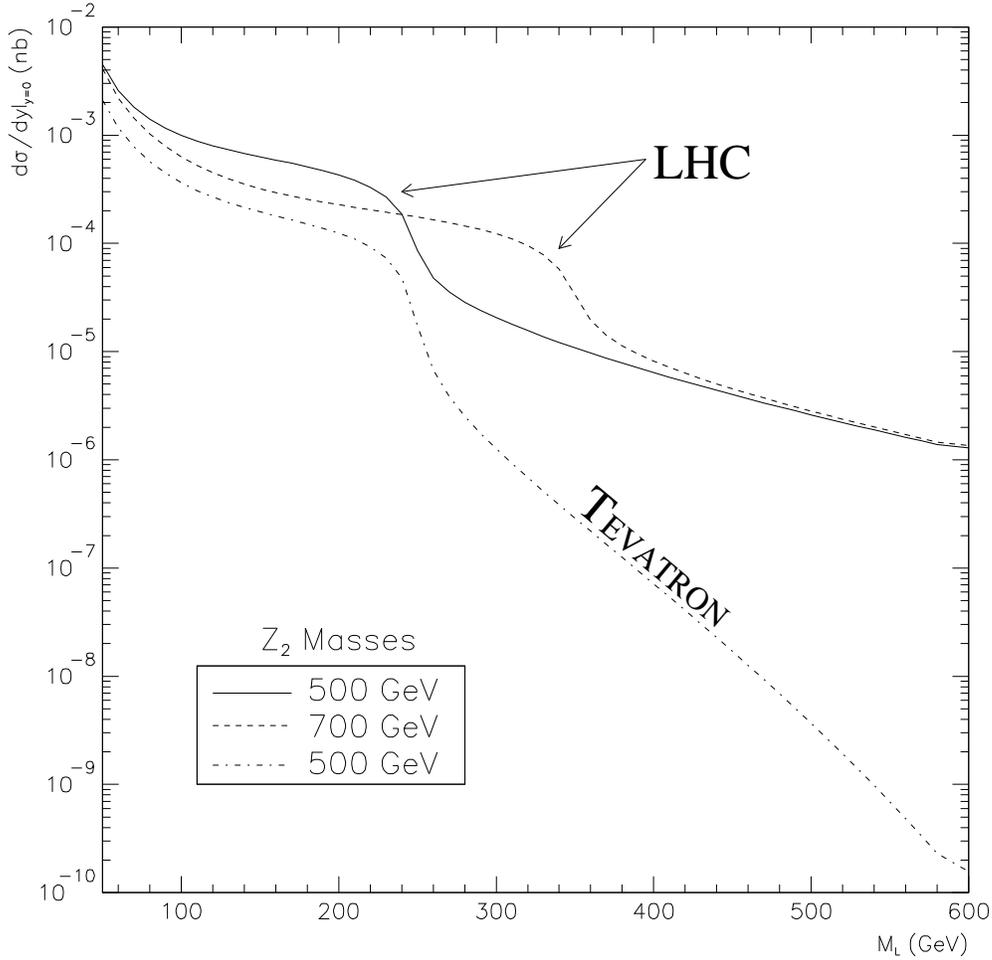}}$$
\caption[foo]{\footnotesize Drell-Yan rapidity distribution at  $y=0\,$,
              as  a   function   of $m_L^{\mbox{}}\,$,     for  $L^+L^-$
              production at $LHC$ and the  $\mbox{T{\sc evatron}}$.  The
              $DO1.1$    \cite{kn:Duke,kn:Owens} quark and    anti-quark
              parton distribution  functions  were used to obtain  these
              results.}
\label{fig:drelled}
\end{figure}

  Fig.~\ref{fig:drelled}  shows $d\sigma/dy|_{y=0}^{\mbox{}}\,$, as    a
function of  $m_L^{\mbox{}}\,$, for Drell-Yan  production of $L^+L^-$ at
$LHC$ and the $\mbox{T{\sc evatron}}\,$. These results are shown for the
regions of the \mbox{${\rm E}_6$} parameter space  that was explored for
gluon-gluon    fusion,  in    the    previous  section:    {\it   i.e.},
$m^{\mbox{}}_{Z_2}=500\,GeV$  and   $700\,GeV$      for  $LHC$,      and
$m^{\mbox{}}_{Z_2}=500\,GeV$  for the  $\mbox{T{\sc evatron}}$.   Notice
that  Drell-Yan   production     becomes   rapidly    suppressed     for
$2m^{\mbox{}}_{L} >   m^{\mbox{}}_{Z_2}\,$:   the $Z_2$   must  now   go
off-shell to produce the heavy lepton pairs.

  In contrast to  gluon-gluon  fusion, the  $LHC$ results for  Drell-Yan
production  are in general, higher  by an  order  of magnitude.  For the
$\mbox{T{\sc evatron}}$ results  the difference is   quite dramatic!  At
the $\mbox{T{\sc evatron}}$,  a $p\bar p$ collider, Drell-Yan production
occurs mainly  {\it via} the $q$  and $\bar q$  valence partons from the
$p$ and $\bar p$, respectively; at the $LHC$, a $pp$ collider, the $\bar
q\;\varepsilon\;p$ must  come from the  sea.  The  gluon distribution is
more similar   to    sea   quark  distributions  than      valence quark
distributions, which  explains why gluon-gluon  fusion is  comparable to
Drell-Yan   production at   the $LHC$,  bit   not  at  the  $\mbox{T{\sc
evatron}}$.

\subsection{Results} 

\begin{figure}[hbtp]
$$\mbox{\epsfxsize=144mm\epsffile{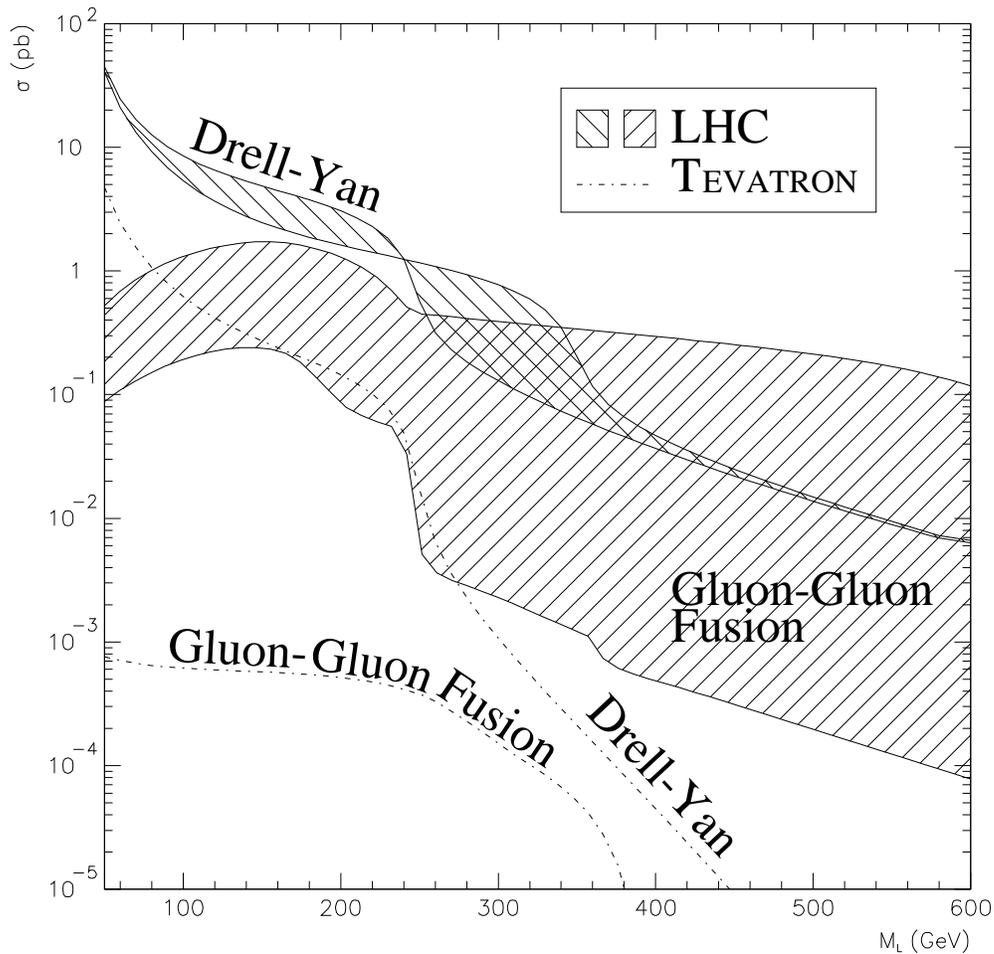}}$$
\caption[Summary             plot       of  $L^+L^-$          production
              cross-sections]{\footnotesize Summary  plot of results for
              the  total  $L^+L^-$  production  cross-section   at $LHC$
              ($\sqrt{s}=14\,TeV$, ${\cal L}\sim 10^5 pb^{-1}/yr$)   and
              $\mbox{T{\sc  evatron}}$    ($\sqrt{s}=1.8\,TeV$,   ${\cal
              L}\sim   10^2  pb^{-1}/yr$)  energies   as a   function of
              $m^{\mbox{}}_{L}\,$.   The hatched  regions  are the $LHC$
              results             for     gluon-gluon    fusion,    from
              Figs.~\ref{fig:decaya}-\ref{fig:decayf},  and    Drell-Yan
              production,  from Fig.~\ref{fig:drelled}.  The    dash-dot
              lines  are    the $\mbox{T{\sc   evatron}}$    results for
              gluon-gluon    fusion,    from  Fig.~\ref{fig:decayg}, and
              Drell-Yan production, from Fig.~\ref{fig:drelled}.}
\label{fig:totally}
\end{figure}

  Fig.~\ref{fig:totally} gives a summary of  the total cross-section for
$L^+L^-$   production for the  \mbox{${\rm E}_6$}  model parameter space
studied in the previous sections.  At the $\mbox{T{\sc evatron}}$ ${\cal
O}({10^{1\pm1}})$         $events/yr$         are       expected     for
$m_L^{\mbox{}}\,\raisebox{-0.625ex}{$\stackrel{<}{\sim}$}\,        {\cal
O}({300})\,GeV$, all of it coming from Drell-Yan production ({\it i.e.},
gluon-gluon                         fusion                        yeilds
$\,\raisebox{-0.625ex}{$\stackrel{<}{\sim}$}\,{\cal           O}({0.1})$
$events/yr$).  For  $LHC$, over a  reasonable range of  parameter space,
${\cal O}({10^{4^{+2}_{-1}}})$ $events/yr$ are expected.

  The charged heavy  lepton in the $MSSM$  model \cite{kn:Montalvo} is a
member of a sequential $4^{\rm th}$ generation  added arbitrarily to the
model;  the lepton  masses were  chosen larger  that  $50\,GeV$ and  the
quarks          larger            than           $150\,GeV$.         For
$m_L^{\mbox{}}\,\raisebox{-0.625ex}{$\stackrel{<}{\sim}$}\,        {\cal
O}({250})\,GeV$ the $LHC$ results are  comparable in order of  magnitude
to  the $MSSM$   predictions,  obtained by   Cieza Montalvo   $et$ $al.$
\cite{kn:Montalvo},  which predicts  ${\cal O}({10^{5}})$   $events/yr$.
However, the   dominant mechanism  in  the  $MSSM$ model  is gluon-gluon
fusion  and  for \mbox{${\rm  E}_6$}   this  contribution yields  ${\cal
O}({10^{4\pm 1}})$ $events/yr$, which is a factor of  at least $10$ less
than $MSSM$ results.   This is a rather surprising  result since it  was
expected that the \mbox{${\rm E}_6$} event rate would be enhanced due to
the greater  number of heavy particles running  around in the loop.  The
parameters in each model were varied to study this non-intuitive result.
Unfortunately,  it  turns out  that  the  \mbox{${\rm  E}_6$} parameters
suppress $L^+L^-$   production,  since  $v_3$  is fairly    large.  This
restriction causes the production  to occur mainly through the $Z_{1,2}$
and $H^0_3$   terms.  The  $MSSM$   has two  neutral Higgses   and   one
pseudo-scalar Higgs that are allowed to contribute to  the processes.  A
very simple  test on the  \mbox{${\rm  E}_6$} model was  done by varying
$v_3/v_2$,   about $v_1/v_2=0.2$, that   showed   for $v_3/v_2=2.8$  and
$m_L\,\raisebox{-0.625ex}{$\stackrel{<}{\sim}$} \,100\,GeV$ a  factor of
10 increase was  obtained.  However, this  region  of \mbox{${\rm E}_6$}
parameter space is forbidden, see Fig.~\ref{fig:mztwo}.
 
  It appears that the  Drell-Yan mechanism, used in  \cite{kn:Montalvo},
includes only the the   $s$-channel photon diagram.   We find   that the
inclusion of the  $Z$ leads to a  factor of 10  increase in the $L^+L^-$
rate, which puts Drell-Yan on equal  footing with gluon-gluon fusion for
$m^{\mbox{}}_{L}\,\raisebox{-0.625ex}{$\stackrel{<}{\sim}$}\,
{\cal O}({100})\,GeV$, in the $MSSM$ with a fourth generation.

  The   $L^\pm\rightarrow\nu_{L}^{\mbox{}}     W^\pm,\,\nu_{L}^{\mbox{}}
H^\pm$ decay modes are expected to be similar  for both models, as these
are  SM-like   decays.   These modes  depend   upon  the mass difference
$\Delta=m^{\mbox{}}_{L}-m_{\nu_L^{\mbox{}}}^{\mbox{}}\,$.   For $\Delta<
m^{\mbox{}}_W << m^{\mbox{}}_{H^\pm}$   the  decays  modes  will be   by
virtual   $W$'s, $W^*\rightarrow f\bar f\,$,  and  on shell for $\Delta>
m^{\mbox{}}_W$.  Leptonic decays of  the $W$'s offer the  possibility of
$L^\pm$ detection  by measuring  $\ell^+\ell^-$ production with  missing
transverse  momentum, ${p\!\!\!/\,}_{T}^{\mbox{}}\,$~\cite{kn:Montalvo}.
The   competing     SM   backgrounds     with   these   processes    are
$pp\rightarrow\tau^+\tau^-,\, W^+W^-,\,  Z^0Z^0$.   Studies have  shown,
using SM couplings, that it  is possible to   pull the $L^+L^-$  signals
from background  for $\Delta>  m^{\mbox{}}_W$  given sufficiently  large
event  rates;  it       is   much  more  difficult      for     $\Delta<
m^{\mbox{}}_W$~\cite{kn:Montalvo,kn:BargerB,kn:Hinchliffe}.   In general
the $MSSM$ event rate is  higher than \mbox{${\rm E}_6$}, and  therefore
detection would   more    likely  indicate  a   $MSSM$  candidate.    If
$m^{\mbox{}}_{H^\pm}\approx{\cal   O}({     m^{\mbox{}}_W})\,$,     then
$H^\pm\rightarrow f_i\bar f_j$ dominates,  for naturally large values of
$\tan\beta$.  Since the Higgs likes to couple to massive particles, this
would  lead multiple heavy jet  events  which in  general would be  very
difficult to pull out of background in  either the $MSSM$ or \mbox{${\rm
E}_6$}.  Similar  processes  are expected  to  occur  for  the  cases  $
m^{\mbox{}}_W<\Delta<m^{\mbox{}}_{H^\pm}$              and             $
m^{\mbox{}}_W\,,m^{\mbox{}}_{H^\pm}<\Delta\,$.     For   large    enough
$m^{\mbox{}}_{H^\pm}$,  the  sfermion  channels  also open,  {\it i.e.},
$H^\pm\rightarrow\tilde f_i\tilde f_j^*$  ({\it e.g.},  $\tilde u \tilde
d^*$).   The sfermions  would eventually  decay  out  leaving  only  the
lightest  supersymmetric    particles  ($LSP$'s),    which  will  escape
undetected along    with   the $\nu_{L}^{\mbox{}}$'s   leaving   lots of
${p\!\!\!/\,}_{T}^{\mbox{}}\,$.   In  fact,  all  of the  aforementioned
process will lead  to events with ${p\!\!\!/\,}_{T}^{\mbox{}}\,$, as the
$\nu_{L}^{\mbox{}}$'s will pass through the detector.

  In  certain regions  of the   $MSSM$   and \mbox{${\rm E}_6$}    model
parameter spaces it   may be possible to  distinguished  between the two
models.  If  $L^+L^-$  event rates  are larger than  those  predicted by
\mbox{${\rm E}_6$}, then the likely candidate is the $MSSM$.  Unlike the
$MSSM$,  it  is  possible that  $m^{\mbox{}}_{H^\pm}< m^{\mbox{}}_W$  in
\mbox{${\rm  E}_6$}~\cite{kn:Hewett},  and  therefore   if $H^\pm$'s are
found in this mass range the more  likely candidate would be \mbox{${\rm
E}_6$}.  Another possible way of telling the models apart is to look for
sfermion  production, $L^\pm\rightarrow f\tilde f^*\,,\tilde f\bar f\,$,
which is  unique to \mbox{${\rm  E}_6$}, since  $L^\pm$ has opposite $R$
parity  to   the    other  $SM$-like fermions      in   the {\bf   27}'s
(Fig.~\ref{fig:esix}).  The sfermion would eventually  decay to an $LSP$
which  is    stable  (assuming   $R$    parity  conservation),  yielding
$jets$+${p\!\!\!/\,}_{T}^{\mbox{}}\,$, in general.  Whether or not it is
possible   to  distinguish  them  from  the  $MSSM$   and the  $SM$-like
backgrounds would  require a much more  detailed study, as the allowable
parameter  space   for sfermions masses   and Yukawa  couplings is quite
large.    Finally, the $MSSM$    does  have fairly   stringent unitarity
constraints on the heavy lepton and heavy quark masses  as a function of
$\tan\beta$~\cite{kn:Montalvo},             in                particular
$m^{\mbox{}}_{L}\,\raisebox{-0.625ex}{$\stackrel{<}{\sim}$}
\,(1200\,GeV)\cos\beta\,$.    Therefore, it  should    be  possible   to
eliminate  $(m^{\mbox{}}_{L},\tan\beta)$ regions  in the $MSSM$ $L^-L^+$
production cross-section plots,  as a  function of  $m^{\mbox{}}_{L}\,$,
such   that only \mbox{${\rm  E}_6$}  models  are allowed.  For example,
assuming $m^{\mbox{}}_{H^0_1}\,\raisebox{-0.625ex}{$\stackrel{>}{\sim}$}
\,{\cal      O}({600})\,GeV$      rules    out     the       $MSSM$  for
$m^{\mbox{}}_{L}\,\raisebox{-0.625ex}{$\stackrel{>}{\sim}$}\,242\,GeV$
and $\tan\beta\,\raisebox{-0.625ex}{$\stackrel{>}{\sim}$}\,5$.  In   the
allowed  $MSSM$ region this    gives  an upper  limit  on  the  $L^+L^-$
production cross-section  of ${\cal O}({10})pb$, at  $LHC$.  Also in the
$MSSM$ there are phenomenological constraints on $\tan\beta$ which could
allow  for   further  restrictions.  A  more  detailed  study   of these
constraints has not been carried out.

  In  closing, it should  be  pointed out that   only a simple model  of
\mbox{${\rm  E}_6$} has been  considered.    It  is possible  for  other
\mbox{${\rm  E}_6$}   models to produce  results similar    to the model
studied   here  or to  the    $MSSM$.  Therefore,  in general,  $L^+L^-$
production  by gluon-gluon fusion should not  be considered a definitive
means of separating out the different  models; several experiments would
be required.

\section{Conclusions}

   The $p{\!\!\! \raisebox{-0.35ex}{ $\stackrel{\,\mbox{{\tiny (}-{\tiny
)}}}{p}$    }\!\!     }\rightarrow   gg\rightarrow    L^+L^-$ production
cross-section was computed for a simple rank-5 \mbox{${\rm E}_6$} model.
For a fairly conservative survey of the  various parameters in the model
we  expect  ${\cal O}({10^{4^{+2}_{  -1}}})$   $events/yr$ at $LHC$, and
${\cal O}({10^{1\pm1}})$  $events/yr$  at  the  $\mbox{T{\sc evatron}}$.
For  $LHC$ and the $\mbox{T{\sc  evatron}}$  it was found that Drell-Yan
production  dominated over  gluon-gluon  fusion  for  $2m_L^{\mbox{}}\le
m^{\mbox{}}_{Z_2}\,$.  For the  $\mbox{T{\sc evatron}}$  events are only
expected             to         be               seen                for
$2m_L^{\mbox{}}\,\raisebox{-0.625ex}{$\stackrel{<}{\sim}$}\,       {\cal
O}({m^{\mbox{}}_{Z_2}})\,$, as   the  Drell-Yan and  gluon-gluon  fusion
rates drop rapidly  beyond this point.   The $LHC$ results were compared
to the  $MSSM$'s (${\cal O}({10^5})$ $events/yr$ \cite{kn:Montalvo}), in
which   gluon-gluon  fusion  is the   dominant   production  mode.   The
gluon-gluon fusion contribution to  $L^+L^-$ production at $LHC$ (${\cal
O}({10^{4\pm1}})$) was found to be at least a factor of 10 less than the
event rates predicted for the $MSSM$, due the $CDF$ and $D0\!\!\!/$ soft
limits     ({\it     i.e.},  assuming    SM   couplings)   placed     on
$m^{\mbox{}}_{Z_2}$~\cite{kn:Shochet}.  These soft constraints  resulted
in the $H^0_{1,2}$  and $P^0$ contributions  to the $L^+L^-$  production
rate  to be   suppressed  leaving only  the   $H^0_3$ and $Z_{1,2}$   to
contribute.   For certain regions  in  the $MSSM$ and \mbox{${\rm E}_6$}
parameter spaces it was demonstrated that  it is possible to distinguish
between the two models, in principle.  However, it should be pointed out
that  there  are many candidate \mbox{${\rm    E}_6$} models which could
yield overlapping results.

\section{Acknowledgements}

The   authors would  like  to thank   Alan  Dekok  for proof-reading our
pre-print  drafts.  The     computing facilities utilized    herein were
provided  by Carleton University's,  Department  of Physics, CRPP, OPAL,
and \mbox{T{\scriptsize HEORY}}  groups, as well as UQAM's D\'epartement
de Physique.  This  research was funded by $NSERC$  of Canada and $FCAR$
du Quebec.

%
%
\appendix
\section{Couplings and Widths for $\hat\sigma(gg\rightarrow L^+L^-)$}
\label{sec-appc}

   This appendix gives a summary  of the calculations  that were used to
obtain  the couplings and the  widths  for the $\hat\sigma(gg\rightarrow
L^+L^-)$ matrix elements given in \S~\ref{sec-cross}.

\subsection{The Couplings}
\label{sec-crscup}
 
   In this section the calculations of the vertex factors used to obtain
the $gg\rightarrow L^+L^-$ matrix elements, given in \S~\ref{sec-cross},
are summarized.

   For the $Z_{1,2}$ exchange diagrams shown in Fig.~\ref{fig:fusion}(a)
the        following          vertex       factors         were     used
\eqnpict{\gaugeff{Z^\mu_{i}}{f}{\bar                                f}}{
\frac{-g}{\sqrt{1-x_W}}\,\gamma^\mu\,     [\tilde   C^{fZ_i}_LP_L+\tilde
C^{fZ_i}_RP_R]\,,\label{eq:cupa} }
\noindent
where $i=1,2$,
\begin{eqnarray}
P_L&=&\frac{1}{2}(1-\gamma_5)\,,\\
P_R&=&\frac{1}{2}(1+\gamma_5)\,,
\end{eqnarray}
and  $C^{qZ_i}_L$ and   $C^{qZ_i}_R$     were the couplings   used    in
Eq.~(\ref{eq:hatqz}).   The  gauge-fermion interaction  Lagrangian   for
$\mbox{${\rm      SU}({2})_{{\rm   L}}$}\otimes\mbox{${\rm U}({1})_{{\rm
Y}}$}\otimes\mbox{${\rm U}({1})_{{\rm E}}$}$ is given by
\begin{equation}
{\cal L}_{\rm int}\supseteq\,-\,\frac{1}{2}\,
(g_L^{\mbox{}}\hat\tau^a_{ij}L_\mu^a+
g_Y^{\mbox{}}\delta_{ij}\hat Y_{Y_i}Y_\mu
+g_E^{\mbox{}}\hat Y_{E_i}E_\mu)\,\bar\psi_i\bar\sigma^\mu\psi_j\,,
\label{eq:appca}
\end{equation}
where the $\psi_i$'s are  two-component spinors, see Eq.~(B.2)  of Haber
and Kane  (HK)~\cite{kn:Haber}.  Defining $g=g_L\,$, $g^\prime=g_Y\,$, $
g^{\prime\prime}=g_E\,$, and $\hat Y=\hat Y_{Y_i}(=2\hat Q-\hat\tau_3)$,
and using the identities
\begin{eqnarray}
aL^3_\mu+bY_\mu&=& (a\cos\theta_W-b\sin\theta_W)Z_\mu
                  +(a\sin\theta_W+b\cos\theta_W)A_\mu\,,\\
E_\mu&=&Z_\mu^\prime\,,
\end{eqnarray}
then Eq.~(\ref{eq:appca}) becomes
\begin{equation}
{\cal L}_{\rm int}\supseteq\,-\,\left\{
\frac{g}{\cos\theta_W}(\hat T_3 - \hat Q\,x_W)Z_\mu
+\,\frac{1}{2}\, g^{\prime\prime}\hat Y_{E}Z^\prime_\mu
\right\}_{ij}[\bar\psi_{(f_L)_i}\bar\sigma^\mu\psi_{(f_L)_j}+
\bar\psi_{(f_L^c)_i}\bar\sigma^\mu\psi_{(f_L^c)_j}]\,,
\end{equation}
where     $\hat       T_3=\hat\tau_3/2\,$,          $\tau_i=\sigma_i\,$,
$\tan\theta_W=g^\prime/g\,$, and $x_W=\sin^2\theta_W\,$. Noting that
\begin{eqnarray}
(\hat T_3 - \hat Q\,x_W)|f^c_L>&=&-(\hat T_3 - \hat Q\,x_W)|f_R>\,,\\
\hat Y_E|f^c_L>&=&-\hat Y_E|f_R>\,,
\end{eqnarray}
yields
\begin{equation}
{\cal L}_{\rm int}\supseteq\,-\,\left\{
\frac{g}{\cos\theta_W}(\hat T_3 - \hat Q\,x_W)Z_\mu
+\,\frac{1}{2}\, g^{\prime\prime}\hat Y_{E}Z^\prime_\mu
\right\}_{ij}[\bar\psi_{(f_L)_i}\bar\sigma^\mu\psi_{(f_L)_j}-
\bar\psi_{(f_R)_i}\bar\sigma^\mu\psi_{(f_R)_j}]\,.
\end{equation}
Using the following identities
\begin{eqnarray}
\;\;\bar\psi_{(f_L)_i}\bar\sigma^\mu\psi_{(f_L)_j}&=&
\bar f_i\gamma^\mu P_Lf_j\,,\\
-\bar\psi_{(f_R)_i}\bar\sigma^\mu\psi_{(f_R)_j}&=&
\bar f_i\gamma^\mu P_Rf_j\,,
\end{eqnarray}
to convert from two-component to four-component spinor notation yields
\begin{equation}
{\cal L}_{\rm int}\supseteq
\frac{-g}{\sqrt{1-x_W}}\sum_{A=L,R}\bar f\left\{
   (T_{3_A}-e_fx_W)\not\!Z
   +\frac{1}{2}\left(\frac{ g^{\prime\prime}}{g}\right)
   y^\prime_{f_A}\,\sqrt{1-x_W}\not\!Z^\prime
\right\}P_A\,f\,,
\end{equation}
see  Eqs.~(A.28) of HK  \cite{kn:Haber}.  Then the $Z$-$Z^\prime$-vertex
factor is
\eqnpict{\gaugeff{Z^\mu,Z^{\prime^\mu}}{f}{\bar f}}{
   \frac{-g}{\sqrt{1-x_W}}\,\gamma^\mu\,
   [C^{fZ,fZ^\prime}_LP_L+ C^{fZ,fZ^\prime}_RP_R]\,,
}
where    the       $C^{fZ,fZ^\prime}_{L,R}\,$'s      are    defined   by
Eqs.~(\ref{eq:hatqzc})   and~(\ref{eq:hatqzd}).  Using  the inverse   of
transformations~(\ref{eq:zmixa}) and~(\ref{eq:zmixb}),
\begin{equation}
\left(\begin{array}{l}
   \tilde Z^\prime \\ Z
\end{array}\right)=
\left(\begin{array}{rr}
   \cos\phi & -\sin\phi \\
   \sin\phi &  \cos\phi
\end{array}\right)
\left(\begin{array}{c}
   Z_1\\ Z_2
\end{array}\right)\,,\label{eq:sleep}
\end{equation}
yields the desired result
\begin{equation}
{\cal L}_{\rm int}\supseteq
\frac{-g}{\sqrt{1-x_W}}\sum_{i=1}^2\sum_{A=L,R}\bar f
   \!\not\!Z_i\,\tilde C^{fZ_i}_AP_A\,f\,,\label{eq:zabvrt}
\end{equation}
{\it i.e.}, vertex factor~\ref{eq:cupa}.

   For  the  $H^0_{1,2,3}$ and    $P^0$   exchange  diagrams  shown   in
Fig.~\ref{fig:fusion}(b) the following vertex factors were used
\eqnpict{\scalarff{H^0_i}{f}{\bar f}}{
   -g\,\frac{m_f}{2m_W}\,K^{fH^0_i}\,,\label{eq:cupb}
}
and
\eqnpict{\scalarff{P^0}{f}{\bar f}}{
   ig\,\frac{m_f}{2m_W}\,\gamma_5\,K^{fP^0}\,,\label{eq:cupc}
}
\noindent
respectively, where  $i=1,2,3\,$.    The  $K^{fH^0_i}$  and   $K^{fP^0}$
couplings  are obtained   from   the Yukawa   interaction part   of  the
Lagrangian given by  Eq.~(\ref{eq:yuk}): noting that $\varepsilon_{ij} =
(i\tau_3)_{ij}$   and    plugging  W,       Eq.~(\ref{eq:soup}),    into
Eq.~(\ref{eq:yuk}) yields
\begin{eqnarray}
{\cal L}_{Yuk}&\supseteq&
-\mbox{$\textstyle\frac{{1}}{{2}}$}\varepsilon_{ij}\{-\lambda_1[
  \Phi_{2_i}^{\mbox{}}(\psi_{Q_j}\psi_{u^c_L}+\psi_{u^c_L}\psi_{Q_j})+
  \Phi_{2_i}^*(\bar\psi_{Q_j}\bar\psi_{u^c_L}+
    \bar\psi_{u^c_L}\bar\psi_{Q_j}
  )
]
\nonumber\\&&+\,
\lambda_2[
  \Phi_{1_i}^{\mbox{}}(\psi_{Q_j}\psi_{d^c_L}+\psi_{d^c_L}\psi_{Q_j})+
  \Phi_{1_i}^*(\bar\psi_{Q_j}\bar\psi_{d^c_L}+
    \bar\psi_{d^c_L}\bar\psi_{Q_j}
  )
]
\nonumber\\&&+\,
\lambda_3[
  \Phi_{1_i}^{\mbox{}}(\psi_{L_j}\psi_{e^c_L}+\psi_{e^c_L}\psi_{L_j})+
  \Phi_{1_i}^*(\bar\psi_{L_j}\bar\psi_{e^c_L}+
    \bar\psi_{e^c_L}\bar\psi_{L_j}
  )
]
\nonumber\\&&+\,
\lambda_4[
  \Phi_{3}^{\mbox{}}(\psi_{R^\prime_i}\psi_{L^\prime_j}+
    \psi_{L^\prime_j}\psi_{R^\prime_i})+
  \Phi_{3}^*(\bar\psi_{R^\prime_i}\bar\psi_{L^\prime_j}+
    \bar\psi_{L^\prime_j}\bar\psi_{R^\prime_i}
  )
]
\nonumber\\&&+\,
\lambda_5[
  \Phi_{3}^{\mbox{}}(\psi_{d^{\prime^c}_L}\psi_{d^\prime_L}+
    \psi_{d^\prime_L}\psi_{d^{\prime^c}_L})+
  \Phi_{3}^*(\bar\psi_{d^{\prime^c}_L}\bar\psi_{d^\prime_L}+
    \bar\psi_{d^\prime_L}\bar\psi_{d^{\prime^c}_L}
  )
]\}\,,\\
&\supseteq&
-\mbox{$\textstyle\frac{{1}}{{2}}$}\{\lambda_1[
  \phi^0_2(\psi_{u_L}\psi_{u^c_L}+\psi_{u^c_L}\psi_{u_L})+
  \phi^{0^*}_2(\bar\psi_{u_L}\bar\psi_{u^c_L}+
    \bar\psi_{u^c_L}\bar\psi_{u_L}
  )
]
\nonumber\\&&+\,
\lambda_2[
  \phi^0_1(\psi_{d_L}\psi_{d^c_L}+\psi_{d^c_L}\psi_{d_L})+
  \phi^{0^*}_1(\bar\psi_{d_L}\bar\psi_{d^c_L}+
    \bar\psi_{d^c_L}\bar\psi_{d_L}
  )
]
\nonumber\\&&+\,
\lambda_3[
  \phi^0_1(\psi_{e_L}\psi_{e^c_L}+\psi_{e^c_L}\psi_{e_L})+
  \phi^{0^*}_1(\bar\psi_{e_L}\bar\psi_{e^c_L}+
    \bar\psi_{e^c_L}\bar\psi_{e_L}
  )
]
\nonumber\\&&+\,
\lambda_4[
  \phi^0_3(\psi_{e^{\prime^c}_L}\psi_{e^\prime_L}+
    \psi_{e^\prime_L}\psi_{e^{\prime^c}_L})+
  \phi^{0^*}_3(\bar\psi_{e^{\prime^c}_L}\bar\psi_{e^\prime_L}+
    \bar\psi_{e^\prime_L}\bar\psi_{e^{\prime^c}_L}
  )
]
\nonumber\\&&+\,
\lambda_5[
  \phi^0_3(\psi_{d^{\prime^c}_L}\psi_{d^\prime_L}+
    \psi_{d^\prime_L}\psi_{d^{\prime^c}_L})+
  \phi^{0^*}_3(\bar\psi_{d^{\prime^c}_L}\bar\psi_{d^\prime_L}+
    \bar\psi_{d^\prime_L}\bar\psi_{d^{\prime^c}_L}
  )
]\}\,,
\end{eqnarray}
and similarly for the other generations. Defining
\begin{equation}
f=\left(\begin{array}{@{}c@{}}
   {\psi_{f_L}}\\{\bar\psi_{f^c_L}}\end{array}\right)
\end{equation}
and using the following identities
\begin{eqnarray}
\psi_{f^c_{1_L}}\psi_{f_{2_L}^{\mbox{}}}=
\psi_{f_{2_L}^{\mbox{}}}\psi_{f^c_{1_L}}
&=&\bar f_1P_Lf_2\,,\\
\bar\psi_{f^c_{1_L}}\bar\psi_{f_{2_L}^{\mbox{}}}=
\bar\psi_{f_{2_L}^{\mbox{}}}\bar\psi_{f^c_{1_L}}
&=&\bar f_2P_Rf_1\,,
\end{eqnarray}
see Eqs.~(A.24), (A.25), and (A.28) of HK \cite{kn:Haber}, implies
\begin{eqnarray}
{\cal L}_{Yuk}&\sim&-\lambda_i(
  \phi^0_j\psi_{f^c_L}\psi_{f_L}+
  \phi^{0^*}_j\bar\psi_{f^c_L}\bar\psi_{f_L}
)\,,\\
&=&-\mbox{$\textstyle\frac{{1}}{{2}}$}\,\lambda_i\,
[\phi^0_j\bar f(1-\gamma_5)f+\phi^{0^*}_j\bar f(1+\gamma_5)f]\,,\\
&=&-\lambda_i\,
[\mbox{Re}\,(\phi^0_j)\bar ff-i\,\mbox{Im}\,
(\phi^0_j)\bar f\gamma_5f]\,,\\
&=&-\mbox{$\textstyle\frac{{1}}{{\sqrt{2}}}$}\,\lambda_i\,
(\phi^0_{jR}\bar ff-i\,\phi^0_{jI}\bar f\gamma_5f)\,.
\end{eqnarray}
Expanding   the   $\phi^0_i$'s in  terms   of    their physical  fields,
Eqs.~(\ref{eq:basa})-(\ref{eq:basb}), yields
\begin{eqnarray}
{\cal L}_{Yuk}&\supseteq&
\underbrace{-\frac{1}{\sqrt{2}}\,\left\{
\lambda_1\nu_2\,\bar u u+
\lambda_2\nu_1\,\bar d d+
\lambda_3\nu_1\,\bar e e+
\lambda_4\nu_3\,\bar e^\prime e^\prime+
\lambda_5\nu_3\,\bar d^\prime d^\prime
\right\}}_{{\bf Eq.~(\ref{eq:yucky})}}
\nonumber\\&&\nonumber\\&&
-\frac{1}{\sqrt{2}}\sum_{j=1}^3\left\{
\lambda_1U_{2j}\,\bar u u+
U_{1j}(\lambda_2\,\bar d d+
\lambda_3\,\bar e e)+
U_{3j}(\lambda_4\,\bar e^\prime e^\prime+
\lambda_5\,\bar d^\prime d^\prime)
\right\}H^0_j\nonumber\\&&\nonumber\\&&
+\frac{i\kappa}{\sqrt{2}}
\{
\lambda_1 v_{13}
\,\bar u\gamma_5 u+
v_{23}
(\lambda_2\,\bar d\gamma_5 d+
\lambda_3\,\bar e\gamma_5 e)
+
v_{12}
(\lambda_4\,\bar e^\prime\gamma_5 e^\prime+
\lambda_5\,\bar d^\prime\gamma_5 d^\prime)
\}
P^0\,,\label{eq:yields}
\end{eqnarray}
The   couplings  can  now   be  read   directly and   give,    {\it via}
Eqs.~(\ref{eq:ycpa})-(\ref{eq:ycpd}),
\begin{eqnarray}
K^{uH^0_i}&=&\frac{1}{\sin\beta}U_{2i}\,,\label{eq:rata}\\
\nonumber\\
K^{dH^0_i}&=&\frac{1}{\cos\beta}U_{1i}\,,\\
\nonumber\\
K^{d^\prime H^0_i}&=&\frac{2 m^{\mbox{}}_W}{g\nu_3}U_{3i}\,,\\
\nonumber\\
K^{eH^0_i}&=&\frac{1}{\cos\beta}U_{1i}\,,\\
\nonumber\\
K^{e^\prime H^0_i}&=&\frac{2 m^{\mbox{}}_W}{g\nu_3}U_{3i}\,,
\label{eq:ratae}
\end{eqnarray}
for the scalar Higgs fields, $H^0_i\,$, and
\begin{eqnarray}
K^{uP^0}&=&\label{eq:ratbb}
\frac{1}{\sin\beta}\,\kappa v_{13}\,,\\
\nonumber\\
K^{dP^0}&=&
\frac{1}{\cos\beta}\,\kappa v_{23}\,,\\
\nonumber\\
K^{d^\prime P^0}&=&
\frac{2 m^{\mbox{}}_W}{gv_3}\,\kappa v_{12}\,,\\
\nonumber\\
K^{eP^0}&=&
\frac{1}{\cos\beta}\,\kappa v_{23}\,,\\
\nonumber\\
K^{e^\prime P^0}&=&
\frac{2 m^{\mbox{}}_W}{gv_3}\,\kappa v_{12}\,,
\label{eq:ratb}
\end{eqnarray}
for pseudo-scalar Higgs fields, $P^0\,$.

  For     the    $H^0_{1,2,3}$    exchange    diagrams     shown      in
Figs.~\ref{fig:fusion}(c)  and \ref{fig:fusion}(d) the following  vertex
factors were used
\eqnpict{\scalarss{H^0_i}{\tilde f_A}{\tilde f_B^{\mbox{}^*}}}{
  \kappa^{\tilde f H_i^0}_{AB} = 
  -\,\frac{g m_Z}{\sqrt{1-x_W}}\,K^{\tilde f H_i^0}_{AB}
  \,,\label{eq:cupd}
}
\noindent
where $A,B=L,R\,$.   The $\kappa^{\tilde q H_i^0}_{AB}$ couplings, which
were obtained from Eq.~(\ref{eq:scalpot}), are as follows:
\begin{eqnarray}
%
%
%
\kappa^{\tilde u H_i^0}_{LL}&=&
(\mbox{$\textstyle\frac{{1}}{{18}}$} g^{\prime\prime^2}-
\mbox{$\textstyle\frac{{1}}{{4}}$}g+
\mbox{$\textstyle\frac{{1}}{{12}}$} g^{\prime^2})\,
U_{1i}\,\nu_1
+(\mbox{$\textstyle\frac{{2}}{{9}}$} g^{\prime\prime^2}+
\mbox{$\textstyle\frac{{1}}{{4}}$}g-
\mbox{$\textstyle\frac{{1}}{{12}}$} g^{\prime^2}-\lambda_1^2)\,
U_{2i}\,\nu_2\nonumber\\&&
-\mbox{$\textstyle\frac{{5}}{{18}}$}\, 
g^{\prime\prime^2} U_{3i}\,\nu_3\,,\\
%
%
%
\kappa^{\tilde u H_i^0}_{RR}&=&
(\mbox{$\textstyle\frac{{1}}{{18}}$} g^{\prime\prime^2}-
\mbox{$\textstyle\frac{{1}}{{3}}$} g^{\prime^2})\,
U_{1i}\,\nu_1
+(\mbox{$\textstyle\frac{{2}}{{9}}$} g^{\prime\prime^2}+
\mbox{$\textstyle\frac{{1}}{{3}}$} g^{\prime^2}-\lambda_1^2)\,
U_{2i}\,\nu_2\nonumber\\&&
-\mbox{$\textstyle\frac{{5}}{{18}}$} 
g^{\prime\prime^2} U_{3i}\,\nu_3\,,\\
%
%
%
\kappa^{\tilde u H_i^0}_{LR}&=&
\mbox{$\textstyle\frac{{1}}{{2}}$}\,[
  (U_{3i}\,\nu_1+U_{1i}\,\nu_3)\lambda-\sqrt{2}\,U_{2i}A_u
]\lambda_1\,,\\
%
%
%
\kappa^{\tilde d H_i^0}_{LL}&=&
(\mbox{$\textstyle\frac{{1}}{{18}}$} g^{\prime\prime^2}+
\mbox{$\textstyle\frac{{1}}{{4}}$}g+
\mbox{$\textstyle\frac{{1}}{{12}}$} g^{\prime^2}-\lambda_2^2)\,
U_{1i}\,\nu_1
+(\mbox{$\textstyle\frac{{2}}{{9}}$} g^{\prime\prime^2}-
\mbox{$\textstyle\frac{{1}}{{4}}$}g-
\mbox{$\textstyle\frac{{1}}{{12}}$} g^{\prime^2})\,
U_{2i}\,\nu_2\nonumber\\&&
-\mbox{$\textstyle\frac{{5}}{{18}}$}\, 
g^{\prime\prime^2} U_{3i}\,\nu_3\,,\\
%
%
%
\kappa^{\tilde d H_i^0}_{RR}&=&
-(\mbox{$\textstyle\frac{{1}}{{36}}$} g^{\prime\prime^2}-
\mbox{$\textstyle\frac{{1}}{{6}}$} g^{\prime^2}+\lambda_2^2)\,
U_{1i}\,\nu_1
-(\mbox{$\textstyle\frac{{1}}{{9}}$} g^{\prime\prime^2}+
\mbox{$\textstyle\frac{{1}}{{6}}$} g^{\prime^2})\,
U_{2i}\,\nu_2\nonumber\\&&
+\mbox{$\textstyle\frac{{5}}{{36}}$} 
g^{\prime\prime^2} U_{3i}\,\nu_3\,,\\
%
%
%
\kappa^{\tilde d H_i^0}_{LR}&=&
\mbox{$\textstyle\frac{{1}}{{2}}$}\,[
  (U_{3i}\,\nu_2+U_{2i}\,\nu_3)\lambda-\sqrt{2}\,U_{1i}A_d
]\lambda_2\,,\\
%
%
%
%
\kappa^{\tilde d^\prime H_i^0}_{LL}&=&
-(\mbox{$\textstyle\frac{{1}}{{9}}$} g^{\prime\prime^2}+
\mbox{$\textstyle\frac{{1}}{{6}}$} g^{\prime^2})\,U_{1i}\,\nu_1
-(\mbox{$\textstyle\frac{{4}}{{9}}$} g^{\prime\prime^2}-
\mbox{$\textstyle\frac{{1}}{{6}}$} g^{\prime^2})\,U_{2i}\,\nu_2
\nonumber\\&&
+(\mbox{$\textstyle\frac{{5}}{{9}}$} g^{\prime\prime^2}-
\lambda_5^2)\,U_{3i}\,\nu_3\,,\\
%
%
%
\kappa^{\tilde d^\prime H_i^0}_{RR}&=&
-(\mbox{$\textstyle\frac{{1}}{{36}}$} g^{\prime\prime^2}-
\mbox{$\textstyle\frac{{1}}{{6}}$} g^{\prime^2})\,U_{1i}\,\nu_1
-(\mbox{$\textstyle\frac{{1}}{{9}}$} g^{\prime\prime^2}+
\mbox{$\textstyle\frac{{1}}{{6}}$} g^{\prime^2})\,U_{2i}\,\nu_2
\nonumber\\&&
+(\mbox{$\textstyle\frac{{5}}{{36}}$} g^{\prime\prime^2}-
\lambda_5^2)\,U_{3i}\,\nu_3\,,\\
%
%
%
\kappa^{\tilde d^\prime H_i^0}_{LR}&=&
\mbox{$\textstyle\frac{{1}}{{2}}$}\,[
  (U_{2i}\,\nu_1+U_{1i}\,\nu_2)\lambda-\sqrt{2}\,U_{3i}A_{d^\prime}
]\lambda_5\,,
\end{eqnarray}
for the squark-Higgs couplings, and
\begin{eqnarray}
%
%
%
\kappa^{\tilde e H_i^0}_{LL}\;&=&
-(\mbox{$\textstyle\frac{{1}}{{18}}$} g^{\prime\prime^2}-
\mbox{$\textstyle\frac{{1}}{{4}}$}g+
\mbox{$\textstyle\frac{{1}}{{4}}$} g^{\prime^2}+\lambda_3^2)\,
U_{1i}\,\nu_1
-(\mbox{$\textstyle\frac{{2}}{{9}}$} g^{\prime\prime^2}+
\mbox{$\textstyle\frac{{1}}{{4}}$}g-
\mbox{$\textstyle\frac{{1}}{{4}}$} g^{\prime^2})\,
U_{2i}\,\nu_2\nonumber\\&&
+\mbox{$\textstyle\frac{{5}}{{36}}$}\, 
g^{\prime\prime^2} U_{3i}\,\nu_3\,,\\
%
%
%
\kappa^{\tilde e H_i^0}_{RR}\;&=&
(\mbox{$\textstyle\frac{{1}}{{18}}$} g^{\prime\prime^2}+
\mbox{$\textstyle\frac{{1}}{{2}}$} g^{\prime^2}-\lambda_3^2)\,
U_{1i}\,\nu_1
+(\mbox{$\textstyle\frac{{2}}{{9}}$} g^{\prime\prime^2}-
\mbox{$\textstyle\frac{{1}}{{2}}$} g^{\prime^2})\,
U_{2i}\,\nu_2\nonumber\\&&
-\mbox{$\textstyle\frac{{5}}{{18}}$} 
g^{\prime\prime^2} U_{3i}\,\nu_3\,,\\
%
%
%
\kappa^{\tilde e H_i^0}_{LR}\;&=&
\mbox{$\textstyle\frac{{1}}{{2}}$}\,[
  (U_{3i}\,\nu_2+U_{2i}\,\nu_3)\lambda-\sqrt{2}\,U_{1i}A_e
]\lambda_2\,,\\
%
%
%
\kappa^{\tilde \nu_e H_i^0}_{LL}&=&
-(\mbox{$\textstyle\frac{{1}}{{36}}$} g^{\prime\prime^2}+
\mbox{$\textstyle\frac{{1}}{{4}}$}g+
\mbox{$\textstyle\frac{{1}}{{4}}$} g^{\prime^2})\, U_{1i}\,\nu_1
-(\mbox{$\textstyle\frac{{1}}{{9}}$} g^{\prime\prime^2}-
\mbox{$\textstyle\frac{{1}}{{4}}$}g-
\mbox{$\textstyle\frac{{1}}{{4}}$} g^{\prime^2})\,
U_{2i}\,\nu_2\nonumber\\&&
+\mbox{$\textstyle\frac{{5}}{{36}}$}\, g
^{\prime\prime^2} U_{3i}\,\nu_3\,,\\
%
%
%
\kappa^{\tilde \nu_e H_i^0}_{RR}&=&
\mbox{$\textstyle\frac{{5}}{{36}}$} 
g^{\prime\prime^2} (U_{1i}\,\nu_1+4U_{2i}\,\nu_2-5U_{3i}\,\nu_3)\,,\\
%
%
%
%
\kappa^{\tilde \nu_e H_i^0}_{LR}&=&0\,,\label{eq:cupdy}
\end{eqnarray}
for the slepton-Higgs  couplings.  The mass eigenstate couplings $\tilde
K^{\tilde  f    H^0_i}_{1,2}$,   given      by     Eqs.~(\ref{eq:kcupa})
and~(\ref{eq:kcupb}), were obtained by inserting
\begin{equation}
\left(\begin{array}{c}
\tilde f_L\\\tilde f_R
\end{array}\right)=
\left(\begin{array}{rr}
\cos\theta_{\tilde f}&-\sin\theta_{\tilde f}\\
\sin\theta_{\tilde f}&\cos\theta_{\tilde f}
\end{array}\right)
\left(\begin{array}{c}
\tilde f_1\\\tilde f_2
\end{array}\right)\,,\label{eq:trany}
\end{equation}
which  is just  the inverse   of  Eq.~(\ref{eq:smix}), into  the  scalar
potential.

   The corresponding pseudo-scalar-Higgses couplings
\eqnpict{\scalarss{P^0}{\tilde f_A}{\tilde f_B^{\mbox{}^*}}}{
  i\,\kappa^{\tilde d^\prime P^0}_{AB}=
  -i\,\frac{g m_Z}{\sqrt{1-x_W}}\,K^{\tilde q P^0}_{AB}
  \label{eq:cupe}
}
\noindent
are obtained in a similar fashion as above: {\it i.e.},
\begin{eqnarray}
%
%
%
%
%
\kappa^{\tilde u P^0}_{AB}&=&
-\,\mbox{$\in$\raisebox{-1.0ex}{\scriptsize$AB$}}\,
\frac{\nu_2}{2\,\sqrt{\nu_{12}^2+\nu^2\nu_3^2}}\,
\left[
(\nu_1^2+\nu_3^2)\lambda+\sqrt{2}\,A_u\nu_{3}\cot\beta
\right]
\lambda_1\,,\\&&\nonumber\\
%
%
%
%
\kappa^{\tilde d P^0}_{AB}&=&
-\,\mbox{$\in$\raisebox{-1.0ex}{\scriptsize$AB$}}\,
\frac{\nu_1}{2\,\sqrt{\nu_{12}^2+\nu^2\nu_3^2}}\,
\left[
(\nu_2^2+\nu_3^2)\lambda+\sqrt{2}\,A_d\nu_{3}\tan\beta
\right]
\lambda_2\,,\\&&\nonumber\\
%
%
%
%
\kappa^{\tilde d^\prime P^0}_{AB}&=&
-\,\mbox{$\in$\raisebox{-1.0ex}{\scriptsize$AB$}}\,
\frac{\nu_3}{2\,\sqrt{\nu_{12}^2+\nu^2\nu_3^2}}\,
\left[
\nu^2\lambda+\sqrt{2}\,A_{d^\prime}\frac{\nu_{12}}{\nu_3}
\right]
\lambda_5\,,
\end{eqnarray}
for squark-pseudo-Higgs couplings, and
\begin{eqnarray}
%
%
%
%
\kappa^{\tilde e P^0}_{AB}\;&=&
-\,\mbox{$\in$\raisebox{-1.0ex}{\scriptsize$AB$}}\,
\frac{\nu_1}{2\,\sqrt{\nu_{12}^2+\nu^2\nu_3^2}}\,
\left[
(\nu_2^2+\nu_3^2)\lambda+\sqrt{2}\,A_e\nu_{3}\tan\beta
\right]
\lambda_3\,,\\
%
%
%
\kappa^{\tilde \nu_e P^0}_{AB}&=&0\,,
\end{eqnarray}
for the slepton-pseudo-Higgs couplings, where
\begin{equation}
\mbox{$\in$\raisebox{-1.0ex}{\scriptsize$AB$}}=\left\{
\begin{array}{r@{\;\;\;{\rm if}\;}l}
 1&A=L,\;B=R\\
 0&A=B\\
-1&A=R,\;B=L
\end{array}
\right.\,.
\end{equation}
In general, these couplings  will also mix to  give the  mass eigenstate
couplings  $\tilde K^{\tilde f   P^0}_{1,2}\,$;  which are defined in  a
similar fashion to Eqs.~(\ref{eq:kcupa}) and (\ref{eq:kcupb}).

\subsection{The Widths}
\label{sec-wit}

  In  this  section all  of  the tree level   two-body  decay widths for
$Z_2\,$, $H^0_i\,$, and $P_0$  are computed.  Therefore the generic  two
body decay formula is given by \cite{kn:Griffiths}
\begin{equation}
\begin{array}{cc}
\raisebox{10mm}{$\displaystyle
  \Gamma=\frac{S|{\cal M}_{ab}|^2}{16\pi m_0}\beta_{ab}\,,
$}&
\mbox{\setlength{\unitlength}{1mm}
  \begin{picture}(10,20)
    \put(20,10){\circle{3}}
    \put(20,12){\vector(0,1){8}}
    \put(20,8){\vector(0,-1){8}}
    \put(23,0){$m_b$}
    \put(23,9){$m_0$}
    \put(23,18){$m_a$}
  \end{picture}
}
\end{array}
\label{eq:width}
\end{equation}
where $m_i\,$, $i=0,a,b\,$, are the masses of the particles, $p_i\,$, in
the decay process $p_0\rightarrow p_ap_b\,$,
\begin{equation}
\beta_{ab}=\sqrt{1-\frac{2(m_a^2+m_b^2)}{m_{0}^2}+
\frac{(m_a^2-m_b^2)^2}{m_{0}^4}}\,,
\end{equation}
such that $\beta_{ab}\equiv\beta_a$ if $a=b\,$, $S$ is a symmetry factor
for the out going particles,  $p_a$ and $p_b\,$,  and ${\cal M}_{ab}$ is
the amplitude for the process.

\subsubsection{$\Gamma_{Z_2}$}

For the $Z_2$ width, the following processes need to be computed:
$$
Z_2\longrightarrow \mbox{\footnotesize $W^+W^-,\,Z_1H^0_i,\,
W^\pm H^\mp,$}\,q_i\bar q_i,\,l_i\bar l_i,\,
\tilde\chi^0_i\bar{\tilde\chi}\mbox{}^0_j,\,
\tilde\chi^+_i\tilde\chi^-_j,\,
\tilde q_i\tilde q_j^*,\,
\tilde l_i\tilde l_j^{\mbox{}^*},\,
\mbox{\footnotesize $H^0_iH^0_j,\,H^+H^-,\,
P^0H^0_i$}\,.
$$

  The $Z_2\rightarrow W^+W^-$ width, which can be  found in Hewett and
Rizzo \cite{kn:Hewett}, is given by
\begin{equation}
\Gamma(\mbox{\footnotesize $Z_2\rightarrow W^+W^-$})=
\frac{g^2m^{\mbox{}}_{Z_2}\,\sin^2\phi}{192\pi(1-x_W)}\,
\left(\frac{m^{\mbox{}}_{Z_2}}{m^{\mbox{}}_Z}\right)^4
\beta_W^3
\left[1+
  20\left(\frac{ m^{\mbox{}}_W}{m^{\mbox{}}_Z}\right)^2+
  12\left(\frac{ m^{\mbox{}}_W}{m^{\mbox{}}_Z}\right)^4
\right]\,.
\end{equation}

  The $Z_2\rightarrow q_i\bar   q_i,\, l_i\bar l_i$   vertex factors are
given by
\eqnpict{
   \varepsilon_\mu(q,\lambda)\!\!\!\!\!
   \overlaystuff{
     \gaugeff{Z_2}{u_f(p)}{\bar v_{\bar f}(p^\prime)}
   }{q}{p}{p^\prime}
}{
   -g\gamma_\mu(v_f-a_f\gamma_5)\,,
}
which  were obtained from  Eq.~(\ref{eq:zabvrt})   by converting to  the
$V-A$ basis: {\it i.e.},
\begin{equation}
a\,P_L+b\,P_R=v_f-a_f\gamma_5\,,\label{eq:vabasis}
\end{equation}
where 
\begin{eqnarray}
v_f&=&\mbox{$\textstyle\frac{{1}}{{2}}$}\,(a+b)\,,\\
a_f&=&\mbox{$\textstyle\frac{{1}}{{2}}$}\,(a-b)\,,
\end{eqnarray}
which yields
\begin{eqnarray}
v_f&=&\frac{1}{2\sqrt{1-x_W^{\mbox{}}}}\,
(\tilde C^{fZ_2}_L+\tilde C^{fZ_2}_R)\,,\\
a_f&=&\frac{1}{2\sqrt{1-x_W^{\mbox{}}}}\,
(\tilde C^{fZ_2}_L-\tilde C^{fZ_2}_R)\,.\label{eq:vabasisd}
\end{eqnarray}
Therefore
\begin{eqnarray}
\overline{|{\cal M}_{f\bar f}|^2}&=&
\mbox{$\textstyle\frac{{1}}{{3}}$}\,g^2
\sum_\lambda\sum_{spin}
|\bar v(p^\prime)\gamma_\mu(v_f-a_f\gamma_5)u(p)
\varepsilon^\mu(q,\lambda)|^2\,,
\nonumber\\&=&\mbox{$\textstyle\frac{{4}}{{3}}$}\,g^2
[v_f^2(m^2_{Z_2}+2m_f^2)+a_f^2(m_{Z_2}^2-4m_f^2)]\,.
\end{eqnarray}
Plugging this into Eq.~(\ref{eq:width}) gives
\begin{equation}
\Gamma(Z_2\rightarrow f\bar f)=c_f\,\frac{g^2}{12\pi}\,
m_{Z_2}^{\mbox{}}\beta_f^{\mbox{}}
\left[
v_f^2\left(1+\frac{2m_f^2}{m_{Z_2}^2}\right)+
a_f^2\left(1-\frac{4m_f^2}{m_{Z_2}^2}\right)
\right]\,,
\end{equation}
where $c_f$ is a colour factor which is 3 for quarks and 1 for leptons.

  The  $Z_2\rightarrow  \tilde    q_i\tilde  q_j^*,\, \tilde   l_i\tilde
l_j^{\mbox{}^*}$ vertex factors are given by
\eqnpict{
   \varepsilon_\mu(q,\lambda)\!\!\!\!\!
   \overlaystuff{
     \gaugess{Z_2}{\tilde f_i}{\tilde f_j^*}
   }{q}{\!\!p_i}{\!\!p_j^\prime}
}{
   -ig\kappa_{ij}(p_j^\prime-p_i)^\mu\,,
}
where $\tilde f_{k=1,2}$ are sfermion mass eigenstates, and
\begin{equation}
\kappa_{ij}=\left\{\begin{array}{r@{\;\;;\;{\rm if\;}}l}
v_f+a_f\,\cos2\theta_{\tilde f}& i=j=1\\
v_f-a_f\,\cos2\theta_{\tilde f}& i=j=2\\
   -a_f\,\sin2\theta_{\tilde f}& i\not =j
\end{array}\right.\,.
\end{equation}
The vertex     factor    is obtained   by   follow  steps    similar  to
Eqs.~(\ref{eq:appca})-(\ref{eq:zabvrt}),
\begin{eqnarray}
{\cal L}_{\rm int}&\supseteq&-\,\frac{i}{2}\,
(g_L^{\mbox{}}\hat\tau^a_{ij}L_\mu^a+
g_Y^{\mbox{}}\delta_{ij}\hat Y_{Y_i}Y_\mu
+g_E^{\mbox{}}\hat Y_{E_i}E_\mu)\,\tilde f_i^*\raisebox{1.9ex}{
   \footnotesize$\leftrightarrow$}\!\!\!\!\!\partial^\mu\tilde f_j\\
&\supseteq&\frac{-ig}{\sqrt{1-x_W}}\sum_{i=1}^2\sum_{A=L,R}
\tilde C^{fZ_i}_A\,\tilde f_A^*\raisebox{1.9ex}{
   \footnotesize$\leftrightarrow$}\!\!\!\!\!\partial_\mu
\tilde f_A\,Z^\mu_i\,,
\end{eqnarray}
followed  by transforming  the sfermions to  their  mass eigenstates  by
using Eq.~(\ref{eq:trany}), 
\begin{eqnarray}
{\cal L}_{\rm int}&\supseteq&\frac{-ig}{\sqrt{1-x_W}}\sum_{i=1}^2\{
[\tilde C^{fZ_i}_L\cos^2\theta_{\tilde f}+
\tilde C^{fZ_i}_R\sin^2\theta_{\tilde f}]
\,\tilde f_1^*\raisebox{1.9ex}{
   \footnotesize$\leftrightarrow$}\!\!\!\!\!\partial_\mu
\tilde f_1\nonumber\\&&
+[\tilde C^{fZ_i}_L\sin^2\theta_{\tilde f}+
\tilde C^{fZ_i}_R\cos^2\theta_{\tilde f}]
\,\tilde f_2^*\raisebox{1.9ex}{
   \footnotesize$\leftrightarrow$}\!\!\!\!\!\partial_\mu
\tilde f_2\nonumber\\&&
-\,\mbox{$\textstyle\frac{{1}}{{2}}$}\,[\tilde C^{fZ_i}_L-
\tilde C^{fZ_i}_R]
\sin2\theta_{\tilde f}\,(
\tilde f_1^*\raisebox{1.9ex}{
   \footnotesize$\leftrightarrow$}\!\!\!\!\!\partial_\mu\tilde f_2+
\tilde f_2^*\raisebox{1.9ex}{
   \footnotesize$\leftrightarrow$}\!\!\!\!\!\partial_\mu\tilde f_1
)
\}\,Z^\mu_i\,,
\end{eqnarray}
and     then    changing   to    the     $V-A$     basis,   {\it    via}
Eqs.~(\ref{eq:vabasis})-(\ref{eq:vabasisd}), to get
\begin{equation}
{\cal L}_{\rm int}\supseteq\,-ig\sum_{i,j,k=1}^2\kappa_{ij}\,
\tilde f_i^*\raisebox{1.9ex}{
   \footnotesize$\leftrightarrow$}\!\!\!\!\!\partial_\mu
\tilde f_j\,Z_k^\mu\,.
\end{equation}
Therefore
\begin{eqnarray}
\overline{|{\cal M}_{\tilde f_i^{\mbox{}}\tilde f_j^*}|^2}&=&
\mbox{$\textstyle\frac{{1}}{{3}}$}\,g^2\sum_\lambda
|\varepsilon_\mu(q,\lambda)\kappa_{ij}(p_j^\prime-p_i)^\mu|^2\,,
\nonumber\\
&=&
\mbox{$\textstyle\frac{{1}}{{3}}$}\,g^2\,m_{Z_2}^2\kappa_{ij}^2
\beta_{\tilde f_i\tilde f_j}^2\,.
\end{eqnarray}
Plugging this into Eq.~(\ref{eq:width}) gives
\begin{equation}
\Gamma(Z_2\rightarrow \tilde f_i^{\mbox{}}\tilde f_j^*)=
c_f\frac{g^2\,m_{Z_2}}{48\pi}\,\kappa_{ij}^2\,
\beta_{\tilde f_i\tilde f_j}^2\,.
\end{equation}
   
  For the range  of VEV's that will  be consider here ({\it i.e.}, large
$v_3$  in    particular)   the   $   Z_2\rightarrow  \mbox{\footnotesize
$Z_1H^0_i,\, W^\pm H^\mp,\,H^0_iH^0_j,\,H^+H^-,\,P^0H^0_i$}\,  $  widths
can be approximated by
\begin{equation}
\Gamma(Z_2\rightarrow V+S)\approx
\frac{17g^2 x^{\mbox{}}_W}{864\pi(1- x^{\mbox{}}_W)}
\,m^{\mbox{}}_{Z_2}\,,
\end{equation}
where  the $H^0_iH^0_j$  contributions  are  kinematically forbidden  or
suppressed~\cite{kn:Hewett}.

  The    $Z_2\rightarrow    \tilde\chi^0_i\bar{\tilde\chi}\mbox{}^0_j,\,
\tilde\chi^+_i\tilde\chi^-_j$ widths are quite difficult to compute, due
to the complex nature  of the mass  matrices, and can contribute as much
as  10-20\%     to    the   total   width,    neglecting     phase space
suppression~\cite{kn:Hewett}.  Here its contribution   will be taken  as
15\%; this approximation proved to have no noticeable impact on $L^+L^-$
production.

\subsubsection{$\Gamma_{H^0_i}$}
For the $H^0_i$ widths the following processes need to be computed:
\begin{eqnarray}
H^0_i&\longrightarrow&\mbox{\footnotesize $Z_jZ_k,\,W^+W^-,$}\,
q_j\bar q_j,\,l_j\bar l_j,\,
\tilde\chi^0_j\bar{\tilde\chi}\mbox{}^0_k,\,
\tilde\chi^+_j\tilde\chi^-_k,\,
\tilde q_j\tilde q_k^*,\,
\tilde l_j\tilde l_k^{\mbox{}^*},\,
\mbox{\footnotesize $H^0_jH^0_k,\,H^+H^-,\,P^0P^0$}
\,.\nonumber
\end{eqnarray}

  The $H^0_i\rightarrow Z_jZ_k,\,W^+W^-$ vertex factors are given by
\eqnpict{
   \overlaystuff{
     \scalarvv{H^0_i}{V_a^\mu}{V_b^{\nu^*}}
   }{q}{\!p_a^{\mbox{}}}{\!p_b^{\mbox{}}}
}{
   iC^{H^0_i}_{V_aV_b}\,g_{\mu\nu}\,,
}
where
\begin{eqnarray}
%
%
%
%
%
C^{H^0_i}_{Z_1Z_1}&=&C^{H^0_i}_{ZZ}\,\cos^2\phi-
2\,C^{H^0_i}_{Z Z^\prime}\sin2\phi+
C^{H^0_i}_{Z^\prime Z^\prime}\,\sin^2\phi\,,\\
%
%
%
%
%
%
%
C^{H^0_i}_{Z_1Z_2}&=&
C^{H^0_i}_{Z Z^\prime}\cos2\phi
-(C^{H^0_i}_{Z\,Z\,}-C^{H^0_i}_{Z^\prime Z^\prime})\sin2\phi\,,\\
%
%
%
%
%
C^{H^0_i}_{Z_2Z_2}&=&C^{H^0_i}_{ZZ}\,\sin^2\phi+
2\,C^{H^0_i}_{Z Z^\prime}\sin2\phi+
C^{H^0_i}_{Z^\prime Z^\prime}\,\cos^2\phi\,,
\end{eqnarray}
for the $Z_i$'s, with
\begin{eqnarray}
%
%
%
C^{H^0_i}_{Z\,Z\,}&=&\frac{1}{4}\,
(g\cos\theta^{\mbox{}}_W+g^\prime\sin\theta^{\mbox{}}_W)^2\,
(U_{1i} v_1 + U_{2i} v_2)\,,\\
C^{H^0_i}_{Z Z^\prime}&=&\frac{ g^{\prime\prime}}{6}\,
(g\cos\theta^{\mbox{}}_W+g^\prime\sin\theta^{\mbox{}}_W)\,
(U_{1i} v_1 - 4 U_{2i} v_2)\,,\\
%
%
%
C^{H^0_i}_{Z^\prime Z^\prime}&=&\frac{ g^{\prime\prime^2}}{36}\,
(U_{1i} v_1 + 16 U_{2i} v_2 + 25 U_{3i} v_3)\,,
\end{eqnarray}
and
%
%
\begin{equation}
C^{H^0_i}_{W^+W^-}=\frac{g^2}{2}\,(U_{1i}v_1+U_{2i}v_2)\,,
\end{equation}
for  the  $W$'s.   The   vertex   factors,  $C^{H^0_i}_{V_aV_b}\,$, were
obtained     by      plugging          Eq.~(\ref{eq:sleep})          and
Eqs.~(\ref{eq:basa})-(\ref{eq:basb})   into  the kinetic   terms for the
scalar-Higgs fields, Eq.~(\ref{eq:hke}). Therefore
\begin{eqnarray}
\overline{|{\cal M}_{ab}^i|^2}&=&
\sum_{\lambda_a\lambda_b}|\varepsilon_\mu(p_a,\lambda_a)
C^{H^0_i}_{V_aV_b}\varepsilon_\nu^*(p_b,\lambda_b)g^{\mu\nu}|^2
\nonumber\\&=&
\frac{m^4_{H^0_i}\,
C^{H^0_i}_{V_aV_b^*}\raisebox{2.5ex}{\scriptsize2}}{4(m_am_b)^2}
\left[1-\frac{2(m_a^2+m_b^2)}{m_{H^0_i}^2}+
\frac{(m_a^2+m_b^2)^2+8(m_am_b)^2}{m_{H^0_i}^4}
\right]\,,\;\;\;\;\;
\end{eqnarray}
which yields, {\it via} Eq.~(\ref{eq:width}),
\begin{equation}
\Gamma(H^0_i\rightarrow V_aV_b)=
\frac{S\,C^{H^0_i}_{V_aV_b}\raisebox{2.5ex}{\scriptsize2}
m^3_{H^0_i}\,\beta_{ab}}{64\pi(m_am_b)^2}\,
\left[1-\frac{2(m_a^2+m_b^2)}{m_{H^0_i}^2}+
\frac{(m_a^2+m_b^2)^2+8(m_am_b)^2}{m_{H^0_i}^4}
\right]\,,
\end{equation}
where $S=\mbox{\footnotesize$\frac{{1}}{{2}}$}\,$ for identical $Z_i$'s,
otherwise $S=1$.

  The $H^0_i\rightarrow q_i\bar q_i,\,l_i\bar l_i$ decay width is
\begin{equation}
\Gamma(H^0_i\rightarrow f\bar f)=
\frac{c_fg^2}{32\pi}\left(\frac{m_f}{ m^{\mbox{}}_W}\right)^2K^{fH^0_i}
\raisebox{2.5ex}{\scriptsize2}
\beta_{H^0_i}^3\,m_{H^0_i}^{\mbox{}}\,,
\end{equation}
{\it via} Eq.~(\ref{eq:width}) with amplitude
\begin{equation}
\overline{|{\cal M}_{f\bar f}|^2}=
\frac{g^2}{2}\left(\frac{m_f}{ m^{\mbox{}}_W}\right)^2K^{fH^0_i}
\raisebox{2.5ex}{\scriptsize2}\,
[m_{H^0_i}^2-4m_f^2]\,,
\end{equation}
where the $K^{fH^0_i}$ couplings are defined by Eq.~(\ref{eq:cupb}).

  For the scalar processes $H^0_i\rightarrow\tilde  \phi \phi^*$ the
vertex factor is
\eqnpict{
   \overlaystuff{
     \scalarss{H^0_i}{\phi_a}{\phi_b^*}
   }{q}{\!p_a^{\mbox{}}}{\!p_b^{\mbox{}}}
}{
   C^{H^0_i}_{\phi_a\phi_b}\,,
}
which yields the decay width
\begin{equation}
\Gamma(H^0_i\rightarrow \phi_a\phi_b^*)=
\frac{c_f}{16\pi m_{H^0_i}^{\mbox{}}}\,
|C^{H^0_i}_{\phi_a\phi_b}|^2\beta_{ab}\,,
\label{eq:widdy}
\end{equation}
{\it via} Eq.~(\ref{eq:width}), with amplitude
\begin{equation}
\overline{|{\cal M}_{\phi_a\phi_b}|^2}=
|C^{H^0_i}_{\phi_a\phi_b}|^2\,.
\end{equation}
For  $H^0_i\rightarrow\tilde     q_j\tilde   q_k^*,\,\tilde  l_j\tilde
l_k^{\mbox{}^*}\,$ the vertex factors are
\begin{equation}
C^{H^0_i}_{\tilde f_j\tilde f_k^*}=
\frac{g\,m_Z}{\sqrt{1- x^{\mbox{}}_W}}\,
K^{\tilde f H_i^0}_{jk},
\end{equation}
where    the $K^{\tilde    f   H_i^0}_{jk}$  couplings    are  given  by
Eqs.~(\ref{eq:rata})-(\ref{eq:ratae}).        For      $H^0_i\rightarrow
H^0_jH^0_k,\,H^+H^-,\,P^0P^0$ the vertex factors are:
\begin{eqnarray}
C^{H^0_2}_{H^0_1H^0_1}&=&
\frac{1}{2}\,\lambda A\,
(U_{12} U_{21} U_{31} + U_{11} U_{22} U_{31} + U_{11} U_{21} U_{32})
\nonumber\\&&
+\left\{U_{12} \left[\frac{-1}{24}( g^{\prime\prime^2}+
9g^2+9 g^{\prime^2}) U_{11}^2
-\left(\frac{1}{18} g^{\prime\prime^2}-\frac{1}{8}g^2-
\frac{1}{8} g^{\prime\prime^2}\right) U_{21}^2
\right.\right.
\nonumber\\&&
\left.+\frac{5}{72} g^{\prime\prime^2} U_{31}^2-\frac{1}{2}\lambda^2
(U_{21}^2  + U_{31}^2)\right]
+ U_{11} \left[
\left(\frac{-1}{18} g^{\prime\prime^2}+\frac{1}{8}g^2+
\frac{1}{8} g^{\prime^2}\right)  U_{21}
U_{22}
\right.
\nonumber\\&&
+\left.\left.\frac{5}{36} g^{\prime\prime^2}  U_{31} U_{32}
-\lambda^2(U_{21} U_{22} + U_{31} U_{32})\right]\right\}v_1
\nonumber\\&&
+\left\{U_{22}\left[
\left(\frac{-1}{18} g^{\prime\prime^2}+\frac{1}{8}g^2+
\frac{1}{8} g^{\prime\prime^2}\right) U_{11}^2
-\frac{1}{24}(16 g^{\prime\prime^2}+9g^2+9 
g^{\prime^2}) U_{21}^2\right.\right.
\nonumber\\&&
+\left.\frac{5}{18} g^{\prime\prime^2} U_{31}^2
-\frac{1}{2}\lambda^2(U_{11}^2  + U_{31}^2 )\right]
+U_{21} \left[
\left(\frac{-1}{9} g^{\prime\prime^2}+
\frac{1}{4}g^2+\frac{1}{4} g^{\prime^2}\right)U_{11}U_{12}
\right.
\nonumber\\&&
+\left.\left.\frac{5}{9} g^{\prime\prime^2} U_{31} U_{32}
-\lambda^2 (U_{11} U_{12} + U_{31} U_{32})\right]\right\}v_2
\nonumber\\&&
+\left\{U_{31}\left[
\frac{5}{36} g^{\prime\prime^2}(U_{11}U_{12}+4U_{21}U_{22})
-\lambda^2(U_{11}U_{12}+U_{21} U_{22})\right]\right.
\nonumber\\&&
+\left.U_{32}\left[\frac{5}{72} 
g^{\prime\prime^2}(U_{11}^2 +4U_{21}^2- 15 U_{31}^2 )
-\frac{1}{2} (U_{11}^2   + U_{21}^2  )\right]\right\}v_3
\,,\\
C^{H^0_3}_{H^0_1H^0_1}&=&
\frac{1}{2}\,\lambda A\,
(U_{13} U_{21} U_{31} + U_{11} U_{23} U_{31} + U_{11} U_{21} U_{33})
\nonumber\\&&
+\left\{U_{13}\left[\frac{-1}{24}( g^{\prime\prime^2}+9g^2+
9 g^{\prime^2}) U_{11}^2
-\left(\frac{1}{18} g^{\prime\prime^2}-\frac{1}{8}g^2-
\frac{1}{8} g^{\prime\prime^2}\right) U_{21}^2
\right.\right.
\nonumber\\&&
\left.+\frac{5}{72} g^{\prime\prime^2} U_{31}^2-\frac{1}{2}\lambda^2
(U_{21}^2  + U_{31}^2)\right]
+U_{11} \left[
\left(\frac{-1}{9} g^{\prime\prime^2}+\frac{1}{4}g^2+
\frac{1}{4} g^{\prime^2}\right)  U_{21}
U_{23}
\right.
\nonumber\\&&
+\left.\left.\frac{5}{36} g^{\prime\prime^2}  U_{31} U_{33} 
-\lambda^2(U_{21} U_{23} + U_{31} U_{33})\right]\right\}v_1
\nonumber\\&&
+\left\{U_{23}\left[
\left(\frac{-1}{18} g^{\prime\prime^2}+\frac{1}{8}g^2+
\frac{1}{8} g^{\prime\prime^2}\right) U_{11}^2
-\frac{1}{24}(16 g^{\prime\prime^2}+9g^2+9 
g^{\prime^2}) U_{21}^2\right.\right.
\nonumber\\&&
+\left.\frac{5}{18} g^{\prime\prime^2} U_{31}^2
-\frac{1}{2}\lambda^2(U_{11}^2  + U_{31}^2 )\right]
+U_{21} \left[
\left(\frac{-1}{9} g^{\prime\prime^2}+
\frac{1}{4}g^2+\frac{1}{4} g^{\prime^2}\right)U_{11}U_{13}
\right.
\nonumber\\&&
+\left.\left.\frac{5}{9} g^{\prime\prime^2} U_{31}U_{33}
-\lambda^2 (U_{11} U_{13} + U_{31} U_{33})\right]\right\}v_2
\nonumber\\&&
+\left\{U_{31}\left[
\frac{5}{36} g^{\prime\prime^2}(U_{11}U_{13}+4U_{21}U_{23})
-\lambda^2(U_{11}U_{13}+U_{21} U_{23})\right]\right.
\nonumber\\&&
+\left.U_{33}\left[\frac{5}{72} 
g^{\prime\prime^2}(U_{11}^2 +4U_{21}^2- 15 U_{31}^2 )
-\frac{1}{2} (U_{11}^2   + U_{21}^2  )\right]\right\}v_3
\,,\\
C^{H^0_3}_{H^0_1H^0_2}&=&
\frac{1}{2}\,\lambda A\,
(U_{13} U_{22} U_{31} + U_{12} U_{23} U_{31} + U_{13} U_{21} U_{32} 
\nonumber\\&& 
+ U_{11} U_{23} U_{32} + U_{12} U_{21} U_{33} + U_{11} U_{22} U_{33}) 
\nonumber\\&& 
-\left\{\left(\frac{1}{9} g^{\prime\prime^2}-
\frac{1}{4}g^2-\frac{1}{4} g^{\prime\prime^2}\right)
(U_{12} U_{21} + U_{11} U_{22}) U_{23} \right.
\nonumber\\&& 
+U_{13}\left[\frac{1}{12}( g^{\prime\prime^2}+9g^2+
9 g^{\prime^2})U_{11} U_{12} 
+\left(\frac{1}{9} g^{\prime\prime^2}-
\frac{1}{4}g^2-\frac{1}{4} g^{\prime\prime^2}\right)
U_{21} U_{22} \right.
\nonumber\\&& 
\left.
-\frac{5}{36}\, g^{\prime\prime^2} U_{31} U_{32}
+\lambda^2 (U_{21} U_{22} + U_{31} U_{32})\right]
-\left.\frac{5}{36}\, g^{\prime\prime^2} 
(U_{12} U_{31} + U_{11} U_{32}) 
U_{33}\right\}v_1
\nonumber\\&& 
-\lambda^2
(U_{12} U_{21} U_{23} + U_{11} U_{22} U_{23} + U_{12} U_{31} U_{33} 
+ U_{11} U_{32} U_{33})v_1
\nonumber\\&& 
-\left\{\left(\frac{1}{9} g^{\prime\prime^2}-
\frac{1}{4}g^2-\frac{1}{4} g^{\prime\prime^2}\right)
 U_{13} (U_{12} U_{21} + U_{11} U_{22}) \right.
\nonumber\\&& 
+U_{23}\left[
\left(\frac{1}{9} g^{\prime\prime^2}-\frac{1}{4}g^2-
\frac{1}{4} g^{\prime\prime^2}\right)
U_{11} U_{12}
+\left(\frac{4}{3} g^{\prime\prime^2}+\frac{1}{3}g^2+
\frac{1}{3} g^{\prime\prime^2}\right) 
U_{21} U_{22}
\right.
\nonumber\\&& 
-\left.\frac{5}{9}\, g^{\prime\prime^2} U_{31} U_{32}
+\lambda^2(U_{11} U_{12} + U_{31} U_{32})\right]
-\left.\frac{5}{9}\, g^{\prime\prime^2} 
(U_{22} U_{31} + U_{21} U_{32}) U_{33}\right\}v_2
\nonumber\\&& 
-\lambda^2
(U_{12} U_{13} U_{21} + U_{11} U_{13} U_{22} + 
U_{22} U_{31} U_{33} + U_{21} U_{32} U_{33})v_2
\nonumber\\&& 
+\left\{
\left[\frac{5}{36}\, 
g^{\prime\prime^2}(U_{12} U_{13} + 4 U_{22} U_{23})
-\lambda^2(U_{12} U_{13} + U_{22} U_{23})\right] U_{31}
\right.
\nonumber\\&& 
+\left[\frac{5}{36}\, 
g^{\prime\prime^2}(U_{11} U_{13} + 4 U_{21} U_{23})
-\lambda^2(U_{11} U_{13} + U_{21} U_{23})\right] U_{32}
\nonumber\\&& 
\left.
+\left[\frac{5}{36}\, g^{\prime\prime^2}(U_{11} U_{12} + 
4 U_{21} U_{22} - 15 U_{31} U_{32})
-\lambda^2(U_{11} U_{12} + U_{21} U_{22})\right] U_{33}
\right\}v_3
\,,\\
C^{H^0_3}_{H^0_2H^0_2}&=&
\frac{1}{2}\,\lambda A\,
(U_{13} U_{22} U_{32} + U_{12} U_{23} U_{32} + U_{12} U_{22} U_{33})
\nonumber\\&&
+\left\{U_{13}\left[\frac{-1}{24}
( g^{\prime\prime^2}+9g^2+9 g^{\prime^2}) U_{12}^2
-\left(\frac{1}{18} g^{\prime\prime^2}-
\frac{1}{8}g^2-\frac{1}{8} g^{\prime\prime^2}\right) U_{22}^2
\right.\right.
\nonumber\\&&
\left.+\frac{5}{72} g^{\prime\prime^2} U_{32}^2-\frac{1}{2}\lambda^2
(U_{22}^2  + U_{32}^2)\right]
+U_{12} \left[
\left(\frac{-1}{9} g^{\prime\prime^2}+
\frac{1}{4}g^2+\frac{1}{4} g^{\prime^2}\right)  U_{22}
U_{23}
\right.
\nonumber\\&&
+\left.\left.\frac{5}{36} g^{\prime\prime^2}  U_{32} U_{33} 
-\lambda^2(U_{22} U_{23} + U_{32} U_{33})\right]\right\}v_1
\nonumber\\&&
+\left\{U_{23}\left[
\left(\frac{-1}{18} g^{\prime\prime^2}+\frac{1}{8}g^2+
\frac{1}{8} g^{\prime\prime^2}\right) U_{12}^2
-\frac{1}{24}(16 g^{\prime\prime^2}+9g^2+
9 g^{\prime^2}) U_{22}^2\right.\right.
\nonumber\\&&
+\left.\frac{5}{18} g^{\prime\prime^2} U_{32}^2
-\frac{1}{2}\lambda^2(U_{12}^2  + U_{32}^2 )\right]
+U_{22} \left[
\left(\frac{-1}{9} g^{\prime\prime^2}+
\frac{1}{4}g^2+\frac{1}{4} g^{\prime^2}\right)U_{12}U_{13}
\right.
\nonumber\\&&
+\left.\left.\frac{5}{9} g^{\prime\prime^2} U_{32}U_{33}
-\lambda^2 (U_{12} U_{13} + U_{32} U_{33})\right]\right\}v_2
\nonumber\\&&
+\left\{U_{32}\left[
\frac{5}{36} g^{\prime\prime^2}(U_{12}U_{13}+4U_{22}U_{23})
-\lambda^2(U_{12}U_{13}+U_{22} U_{23})\right]\right.
\nonumber\\&&
+\left.U_{33}\left[\frac{5}{72} 
g^{\prime\prime^2}(U_{12}^2 +4U_{22}^2- 15 U_{32}^2 )
-\frac{1}{2} (U_{12}^2   + U_{22}^2  )\right]\right\}v_3
\,,
\end{eqnarray}
for the neutral-scalar-Higgses;
\begin{eqnarray}
%
%
%
%
%
C^{H^0_i}_{H^+H^-}&=&\frac{-1}{4\,(1+\cot^2\beta)}\left\{
v_1(3g^2+ g^{\prime^2}-4\lambda^2)(U_{1i}+U_{2i}\cot\beta)
\right.
\nonumber\\&&
+\left(g^2- g^{\prime^2}+\frac{4}{9} g^{\prime\prime^2}\right)
(v_1U_{1i}\cot^2\beta+v_2U_{2i})
\nonumber\\&&
+\frac{1}{9} g^{\prime\prime^2}[ v_1(v_1U_{1i}+16U_{2i}\cot\beta)-5v_3
(1+4\cot^2\beta)U_{3i}]
\nonumber\\&&
+4\,(\lambda^2v_3(1+\cot^2\beta)+\lambda \cot\beta)U_{3i}
\,,
\end{eqnarray}
for the charged-scalar-Higgses;
\begin{eqnarray}
C^{H^0_i}_{P^0P^0}&=&\frac{v_3^2}{2(v_1^2v_2^2+v^2v_3^2)}\left\{
\left[m_Z^2\cos2\beta-
\frac{1}{9}\left(\frac{ g^{\prime\prime}}{g}\right)^2m_W^2-\lambda^2
\frac{v_1^2v_2^2}{v_3^2}\right]
(v_1U_{1i}+v_2U_{2i})\right.
\nonumber\\&&
-\lambda(v_1^3U_{1i}+v_2^3U_{2i})
-\lambda^2v_3(v_1^2+v_2^2)U_{3i}
-\lambda A\frac{v_1v_2}{v_3}
(v_1U_{1i}+v_2U_{2i}+v_3U_{3i})
\nonumber\\&&\left.
+\frac{5}{36} g^{\prime\prime^2}\left[\frac{v_1^2v_2^2}{v_3^2}
(v_1U_{1i}+4v_2U_{2i}-5v_3U_{3i})
+v_3(4v_1^2+v_2^2)U_{3i}\right]\right\}\,,
\end{eqnarray}
for the pseudo-scalar-Higgses, which were    all extracted by   plugging
Eqs.~(\ref{eq:basa})-(\ref{eq:basb})  for the physical Higgs fields into
the Higgs potential, Eq.~(\ref{eq:fox}).

  The    $H^0_i\rightarrow  \tilde\chi^0_j\bar{\tilde\chi}\mbox{}^0_k,\,
\tilde\chi^+_j\tilde\chi^-_k$  decay processes are  quite complicated to
compute.   Here  a   simple  approximation   was   made  in  which   for
$m^{\mbox{}}_{P^0}\,\raisebox{-0.625ex}{$\stackrel{<}{\sim}$}\,    {\cal
O}({500})\,GeV$ its contribution to  the width was 15\%, otherwise 50\%.
This  addition  had a  negligible affect  on $L^+L^-$  production, since
$m^{\mbox{}}_{P^0}=200\,GeV$.

\subsubsection{$\Gamma_{P^0}$}

For the $P^0$ width the following processes need to be computed:
\begin{eqnarray}
P^0&\longrightarrow&Z_iZ_j,\,W^\pm H^\mp,\,
q_i\bar q_i,\,l_i\bar l_i,\,
\tilde\chi^0_i\bar{\tilde\chi}\mbox{}^0_j,\,
\tilde\chi^+_i\tilde\chi^-_j,\,
\tilde q_i\tilde q_j^*,\,
\tilde l_i\tilde l_j^{\mbox{}^*}
\,.\nonumber
\end{eqnarray}

  For $P^0\rightarrow Z_iZ_j,\,W^\pm H^\mp$  the  widths are zero  since
here $m^{\mbox{}}_{P^0}<m^{\mbox{}}_{Z_2}$ and $m^{\mbox{}}_{P^0}\approx
m^{\mbox{}}_{H^\pm}\,$: see Figs.~\ref{fig:mztwo}-\ref{fig:mhzerob}  and
discussion therein.

  The $P^0\rightarrow q_i\bar q_i,\,l_i\bar l_i$ decay widths are
\begin{equation}
\Gamma(P^0\rightarrow f\bar f)=
\frac{c_fg^2}{32\pi}\left(\frac{m_f}{ m^{\mbox{}}_W}\right)^2K^{fP^0}
\raisebox{2.5ex}{\scriptsize2}
\beta_{P^0}^{\mbox{}}m_{P^0}^{\mbox{}}\,,
\end{equation}
{\it via} Eq.~(\ref{eq:width}), with amplitudes
\begin{equation}
\overline{|{\cal M}_{f\bar f}|^2}=
\frac{g^2}{2}\left(\frac{m_f}{ m^{\mbox{}}_W}\right)^2K^{fP^0}
\raisebox{2.5ex}{\scriptsize2}\,m_{P^0}^2\,,
\end{equation}
where the $K^{fP^0}$ couplings defined by Eq.~(\ref{eq:cupc}).

  The  $P^0\rightarrow\tilde     q_j\tilde   q_k^*,\,\tilde  l_j\tilde
l_k^{\mbox{}^*}\,$ decay widths are
\begin{equation}
\Gamma(P^0\rightarrow \tilde f_j\tilde f_k^*)=
\frac{c_fg^2\,m_Z^2}{16\pi(1- x^{\mbox{}}_W)m_{P^0}^{\mbox{}}}
K^{\tilde f P^0}_{jk}\raisebox{2.5ex}{\scriptsize2}
\beta_{\tilde f_j\tilde f_k}\,,
\end{equation}
{\it via} Eq.~(\ref{eq:widdy}), with vertex factors
\begin{equation}
C^{P^0}_{\tilde f_j\tilde f_k^*}=
\frac{g\,m_Z}{\sqrt{1- x^{\mbox{}}_W}}\,
K^{\tilde f P^0}_{jk}\,,
\end{equation}
where    the   $K^{\tilde  f  P^0}_{jk}$    couplings    are  given   by
Eqs.~(\ref{eq:ratbb})-(\ref{eq:ratb}).

   In this work $m^{\mbox{}}_{P^0}$  was  fixed at $200\,GeV$.  At  this
mass  $P^0\rightarrow   \tilde   \chi^0_i  \bar{\tilde\chi}  \mbox{}^0_j
,\,\tilde\chi^+_i\tilde\chi^-_j$ decays are suppressed~\cite{kn:Gunion}.

%
%
%
%

\end{document}